\documentclass[10pt,twoside,fleqn]{amsart}

\usepackage{etex}
\reserveinserts{50}
\usepackage[main=english,ngerman]{babel}
\usepackage{scrextend}
\usepackage{import}

\usepackage[latin1]{inputenc}
\usepackage[T1]{fontenc}
\usepackage{nicefrac}

\usepackage[varqu,varl]{zi4}
\usepackage[largesc,tighter]{newpxtext}
\usepackage[defaultsans,scale=0.9]{comfortaa}
\usepackage[varg,slantedGreek,bigdelims,upint,nosymbolsc]{newpxmath}

\usepackage{bm}

\usepackage[sans]{dsfont}
\usepackage{yfonts}
\usepackage{nicefrac}
\usepackage{ragged2e}

\usepackage[babel,german=guillemets,english=american]{csquotes}

\usepackage[
a4paper,
driver=pdftex,
includehead=true,
includefoot=true,
top=1.5cm,
bottom=1.5cm,
inner=2.5cm,
outer=2.5cm,
marginparsep=0.5cm,
marginparwidth=2cm,
footskip=2.5\baselineskip,
headheight=1\baselineskip,
headsep=1.25\baselineskip,
heightrounded,
centering
]{geometry}

\usepackage[
activate={true},
final,
auto=true,
protrusion=true,
expansion=true,
kerning=true,
tracking=true,
spacing=true,
verbose=silent,
babel=true]
{microtype}
\SetTracking[spacing = {25*,166, } ]{ encoding = *, shape = sc }{-5}

\usepackage{setspace}
\allowdisplaybreaks[4]

\usepackage{marginfix} 
\usepackage[multiple,stable]{footmisc}

\deffootnote[1.75em]{0em}{0em}{\makebox[1.5em][l]{\textcolor{Blue}{\rule[-0.5mm]{0.75pt}{8.25pt}}\,{\thefootnotemark}}\tiny{\,}}
\deffootnotemark{\tiny{\,}\textcolor{Blue}{\rule[-0.35mm]{0.75pt}{9.75pt}}\raisebox{2.25pt}{\footnotesize\,\thefootnotemark}}

\usepackage{endnotes}

\usepackage[compact,calcwidth,rubberchapters]{titlesec}
\usepackage{titletoc}

\renewcommand\thesection{\arabic{section}}
\renewcommand\thesubsection{\Alph{subsection}}

\newcommand{\hsp}{\hspace{15pt}}
\newcommand{\hspp}{\hspace{5pt}}

\titleformat{\section}[hang]
{\LARGE}
{\hsp\textcolor{gray}{\raisebox{-3.75pt}{\fontsize{28}{26}\selectfont\sffamily\thesection.}\hspp}}
{0em}
{\sf\bfseries\Large}

\titleformat{\subsection}[hang]
{\slshape\large}
{\hsp\makebox[1.75em][l]{\textcolor{gray}{\sl{\LARGE{\sffamily\S}}\large\sffamily\bfseries\thesubsection}}}
{0em}
{\sf\large\slshape}

\titlespacing*{\section}
{0pt}
{1.75\baselineskip plus 0.125\baselineskip minus 0.0625\baselineskip}
{0.25\baselineskip plus 0.0625\baselineskip}

\titlespacing*{\subsection}
{0pt}
{1\baselineskip plus 0.125\baselineskip minus 0.0625\baselineskip}
{0\baselineskip plus 0.0625\baselineskip}

\titlecontents{section}[1.25em]
{\vspace{0.25\baselineskip}\sf\bfseries}
{\contentslabel[{\textcolor{gray}{\sf\bfseries\thecontentslabel}}\quad]{1.25em}}
{}
{\quad\sf\textit{\contentspage}}[]

\titlecontents*{subsection}[0em]
{\vspace{-0.5\baselineskip}\small\sf\itshape}
{}
{}
{}
[\, {\raisebox{.0ex}{$\diamond$}}\, ]

\contentsmargin[0em]{0em}

\makeatletter
\def\cleardoublepage{\clearpage\if@twoside \ifodd\c@page\else
	\hbox{} 
	\vspace*{\fill}
	\thispagestyle{empty}
	\newpage
	\if@twocolumn\hbox{}\newpage\fi\fi\fi}
\makeatother

\usepackage{etoolbox}
\makeatletter
\patchcmd{\ttl@select}{\strut}{}{}{}
\patchcmd{\ttlh@hang}{\strut}{}{}{}
\patchcmd{\ttlh@hang}{\strut}{}{}{}
\makeatother

 \usepackage{xpatch}
 \xapptocmd\normalsize{%
  \abovedisplayskip=0.5\baselineskip plus 0.25\baselineskip minus 0.125\baselineskip
  \abovedisplayshortskip=0.125\baselineskip plus 0.05\baselineskip minus 0.05\baselineskip
  \belowdisplayskip=0.5\baselineskip plus 0.25\baselineskip minus 0.125\baselineskip
  \belowdisplayshortskip=0.125\baselineskip plus 0.05\baselineskip minus 0.05\baselineskip
 }{}{}

\makeatletter
\patchcmd{\ttl@straight@ii}{\vspace{\@tempskipb}}{\vskip \@tempskipb}{}{}
\makeatother

\usepackage[maxfloats=50]{morefloats}
\usepackage{graphicx}
\usepackage{multirow,longtable,booktabs,tabularx,tabulary}
\usepackage[usenames,dvipsnames,table]{xcolor}

\usepackage{tikz}

\usepackage{pifont}
\usepackage[misc]{ifsym}
\usepackage{bbding}

\usepackage{marvosym}
\usepackage{dictsym}
\usepackage[htt]{hyphenat}

\usepackage[version=3]{mhchem}

\usepackage{fixltx2e}
\MakeRobust{\eqref}

\usepackage{mathtools}
\usepackage{nccmath}
\usepackage{enumitem}
\usepackage{ulem}

\usepackage{multicol}
\usepackage[noorphans,vskip=-0.25\baselineskip plus 0.125\baselineskip minus 0.0625\baselineskip]{quoting}

\usepackage{fontawesome}

\usepackage{caption}
\usepackage{floatrow}

\DeclareNewFloatType{Beispiel}{placement=tbph,within=section,fileext=bsp}
\DeclareNewFloatType{Details}{placement=tbph,within=section,fileext=det}

\addto{\captionsngerman}{%
  \renewcommand*{\contentsname}{Inhalt}

}

\addto{\captionsenglish}{%
  \renewcommand*{\contentsname}{Contents}

}

\usepackage[
backend=bibtex,
natbib=true,
autocite=superscript,
style=numeric-comp,
ibidtracker=true,
idemtracker=true,
maxnames=99,
doi=true,
url=true,
isbn=false,
sorting=none,
autopunct=false,
backref,
mcite]{biblatex}

\DeclareFieldFormat{pages}{#1}
\DeclareFieldFormat{authortype}{\textsc{#1}}
\DeclareFieldFormat{author}{\textsc{#1}}
\DeclareFieldFormat{title}{\mkbibquote{\normalfont #1}}
\DeclareFieldFormat{journaltitle}{\normalfont{#1}}
\DeclareFieldFormat{booktitle}{\normalfont #1}
\DeclareFieldFormat{year}{\normalfont\useosf{#1}}

\renewbibmacro{in:}{\ifentrytype{article}{}{\printtext{\bibstring{in}\intitlepunct}}}

\DeclareFieldFormat{postnote}{\mkpageprefix[pagination]{#1}}

\AtBeginDocument{%

}

\DeclareCiteCommand{\citenum}
  {}
  {\printfield{labelnumber}}
  {}
  {}

\DefineBibliographyStrings{german}{%
  references = {Zitierte Literatur},
}

\DeclareFieldFormat{labelnumberwidth}{\textcolor{OliveGreen}{\sf\bfseries{#1}{\tiny\,}\raisebox{0.625pt}{)}}\hspace*{-5pt}}

\DeclareCiteCommand{\supercite}
[\mkbibsuperscript]%
{\usebibmacro{cite:init}%
\iffieldundef{prenote}%
{}%
{\BibliographyWarning{Ignoring prenote argument}}%
\iffieldundef{postnote}%
{}%
{\BibliographyWarning{Ignoring postnote argument}}%
{\hspace*{-0.75pt}\rotatebox[origin=c]{45}{\textcolor{OliveGreen}{\normalsize\bf\MVRightarrow}}\hspace*{-3.25pt}}
}%
{\sf\bfseries\usebibmacro{citeindex}%
\usebibmacro{cite:comp}}
{}%
{\usebibmacro{cite:dump}
}%
\newbibmacro{string+doiurlisbn}[1]{%
  \iffieldundef{doi}{%
    \iffieldundef{url}{%
      \iffieldundef{isbn}{%
        \iffieldundef{issn}{%
          #1%
        }{%
          \href{http://books.google.com/books?vid=ISSN\thefield{issn}}{#1}%
        }%
      }{%
        \href{http://books.google.com/books?vid=ISBN\thefield{isbn}}{#1}%
      }%
    }{%
      \href{\thefield{url}}{#1}%
    }%
  }{%
    \href{http://dx.doi.org/\thefield{doi}}{#1}%
  }%
}

\newcommand*\extlinktext{\textcolor{BrickRed}{\small\faExternalLink}}
\AtBeginBibliography{\renewcommand*\extlinktext{\textcolor{BrickRed}{\scriptsize\faExternalLink}}}

\DeclareFieldFormat*{url}{\usebibmacro{string+doiurlisbn}\extlinktext\nolinebreak[4]\nopunct}
\DeclareFieldFormat*{doi}{\usebibmacro{string+doiurlisbn}\extlinktext\nolinebreak[4]\nopunct}
\DeclareFieldFormat*{urldate}{}

\usepackage[
colorlinks,
hyperfigures,
breaklinks=true,
raiselinks,
nesting,
hyperindex,
pdfview=Fit,
bookmarksnumbered,
bookmarksopen,
hyperfootnotes=false,
pdfstartview=FitH,
linkcolor=Blue,
citecolor=OliveGreen,
urlcolor=BrickRed,
filecolor=yellow,
pdfauthor={Thomas Grohmann},
pdfcreator={Thomas Grohmann},
pdftitle={A systematic four-dimensional approach to strong field control of molecular torsions},
]{hyperref}

\usepackage{fancyhdr}
\usepackage{calc}
\usepackage{lastpage}

\usepackage{colortbl}
\fancyheadoffset[RO]{0cm}
\fancypagestyle{plain}{\fancyhf{}
\fancyfoot[RO,LE]{\textcolor{gray}{\LARGE{\sf{\thepage}}}}}

\renewcommand{\sectionmark}[1]{\markright{#1}}

\def\be{\begin{equation}}
\def\ee{\end{equation}}
\def\bse{\begin{subequations}}
\def\ese{\end{subequations}}
\def\beo{\begin{equation*}}
\def\eeo{\end{equation*}}
\def\bea{\begin{eqnarray}}
\def\eea{\end{eqnarray}}
\def\beao{\begin{eqnarray*}}
\def\eeao{\end{eqnarray*}}
\def\beqo{\begin{quote}}
\def\enqo{\end{quote}}
\def\ben{\begin{enumerate}}
\def\een{\end{enumerate}}
\def\bit{\begin{itemize}}
\def\eit{\end{itemize}}
\def\bed{\begin{description}}
\def\eed{\end{description}}
\def\berale{\begin{raggedright}}
\def\eerale{\end{raggedright}}

\def\tssc#1{\textsuperscript{#1}}

\def\mal#1{\mathcal #1}
\def\mak#1{\mathfrak #1}

\def\bra#1{ \langle #1 \lvert}
\def\ket#1{\rvert #1\rangle}
\def\braket#1{\langle #1 \rangle}

\definecolor{peru}{RGB}{205,133,63} 
\definecolor{meergruen}{RGB}{46,139,87}
\definecolor{dunkel-blau}{RGB}{0,0,139}
\definecolor{dunkel-magenta}{RGB}{139,0,139}
\definecolor{rot}{rgb}{0.5,0,0}
\definecolor{grun}{rgb}{0,0.3,0}
\definecolor{pale-brown}{RGB}{152,118,54}
\definecolor{copper}{RGB}{184,115,51}
\definecolor{bronze}{RGB}{205,127,50}
\definecolor{wheat}{RGB}{245,222,179}
\definecolor{tan}{RGB}{210,180,140}
\definecolor{orche}{RGB}{204,119,34}
\definecolor{corn}{RGB}{251,236,93}
\definecolor{golden-yellow}{RGB}{255,223,0}
\definecolor{dark-olive}{RGB}{85,107,47}
\definecolor{dark-orange}{RGB}{238,173,14}
\definecolor{dark-brown}{RGB}{92,64,51}

\newcolumntype{C}[1]{>{\centering\arraybackslash}m{#1}}
\newcolumntype{R}[1]{>{\raggedleft\arraybackslash}m{#1}}
\newcolumntype{L}[1]{>{\raggedright\arraybackslash}m{#1}}
\newcolumntype{Z}[1]{>{\arraybackslash}p{#1}}

\newtagform{brackets2}[\sf\color{gray}]{\textcolor{gray}{\sf Eq.{\tiny\,}}{}}{{}}
\usetagform{brackets2}

\let\eqref\ref
\let\cite\autocite

\begin{document}
\setlength{\topskip}{1\baselineskip plus 0.125\baselineskip minus 0.0625\baselineskip}
\setlength{\parskip}{0.75\baselineskip plus 0.05\baselineskip minus 0.05\baselineskip}
\setlength{\parindent}{0pt}
\setlength{\footnotesep}{0.75\baselineskip plus 0.125\baselineskip minus 0.075\baselineskip}
\flushbottom

\setlength\mathindent{4.5em}

\selectlanguage{english} 
\pagestyle{fancy}

\renewcommand{\contentsname}{\RaggedRight\Large\sf\bfseries Contents\vspace*{-0.25\baselineskip}}

\thispagestyle{plain}
\renewcommand{\sectionmark}[1]{\markright{#1}}
\renewcommand{\subsectionmark}[1]{\markleft{#1}}


\setstretch{1.95}
{\Huge\sf A systematic four-dimensional approach to strong field control of molecular torsions}\\[-1.05\baselineskip]

\noindent{\sf\large {Thomas Grohmann}\tssc{1,2}, Monika Leibscher\tssc{3,4}, Tamar Seideman\tssc{1}}

\setstretch{1.125}
{\sf
\makebox[0.5em][l]{\tssc{1}}\textit{\small Department of Chemistry, Northwestern University, 2145 Sheridan Rd, Evanston, Illinois 60208, USA}\\
\makebox[0.5em][l]{\tssc{2}}\textit{\small Institut f\"ur Chemie und Biochemie, Freie Universität Berlin, Takustr. 3, 14195 Berlin}\\
\makebox[0.5em][l]{\tssc{3}}\textit{\small Institut f\"ur Theoretische Physik, Leibniz Universit{\"a}t Hannover, Appelstr. 2, Hannover, Germany}\\
\makebox[0.5em][l]{\tssc{4}}\textit{\small Institut f\"ur Physikalische Chemie, Christian-Albrechts-Universität Kiel, Olshausenstraße 40, 24098 Kiel, Germany}}

\vspace*{0.75\baselineskip}

\hfill\textcolor{gray}{\sf\itshape{\today}}
\vspace*{1.25\baselineskip}

\textit{A modified version of this manuscript will be submitted for publication in The Journal of Chemical Physics soon. Please feel free to send any comments or question to 
thomas.grohmann@fu-berlin.de.}

\vspace*{0.25\baselineskip}

\textcolor{Gray}{\noindent\hfill{\rule{\textwidth}{1.25pt }}\hfill}

\vspace*{-0.25\baselineskip}

We introduce a four-dimensional quantum model for describing the torsional control of $\rm G_{16}$-type molecules in the electronic ground state, based on the 
symmetry-adapted variational method. We define conditions for which lower-dimensional models, commonly used to simulate the strong-field control of molecular torsions, are 
reliable approximations to a four-dimensional treatment. In particular, we study the role of different types of rotational-torsional couplings---the field-free coupling and the 
field-induced coupling---and show that the conclusions recently drawn on the role of rotational-torsional couplings in the process of torsional alignment are not correct. 
Furthermore, we demonstrate how important an adequate description of the molecular polarizability is for reliably predicting the torsional alignment.

\vspace*{-0.25\baselineskip}

\textcolor{Gray}{\noindent\hfill{\rule{\textwidth}{1.25pt }}\hfill}

\vspace*{-0.25\baselineskip}

\tableofcontents

\vspace*{-2.25\baselineskip}

\textcolor{Gray}{\noindent\hfill{\rule{\textwidth}{1.25pt }}\hfill}

\vspace*{0.25\baselineskip}


\section{Torsional control and models of reduced dimensionality}
\label{sec:intro}

\thisfloatsetup{capposition=beside,capbesideposition={inside,bottom},floatwidth=8.25cm}
\begin{figure}[b!]
\vspace*{1\baselineskip plus 0.125\baselineskip minus 0.075\baselineskip}

\centering{\includegraphics[width=8cm]{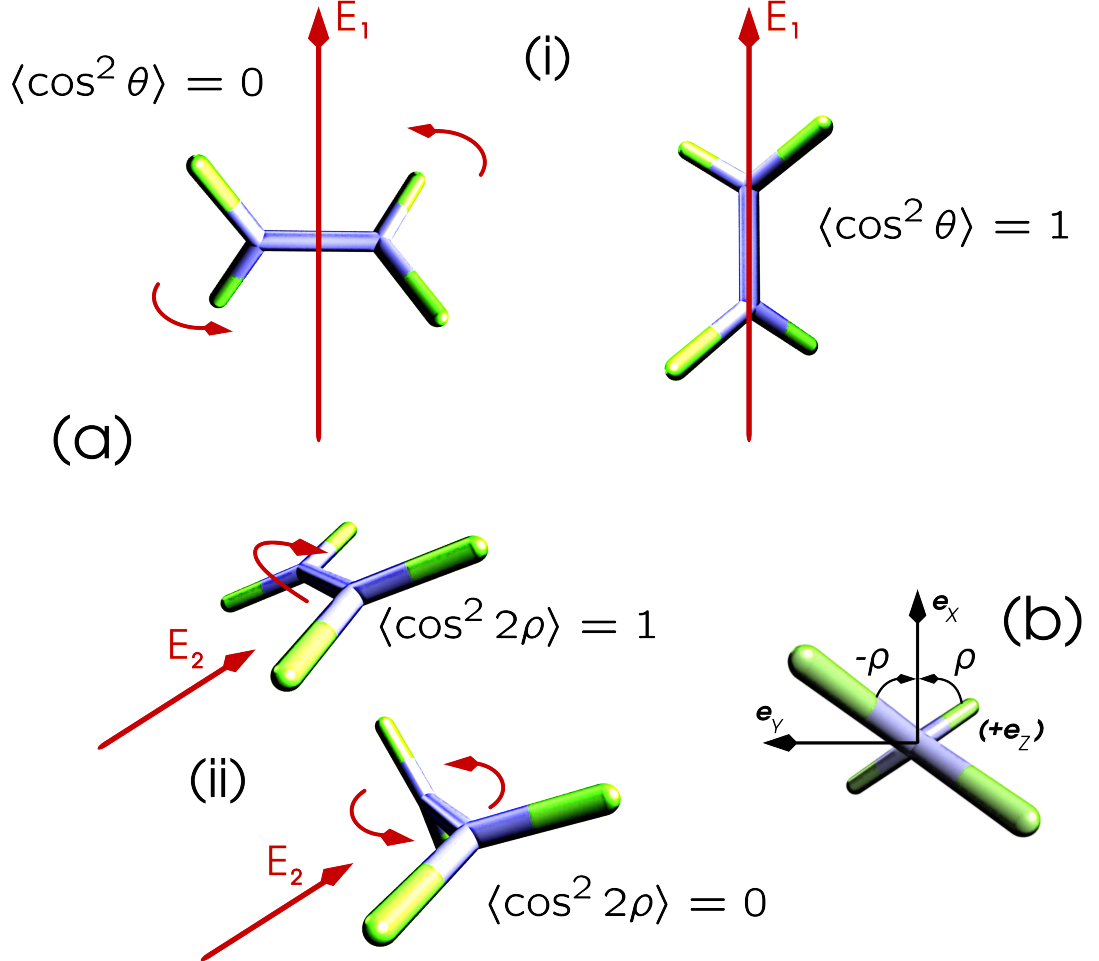}}
\caption{\textit{(a)} Classical depiction of the two-step mechanism for the alignment of molecules with observable torsion: (i) a nanosecond laser pulse ${\bm E}_1$ aligns the 
molecules along their main principal axis and \textit{(ii)} a second laser pulse, having a polarization perpendicular to the first laser pulse, is used to manipulate the torsional 
motions. If the main principal axis of the molecule is perfectly aligned to ${\bm E}_1$, $\braket{\cos^2 \theta}=1$; if the bond axis is perfectly anti-aligned to ${\bm E}_1$, 
$\braket{\cos^2 \theta}=0$. If all molecules have a coplanar structure, for the torsional alignment factor holds $\braket{\cos^2 2\rho}=1$; if all molecules have an 
staggered conformation $\braket{\cos^2 2\rho}=0$. \textit{(b)} Definition of the torsion angle $\rho$. The dihedral angle is $\beta=2\rho$.}
\label{mod-2step}
\end{figure}

Most theoretical or joint theoretical-experimental studies on laser-controlled molecular 
torsions\cite{Fujimura.1999,Hoki.2001,Kroner.2003,Fujimura.2004,Ramakrishna.2007,Kroner.2007,Reuter.2008,Madsen.2009,Madsen.2009b,Parker.2011,Coudert.2011,Parker.2012,Floss.2012,Hansen.2012,Ashwell.2013,Ashwell.2013b,Belz.2013,Ortigoso.2013,Christensen.2014,Coudert.2015,Obaid.2015,AlJabour.2015,Omiste.2017}
rely on a two-step model: A linearly, circularly or elliptically polarized nanosecond laser pulse aligns the molecule adiabatically along its main principle axis, before a 
second femtosecond laser pulse with perpendicular polarization excites the molecular torsion selectively, see Fig. \ref{mod-2step} for an illustration. If this mechanism was 
perfectly true, we were able to simulate the control of molecular torsion considering only two degrees of freedom, the torsion angle $\rho$ and the rotation about the main 
principal axis $\chi$. Several experimental studies underline the validity of this strongly idealized two-dimensional [2D] approach to torsional 
control,\cite{Madsen.2009,Madsen.2009b,Hansen.2012,Christensen.2014} 
which has been the premise of many quantum dynamical
simulations.\cite{Fujimura.1999,Hoki.2001,Kroner.2003,Fujimura.2004,Ramakrishna.2007,Kroner.2007,Madsen.2009,Madsen.2009b,PerezHernandez.2010,Parker.2011,Ashwell.2013,Ashwell.2013b,Belz.2013,Christensen.2014,Obaid.2015,AlJabour.2015} 
Only recently, it was demonstrated experimentally that the torsion of a molecule can be controlled and enhanced by using two moderately strong, time-delayed, 
off-resonant laser pulses with appropriately chosen parameters.\cite{Christensen.2014}

Yet, some theoretical studies pointed out it may be impossible to control molecular torsions separately from other degrees of freedom, 
in particular the three rotational modes of the molecule.\cite{Coudert.2011,Ortigoso.2013,Coudert.2015,Omiste.2017} 
Simulating torsional control within a four-dimensional [4D] quantum dynamical approach by taking into account all rotational degrees of freedom $\theta$, $\phi$, $\chi$ 
and the torsion {angle} $\rho$, they have shown the torsional alignment to be strongly depending on the overall rotation, and as temperature increases the torsional 
alignment may be completely destroyed. As the main reason for the potential uncontrollability of torsions, these studies identified the coupling between rotational and 
torsional modes.\cite{Coudert.2011,Ortigoso.2013}
Furthermore, they found the torsional alignment to be overestimated, if simulated with the 2D approach.\cite{Coudert.2015}

In a very recent publication\cite{Grohmann.2017}, 
however, we disputed these conclusions. First, the scenario considered in Ref. \citenum{Coudert.2011,Ortigoso.2013,Coudert.2015,Omiste.2017} is substantially different from 
the two-step mechanism from Fig. \ref{mod-2step}. These studies consider the simplest approach to torsional control: one linear-polarized laser pulse is used to steer the torsion 
without aligning the molecule along its main principal axis first. As opposed to this, we were able to show for several examples that there are no considerable differences 
between the 2D and 4D description of the two-step model, if the parameters of the two laser pulses are appropriately chosen. 

Our simulations further suggest that the validity of the 2D approach to torsional control depends on the molecule, {in particular on} the polarizability and the ratio of the 
rotational constants of the molecule. Low-dimensional models may even underestimate the torsional alignment achieved by the two-pulse scenario from Fig. \ref{mod-2step}. 

\thisfloatsetup{capposition=beside,capbesideposition={inside,bottom},floatwidth=8.25cm}
\begin{figure}[t!]
\vspace*{1\baselineskip plus 0.125\baselineskip minus 0.075\baselineskip}

\centering
{\includegraphics[width=8cm]{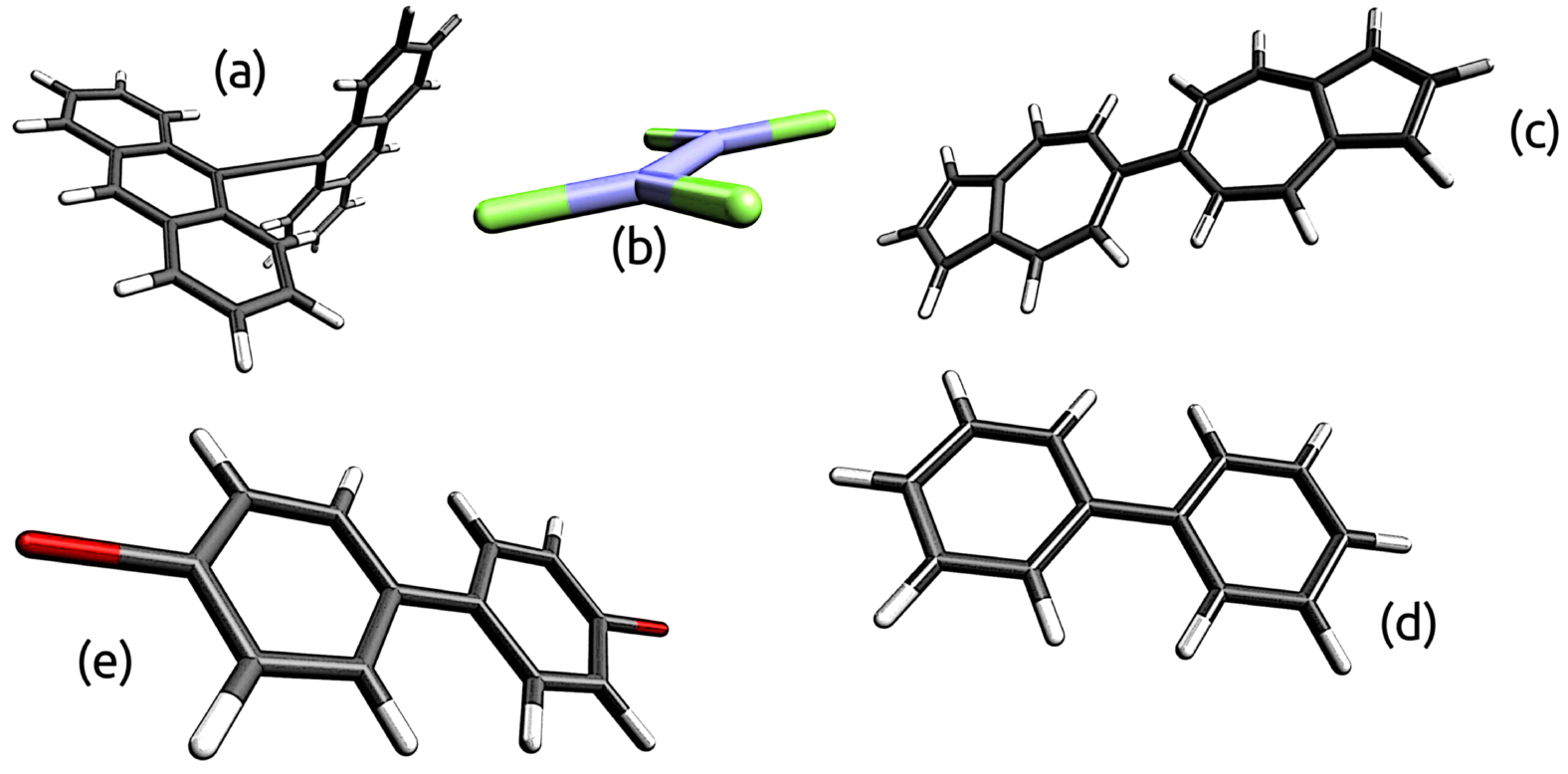}}
\caption{A selection of $\rm G_{16}$-type molecules with feasible torsion in the electronic ground state: \textit{(a)} 9-[2-(anthracen-9-yl)ethynyl]anthracene [in the 
following abbreviated as AAC]; \textit{(b)} diboron tetrafluorid [\ce{B2F4}]; \textit{(c)} 6,6'-Biazunlenyle [\ce{Biazu}]; (d) Biphenyl; and \textit{(e)} 4,4'-Dibromobiphenyl 
[\ce{DBBP}].}
\label{mol-gal}
\end{figure}

Nevertheless, our insights into the mechanisms of torsional alignment do not allow for discarding the conclusions of earlier works. Indeed, torsions and rotations are coupled, 
making it necessary to rethink the premises of low-dimensional models used to describe torsional control. In general, the torsion of a molecule cannot be manipulated 
independently of its rotational modes and consequently, aligning the molecule along the main principal axis without changing the torsional state is not possible. To discuss the 
conditions that have to be met for the 2D model to be a good approximation, we unfold here our 4D approach to strong-field control of torsions of molecules in the electronic 
ground state in full detail. 

We also clarify what ``{rotational-torsional coupling}'' actually means---there are different types of couplings, each having different origins and consequences. The various 
types of rotational-torsional couplings are in particular interesting, because they directly compromise the picture quantum dynamical models conventionally convey: Molecular 
motions being faster than the relevant process can be adiabatically separated, while slower modes can be considered to be frozen. When employing the two-step model from 
Fig. \ref{mod-2step}, theoreticians therefore usually assume torsions and rotations can be adiabatically separated. We show hereafter how these assumptions need to be 
modified to consistently describe the process of aligning molecules with feasible torsion. 

To illustrate our arguments, we focus in the present study on molecules consisting of two identical rotors with $\rm C_{2v}$-sym\-me\-try; we call them $\rm G_{16}$-type 
molecules, in conformity with their molecular symmetry [MS] group.\cite{Merer.1973,Bunker.1998}
We demonstrate our approach for selected representatives of this class of molecules; their classical structures are depicted in Fig. \ref{mol-gal}. Using the symmetry-adapted 
variational method, quantum chemistry and the adiabatic and sudden-approximation, we solve the time-dependent Schr\"odinger equation for all three rotations and
the torsion of these molecules, to simulate the control of the torsion according to the two-step mechanism from Fig. \ref{mod-2step}. Based on the knowledge we have acquired 
through our 4D simulations, we define the conditions the conventional 2D model has to fulfill to be a reliable approximation. For cases where the this 2D model fails, we briefly 
discuss how the model can be modified to still account for the relevant effects. We are thus able to show here: Yes, low-dimensional models can successfully describe the 
strong-field control of torsions if certain conditions for  the properties of the studied molecules are met.


\section{Three types of rotational-torsional coupling}
\label{sec:coupling}

The main point that was made in earlier works why torsional control of non-rigid molecules cannot be described with one- or two-dimensional models, is the (strong) 
coupling of the rotations with the torsion.\cite{Coudert.2011,Ortigoso.2013,Coudert.2015} 
An analysis of the coupling, however, is missing in these studies; they do not give a systematic account of how and why torsional and rotational motions correlate. Here, we 
intend to bridge this gap. 

We have identified three types of rotational-torsional coupling: the field-free, the field-induced and the symmetry-induced coupling. Each of them affects the 
rotational-torsional motions in a different way; not all of them have negative consequences, as the term ``coupling'' might imply. Since they allow for a better understanding of 
the rotational-torsional quantum dynamics and they are the cornerstone of our further deliberations, we discuss them in detail in the following.

\subsection{Field-free rotational-torsional couplings}
\label{subsec:field-free-coupling}

A common approach to describing the field-free rotational-torsional motion of molecules with observable torsion in the electronic ground state is the internal-axis 
method [IAM]. Using this method, we are able to write the Hamiltonian for the rotational-torsional motion as\cite{Bunker.1998,Soldan.1996}
\begin{equation}
\label{h-rt}
\hat{H}^{\rm rt}
= 
\hat{H}^{\rm rot}_{\rho} + \hat{H}^{\rm tor}\;,
\end{equation}
where 
\addtocounter{equation}{-1}
\begin{subequations}
\bea
\label{h-rot}
\hat{H}^{\rm rot}_{\rho}
=&
\frac{{\mak B}_{X^2+Y^2}}{\hbar^2}\left(\hat{J}_X^2+\hat{J}_Y^2\right) 
+
\frac{{\mak B}_{X^2-Y^2}}{\hbar^2}\left(\hat{J}_X^2-\hat{J}_Y^2\right)
+
\frac{\mak A}{\hbar^2}\hat{J}_Z^2
\eea
denotes the Hamiltonian for the rotational motion and
\begin{equation}
\label{h-tor}
\hat{H}^{\rm tor}
=
\frac{\mak F}{\hbar^2} \hat{J}^2_{\rho}
+
E^{\rm el}_{0}(\rho)
\end{equation}
\end{subequations}
is the Hamiltonian for the pure torsion; for $\rm G_{16}$-type molecules, ${\mak A}={\mak F}$.

In Eq. \eqref{h-rot}, $\hat{J}_Q$, $Q=X,Y,Z$\footnote{We follow here the convention of scattering theory for denoting molecule-fixed and space-fixed coordinates, 
\textit{i.e.} we use capital letters for body-fixed and small letters for space-fixed coordinates. In the literature of molecular spectroscopy, however, the convention is 
exactly the opposite.} are the molecule-fixed angular momenta, $\mak A$ is the rotational constant with respect to the main principal axis and\cite{Merer.1973}
\bse
\label{B-const-rho}
\begin{align}
\frac{{\mak B}_{X^2+Y^2}}{\mak B}
&=
\frac{1}{1-{\mak B}_{\rm red}^2\cos^2(2\rho)}	\\
\frac{{\mak B}_{X^2-Y^2}}{\mak B}
&= 
\frac{{\mak B_{\rm red}}\cos(2\rho)}{\left(1- {\mak B}^2_{\rm red} \cos^2(2\rho)\right)}\;.
\end{align}
\ese
As the quantities ${\mak B}_{X^2\pm Y^2}$ are functions of the torsion angle $\rho$, the rotational motions and the torsion of $\rm G_{16}$-type molecules are 
inherently coupled. In the following, we call this type of coupling the field-free rotational-torsional coupling. In Eqs. \eqref{B-const-rho}, we have introduced the reduced 
rotational constant 
\be
\label{def-Bred}
{\mak B_{\rm red}} = \frac{\mak B}{2{\mak A}}\;,
\ee
with ${\mak B}$ denoting the rotational constant for the $\rm D_{\rm 2d}$ structure of the molecules, i.e. for $2\rho = \nicefrac{\pi}{2}$. 

Conversely, the Hamiltonian for the pure torsion, Eq. \eqref{h-tor} does not depend on the rotational degrees of freedom. Besides the torsional constant $\mak F$ and the 
angular momentum for the torsion $\hat{J}_{\rho}$, $\hat{H}^{\rm tor}$ only contains the energy of the electronic ground state $E^{\rm el}_{0}(\rho)$ as a function of the 
torsion angle $\rho$, therefore being independent of the rotational coordinates $\theta,\phi,\chi$.

Defining the reduced rotational constant ${\mak B}_{\rm red}$, Eq. \eqref{def-Bred}, gives us a quantitative measure for the field-free rotational-torsional coupling. If we 
scale the Hamiltonian $\hat{H}^{\rm rt}$, Eq. \eqref{h-rt}, with respect to $\mak B$, the field-free coupling is determined by $\mak B_{\rm red}$ alone, and the larger 
$\mak B_{\rm red}$, the larger is the coupling between rotations and torsion. Expanding the functions ${\mak B}_{X^2\pm Y^2}$ from Eq. \eqref{B-const-rho} in terms of a 
power series underlines our argument. We then obtain
\bse
\label{exp-rot-konst}
\begin{align}
\frac{{\mak B}_{X^2 + Y^2}}{\mak B} 
&= 
1 + {\mak B}^2_{\rm red}\cos^2(2\rho) + {\mak B}^4_{\rm red}\cos^4(2\rho)  + ... \\
\frac{{\mak B}_{X^2 - Y^2}}{\mak B}
 &= {\mak B}_{\rm red}\cos(2\rho) + {\mak B}^3_{\rm red}\cos^3(2\rho) +  ... \quad.
\end{align}
\ese

If ${\mak B}_{\rm red} \rightarrow 0$, Eqs. \eqref{exp-rot-konst} show us, we can write 
\bse
\label{approx-rot-konst}
\begin{align}
{\mak B}_{X^2+Y^2}(\rho)
& 
\approx {\mak B} \\
{\mak B}_{X^2-Y^2}(\rho)
&\approx 0 \quad.
\end{align}
\ese
Hence, in this approximation, the rotational constants are all independent of $\rho$, and the torsion and rotations of the molecule decouple. Then, Eq. \eqref{h-rot} 
reduces to
\be
\hat{H}^{\rm rot}_{0} = \frac{{\mak B}}{\hbar^2}\,\hat{J}{}^2+\frac{{\mak A}-{\mak B}}{\hbar^2}\hat{J}_Z^2\,,
\ee
which is the rotational Hamiltonian of a symmetric top. This result implies that the asymmetry splitting, which is specific to asymmetric top molecules, vanishes for 
decoupled rotational and torsional motions as well, and within the limit $\mak B_{\rm red} \rightarrow 0$, $\rm G_{16}$-type molecules can be treated as a symmetric 
top with decoupled torsion.

Why is the field-free rotational-torsional coupling different from the rotational-vibrational coupling present in rigid molecules? For molecules with no observable internal motions, 
too, the rotational constants contained in the field-free Hamiltonian are depending on the internal coordinates, leading to an inherent coupling of rotations and vibrations. Yet, for 
a molecule without feasible torsion the magnitude of this coupling is small, causing a just as small energy correction to the rotational energy. In case of a molecule with 
observable torsion, however, the rotational parameters $\mak B_{X^2 \pm Y^2}$ are strongly depending on the torsional angle, as Eqs. \eqref{B-const-rho} clearly show. Thus, 
we cannot follow the traditional line and treat the coupling of rotations and torsions as a correction to an uncoupled system, unless the reduced rotational constant 
$\mak B_{\rm red}$ is very small. How strongly the field-free coupling actually influences the quantum dynamics of the studied non-rigid molecule is therefore depending on the 
particular molecule.

A last point we find worth to mention here is concerning the method of setting up the rotational-torsional Hamiltonian $\hat{H}^{\rm rt}$ of the isolated molecule. 
Although the IAM is used in all quantum dynamical studies on torsional control we are aware 
of,\cite{Fujimura.1999,Hoki.2001,Kroner.2003,Fujimura.2004,Ramakrishna.2007,Kroner.2007,Reuter.2008,Madsen.2009,Madsen.2009b,PerezHernandez.2010,Parker.2011,Ashwell.2013,Ashwell.2013b,Belz.2013,Christensen.2014,Obaid.2015,AlJabour.2015}
it is not the only strategy for deriving the rotational-torsional Hamiltonian $\hat{H}^{\rm rt}$.\cite{Bunker.1998}
When applied, the advantage of the IAM is the absence of kinetic energy cross-terms in $\hat{H}^{\rm rt}$, \textit{i.e.} product terms of angular momenta conjugated to 
different coordinates. The disadvantages, however, are not only $\mak B_{X^2 \pm Y^2}$ to be depending on the torsion angle and the symmetry correlations of the 
rotational and torsional eigenfunctions we address in Subsection \ref{subsec:sym-coupling} of this Section. More importantly for practical applications, the torsion angle needs to 
be defined according to Fig. \ref{mod-2step}. Contrarily, most studies on torsional control using the IAM define the torsion as being identical to the dihedral angle. Yet, this 
definition is incorrect when employing the IAM, as it leads to a violation of energy invariance\cite{Grohmann.2012}
and to incorrect symmetry properties of the torsional and rotational eigenfunctions.\cite{Hougen.1964,Bunker.1998}

\subsection{Field-induced rotational-torsional couplings}
\label{subsec:field-coupling}

In the scenario we envision in the present study, two moderate intense laser pulses being off-resonant to any molecular transition are used to control the 
rotational-torsional motions of a molecule. As it was shown and validated in countless studies, the Hamiltonian for the interaction in this case is given 
by\cite{Seideman.2005,Stapelfeldt.2003}
\be
\label{h-int}
\hat{H}^{\rm int}
=
-\frac{1}{4}\sum_{q,q'} \epsilon^*_{q} (t)\cdotp\alpha_{qq'} (\rho)\cdotp \epsilon_{q'}(t) \;,
\ee
where $q,q' = x,y,z$, ${\epsilon}_q(t)$ are the space-fixed components of the envelope of the laser field, and $\alpha_{qq'}$ are the space-fixed components of the tensor of 
the dynamic polarizability. In contrast to a rigid molecule, the components $\alpha_{qq'}$ depend on the torsion angle $\rho$ in a molecule-specific way. 

Several approaches are conceivable when realizing the two-step strategy of torsional control, illustrated in Fig. \ref{mod-2step}. In the following, we assume the first laser field to 
be a nanosecond laser pulse and therefore long compared to the timescale of the rotational-torsional motions (adiabatic limit), while the second field is supposed to be a 
femtosecond laser pulse, i.e. short compared to the timescale of the rotational-torsional motions (impulsive limit). In both cases, we confine our discussion to linear-polarized 
laser fields. Then, the Hamiltonian Eq. \eqref{h-int} reduces to
\be
\label{h-int-sce}
\hat{H}^{\rm int}_i = - \frac{ |\epsilon_i (t)|^2}{4} \alpha_{qq} \;,
\ee
where we choose $q=z$ for pulse $i=1$ and $q=x$ for pulse $i=2$, respectively. To quantify the effect of the laser pulses on the molecule, we express the laboratory-fixed 
components of the molecular polarizabilities
\addtocounter{equation}{-1}
\bse
\label{alpha-qq}
\bea
\label{alpha-zz}
\alpha_{zz} 
&=& 
\frac{\alpha^{(0,0)}}{\sqrt{3}}  + \frac{2\,\alpha^{(2,0)}}{\sqrt{6}} {\mak D}^2_{0,0} 
+ 
\frac{\alpha^{(2,2)}}{\sqrt{3}} \left ( {\mak D}^2_{0,2} + \textbf{c.c.} \right ) \\
\label{alpha-xx}
\alpha_{xx} 
&=& 
\frac{\alpha^{(0,0)}}{\sqrt{3}} - \frac{2 \alpha^{(2,0)}}{\sqrt{6}} \left[ {\mak D}^2_{0,0} - \frac{3}{\sqrt{6}}  \left( {\mak D}^2_{2,0} + \textbf{c.c.}\right) \right ] 
- \frac{\alpha^{(2,2)}}{\sqrt{2}}  \left [ \frac{1}{\sqrt{6}} {\mak D}^2_{0,2}  - \frac{1}{2}\left({\mak D}^2_{2,2} + {\mak D}^2_{2,-2}\right) + \textbf{c.c.} 
\right ] 
\eea
\ese
in terms of the elements of the Wigner D-matrices ${\mak D}^J_{m,k}$\cite{Zare.1988} and the symmetry-adapted, molecule-fixed components of the polarizability tensor 
\cite{Bunker.1998} $\alpha^{(0,0)}$, $\alpha^{(2,0)}$ and $\alpha^{(2,2)}$, see Appendix \ref{app:der-H-int} for a derivation. 

The structure of Eqs. \eqref{alpha-zz} and \eqref{alpha-xx} shows us that the Hamiltonian for the interaction with an off-resonant laser field contains products of terms 
depending on the torsional angle $\rho$ and the Euler angles $\theta$, $\phi$, $\chi$---the polarizabilities $\alpha^{(J,K)}$ are functions of the torsion angle, while the Wigner 
D-matrices ${\mak D}^J_{m,k}$ depend on the Euler angles. Consequently, any laser pulse will, at least in principle, always excite both type of motions; rotational excitations 
are {invariably} accompanied by torsional excitations, and \textit{vice versa}. This type of coupling we call hereafter field-induced coupling.

At this point, we can already give a qualitative discussion of the individual parts of Eq. (\ref{h-int-sce}). If  $\alpha^{(0,0)}$ depends on  $\rho$, torsion can be excited 
independently from molecular rotation. The term ${\mak D}^2_{0,0}$ is responsible for the alignment of the main molecular  axis. If the $\rho$-dependence of 
$\alpha^{(2,0)}$ is weak, the molecule can be aligned without changing its torsional state. If $\alpha^{(2,0)}$  strongly depends on $\rho$, alignment of the principal axis and 
torsional alignment cannot be separated. Finally, $\alpha^{(2,2)}$ couples torsion and rotation perpendicular to the principal molecular axis. Additionally, the same arguments 
we explained in Subsection \ref{subsec:field-free-coupling} of this Section apply: Although the polarizabilities of rigid molecules depend on the internal coordinates, too, the 
magnitude the polarizabilities change while the molecules undergo torsion are, in general, much larger. 

Yet, a large field-induced coupling is not necessarily counterproductive for the control of molecular torsion. In our earlier studies based on the 2D model, the torsional and 
rotational motions were also coupled by the field,\cite{Floss.2012} 
but we could not observe a ``field-induced breakdown of the torsional alignment'' as demonstrated in other works.\cite{Coudert.2011,Ortigoso.2013,Coudert.2015}
As we show in Section \ref{sec:two-step}, whether or not the field-induced coupling has negative effects is a matter of which excitation scheme we use, not a feature of torsional 
alignment in general.

\subsection[Symmetry-induced couplings]{Symmetry-induced couplings}
\label{subsec:sym-coupling}

The last type of rotational-torsional coupling we have identified to be important for describing the torsional alignment of $\rm G_{16}$-type molecules is less intuitive than the 
former two: The rotational and torsional states are not only coupled quantitatively through the $\rho$-dependence of molecular properties, but also by symmetry. Here, we focus 
on three aspects: (1) the need for classifying the rotational and torsional basis states according to an extended MS group, an EMS group; (2) the correlation of rotational and 
torsional states; and (3) the coupling of rotational and torsional basis states of different symmetry.  

The first two facets of the symmetry-induced coupling originate from the transformation properties of $\chi$ and $\rho$. As theoretical spectroscopists have discussed in great 
detail,\cite{Hougen.1964,Merer.1973,Hougen.1983,Bunker.1998} the angles $\chi$ and $\rho$ are ``double-valued'' within an IAM treatment. As a consequence, the torsional 
and rotational eigenfunctions---and thus any arbitrary rotational-torsional state of the molecule---have to be classified according to the irreducible representations of an EMS 
group, see in particular the book of \citeauthor{Bunker.1998}\cite{Bunker.1998} pp. 515 for a detailed explanation. The EMS group of molecules with feasible torsion consisting 
of two identical rotors with $\rm C_{2v}$-symmetry is $\rm G_{16}(EM)$ and was first investigated by \citeauthor{Merer.1973}.\cite{Merer.1973}

To illustrate why using an EMS-group leads to a symmetry-induced coupling, we consider an arbitrary rotational-torsional state. This state we can always expand according to
\bse
\be
\label{wf-rt}
\Psi^{\Gamma^{\rm rt}}(t)= 
\sum_{\mathtt{n}_{\rm rt}} c_{\mathtt{n}_{\rm rt}}
\Phi^{\Gamma^{\rm rt}}_{\mathtt{n}_{\rm rt}}\exp\left(-\frac{\rm i}{\hbar} E^{\Gamma^{\rm rt}}_{\mathtt{n}_{\rm rt}}t\right)\;,
\ee
whereby we can calculate the rotational-torsional eigenfunctions using a variational approach with the ansatz 
\be
\label{ef-rt}
\Phi^{\Gamma^{\rm rt}}_{\mathtt{n}_{\rm rt}} 
=
\sum_{\mathtt{n}_{\rm rot}}\sum_{\mathtt{n}_{\rm tor}} c_{\mathtt{n}_{\rm rot},\mathtt{n}_{\rm tor}} 
\Phi^{\Gamma^{\rm rot}}_{\mathtt{n}_{\rm rot}}(\theta,\phi,\chi)
\cdotp
\Phi^{\Gamma^{\rm tor}}_{\mathtt{n}_{\rm tor}}(\rho)\;.
\ee
\ese
In both equations, the expansion coefficients $c_{\mathtt{n}_{\rm rt}}$ and $c_{\mathtt{n}_{\rm rot},\mathtt{n}_{\rm tor}}$, the rotational-torsional eigenenergies 
$E^{\Gamma^{\rm rt}}_{\mathtt{n}_{\rm rt}}$, the rotational-torsional eigenstates $\Phi^{\Gamma^{\rm rt}}_{\mathtt{n}_{\rm rt}}$ and the rotational and 
torsional basis functions, $\Phi^{\Gamma^{\rm rot}}_{\mathtt{n}_{\rm rot}}$, and $\Phi^{\Gamma^{\rm tor}}_{\mathtt{n}_{\rm tor}}$, are fully characterized by the 
rotational-torsional quantum numbers $\mathtt{n}_{\rm rt}$ and the rotational and torsional quantum numbers $\mathtt{n}_{\rm rot}$ and $\mathtt{n}_{\rm tor}$, 
respectively. Additionally, however, we can classify the eigenfunctions and basis states according to the irreducible representations $\Gamma$ of the EMS group 
$\rm G_{16}(EM)$.

The first type of symmetry-induced coupling arises from the transformation properties of $\Phi^{\Gamma^{\rm rt}}_{\mathtt{n}_{\rm rt}}$ within $\rm G_{16}(EM)$ and 
from the characteristics of the irreducible representations $\Gamma$. One feature of EMS groups is that their irreducible representations $\Gamma$ can be grouped into 
single-valued  and double-valued {$\Gamma^{\rm (d)}$ } representations. The rotational-torsional states $\Psi^{\Gamma^{\rm rt}}$ and $\Phi^{\Gamma^{\rm rt}}$ in Eq. 
\eqref{wf-rt} must transform according to a single-valued irreducible representation.\cite{Bunker.1998}
Since
\be
\Gamma^{\rm rt} = \Gamma^{\rm rot} \otimes \Gamma^{\rm tor}\;,
\ee 
the irreducible representations of the torsional and rotational basis states must be therefore either both single-valued or both double-valued.\cite{Merer.1973,Bunker.1998} 
Thus, for $\Gamma^{\rm rt}$ to be single-valued the symmetry of the torsional and rotational states are correlated, or, how we call it hereafter, symmetry-coupled.

The second type of symmetry-induced coupling is a immediate consequence of the irreducible representations $\Gamma^{\rm rot}$ and $\Gamma^{\rm tor}$ being correlated.
As the irreducible representations $\Gamma^{\rm rot}$ and $\Gamma^{\rm tor}$ can be directly related to the quantum numbers $\mathtt{n}_{\rm rot}$ and 
$\mathtt{n}_{\rm tor}$ specifying the rotational and torsional basis states, not every rotational basis state can be combined with every torsional state. Thus, the quantum 
numbers of rotational and torsional basis states have to fulfill certain conditions to be symmetry-allowed. As we illustrate in Subsection \ref{subsec:savm} of this Section, this 
correlation is of great importance when setting up the proper symmetry-adapted basis for solving the time-dependent Schr\"odinger equations.

Moreover, some of the quantities in the Hamiltonians $\hat{H}^{\rm rt}$ and $\hat{H}^{\rm int}$, \textit{c.f.} Eqs. \eqref{h-rt},  \eqref{h-int} and \eqref{alpha-qq}, are not 
totally symmetric in $\rm G_{16}(EM)$, pointing to a third aspect of the symmetry-induced coupling.  Instancing the Hamiltonian $\hat{H}^{\rm rt}$, we can show with the 
character table of $\rm G_{16}(EM)$, see Table III in the work of \citeauthor{Merer.1973},\cite{Merer.1973} 
that
\bse
\label{sym-op-ff}
\begin{align}
\mak B_{X^2-Y^2} & \sim {\rm B^+_{1g}} \\
\hat{J}^2_X - \hat{J}^2_Y  & \sim {\rm B^+_{1g}}
\end{align}
and for the product
\be
\mak B_{X^2-Y^2}\left(\hat{J}^2_X - \hat{J}^2_Y\right)  \sim {\rm A_{1g}^+}\;,
\ee
\ese
reflecting the invariance of the Hamiltonian $\hat{H}^{\rm rt}$ in the group $\rm G_{16}(EM)$. As a consequence, when written in the basis Eq. \eqref{ef-rt}, the matrix 
representation of ${\mak B}_{X^2-Y^2}$ contains non-zero elements between torsional states belonging to different irreducible representations $\Gamma^{\rm tor}$, while  
the matrix representation of $\hat{J}^2_X - \hat{J}^2_Y$ contains non-zero elements between rotational states of different rotational symmetry $\Gamma^{\rm rot}$.
This additional symmetry-induced coupling directly follows from the {vanishing integral rule}, which states that the matrix elements of any operator $\hat{O}$ 
transforming irreducible in the (E)MS group of the molecule
\bse
\be
{O}_{nm} = \int {\rm d}V\; (\Phi^{\Gamma_n})^*\cdotp \hat{O}^{\Gamma}\cdotp \Phi^{\Gamma_m} 
\ee
are only non-zero if
\be
\label{van-int}
\Gamma^*_n\otimes \Gamma \otimes \Gamma_m \supseteq \Gamma_{\rm ts}\;,
\ee
\ese
with $\Gamma_{\rm ts}$ denoting the total symmetric representation of the (E)MS group.\cite{Bunker.1998}
Consequently, the matrix representations of ${\mak B}_{X^2-Y^2}$ and $\hat{J}^2_X - \hat{J}^2_Y$ contain only non-zero elements between basis states belonging to
different irreducible representations. Therefore, the true eigenfunctions of $\hat{H}^{\rm rt}$ of one particular symmetry $\Gamma^{\rm rt}$ contain torsional and 
rotational basis functions of different symmetry.

The same holds true if the molecules are manipulated by an off-resonant laser field, see Eqs. \eqref{h-int} and \eqref{alpha-qq} for the definition of the Hamiltonian for the 
interaction $\hat{H}^{\rm int}$. Here, since
\bse
\label{sym-op-int}
\begin{align}
\alpha^{(2,{2})}(\rho) & \sim {\rm B^+_{1g}} \\
{\mak D}^J_{m,\pm 2}  & \sim {\rm B^+_{1g}}\;,
\end{align}
\ese
the field-induced coupling mediated by $\alpha^{(2,{2})}(\rho)$ and ${\mak D}^J_{m,\pm 2}$ couples also torsional and rotational states of different symmetry. 

Let us clarify this aspect of the symmetry-induced coupling by an example. Throughout this work, we only consider rotational-torsional states with symmetry 
$\Gamma^{\rm rt}={\rm A}^{\rm +}_{\rm 1g}$. As we show in Table 4 and 5 in the supplemental material, these states can be formed by rotational and torsional basis 
states with symmetry $\Gamma^{\rm rot}={\Gamma^{\rm tor}}={\rm A_{1g}^{+}}$ and $\Gamma^{\rm rot}={\Gamma^{\rm tor}}={\rm B_{1g}^{+}}$, respectively.
Hence, taking into account Eqs. \eqref{sym-op-ff} and \eqref{sym-op-int}, rotational and torsional basis states with symmetry ${\rm A_{1g}^{+}}$ must be coupled with 
rotational and torsional basis states with symmetry ${\rm B_{1g}^{+}}$ to fulfill Eq. \eqref{van-int}. As the product of both rotational and torsional basis 
functions must always have ${\rm A_{1g}^{+}}$ symmetry, the coupling of rotational states with ${\rm A_{1g}^{+}}$ and ${\rm B_{1g}^{+}}$ symmetry is always 
accompanied by a coupling of torsional states with ${\rm A_{1g}^{+}}$ and ${\rm B_{1g}^{+}}$ symmetry. Thus, the rotational and torsional states are symmetry coupled 
by the operators Eqs. \eqref{sym-op-ff} and \eqref{sym-op-int}, and we cannot formulate selection rules, \textit{i.e.} find the non-zero matrix elements of $\hat{H}^{\rm rt}$
and $\hat{H}^{\rm int}$, for the rotational states without taking into account the selection rules for the torsional states.

Taken all together, all three aspects of the symmetry-induced coupling prevent that we can model the control of the torsion as being independent of the rotations of 
$\rm G_{16}$-type molecules.  Although we do not systematically study the symmetry-induced coupling in this publication, we stress that most studies on torsional 
alignment ignore all aspects of the symmetry-induced coupling 
discussed,\cite{Fujimura.1999,Hoki.2001,Kroner.2003,Fujimura.2004,Kroner.2007,Madsen.2009,Madsen.2009b,Parker.2011,Parker.2012,Floss.2012,Hansen.2012,Christensen.2014} 
and therefore, their conclusions need to be reevaluated.

\subsection{Rotational-torsional couplings and the two-dimensional model}
\label{subsec:coupling-2D}

The last aspect being important to follow our arguments is which type of couplings occur in a 2D treatment of torsional control. The 2D model that has become so popular to 
describe the rotational-torsional motions molecules during the last 
decades\cite{Hoki.2001,Kroner.2003,Ramakrishna.2007,Parker.2011,Grohmann.2007,Floss.2012,Parker.2012,Ashwell.2013,Ashwell.2013b}
premises the molecules to be perfectly aligned along their axis of torsion. Consequently, if we use this model, we presuppose that the first step of the two-step mechanism
illustrated in Fig. \ref{mod-2step} was realized successfully.

If the molecule-fixed $e_Z$-axis is parallel or anti-parallel to the space-fixed $e_z$-axis, and therefore to the polarization vector of the first laser pulse, 
$\theta = \left\{0,\pi\right\}$. Then, the angle $\phi$ is redundant, and we can choose $\phi = 0$. Calculating the limit $\theta\rightarrow 0$ and $\phi\rightarrow 0$ in Eqs.
\eqref{h-rt}, we obtain the field-free Hamiltonian for the remaining two coordinates, $\chi$ and $\rho$, 
\be
\label{h-2d-ff}
\hat{H}^{\rm 2D}
=
\frac{\mak A}{\hbar^2}\hat{J}_Z^2 
+
\frac{\mak F}{\hbar^2}\hat{J}^2_{\rho}
+
E^{\rm el}_{0}(\rho)\,.
\ee
The eigenfunctions of $\hat{H}^{\rm 2D}$ can be written as 
\be
\label{ef-2d}
\Phi^{\rm 2D}_{k,\mathtt{n}_{\rho}} (\chi,\rho)= \Phi^{\rm rot}_{k}(\chi)\cdotp\Phi^{\rm tor}_{\mathtt n_{\rho}}(\rho)\;,
\ee
where the rotational eigenfunctions are
\addtocounter{equation}{-1}
\bse
\be
\label{ef-2d-rot}
\Phi^{\rm rot}_{k}(\chi) = \frac{1}{\sqrt{2\pi}} \exp\left({\rm i}k\chi \right)
\ee
and the eigenfunctions of the pure torsional Hamiltonian, $\Phi^{\rm tor}_{\mathtt n_{\rho}}(\rho)$, we expand according to
\be
\label{ef-2d-tor}
\Phi^{\rm tor}_{\mathtt n_{\rho}} = \sum_{k_{\rho}} c_{\mathtt{n}_{\rho},k_{\rho}} \Phi_{k_{\rho}} (\rho)
\ee
with
\be
\label{eig-plan-rot}
\Phi_{k_{\rho}}(\rho) = \frac{1}{\sqrt{2\pi}} \exp\left({\rm i}k_{\rho}\rho \right) \;.
\ee
\ese
As the rotational constant $\mak A$ in Eq. \eqref{h-2d-ff} is independent of $\rho$ and $\hat{H}^{\rm 2D}$ contains no explicit cross terms, the rotation and 
torsion are not quantitatively coupled in the 2D case. Hence, no field-free coupling arises when employing 2D model.

Analogously, by setting $\theta=0$ and $\phi=0$ in Eqs. \eqref{alpha-qq}, we obtain for the interaction of the molecule with the second laser pulse
\be
\label{h-int-2d}
\hat{H}^{\rm int}_{2} (t)
=
-\frac{\left|\epsilon_2 (t)\right|^2}{4}\left(\frac{\alpha^{(0,0)}}{\sqrt{3}} - \frac{\alpha^{(2,0)}}{\sqrt{6}} + 2\alpha^{(2,{2})}\cos(2\chi)  \right)\;,
\ee 
see also Appendix \ref{app:der-H-int}. Due to the last term on the right-hand side of Eq. \eqref{h-int-2d}, in the 2D model rotations and torsions are coupled by the 
field. Thus, contrary to the field-free rotational-torsional coupling, the field-induced coupling occurs as well in the 2D treatment of torsional control.

Likewise, the last type of coupling, the correlation of the torsional and rotational symmetries, is also present in the 2D model: As the symmetry-induced coupling originates
from the transformation properties of the angles $\chi$ and $\rho$, the rotational and torsional symmetries are still correlated.\cite{Grohmann.2012}

In summary, rotational-torsional couplings are also present when we use the 2D model to describe the torsional control of non-rigid molecules. Only the field-free 
coupling is vanishing if the molecules are assumed to be perfectly aligned along their main principal axes. We study the influence of the rotational-torsional coupling in 
detail in Section \ref{sec:two-step}.  Yet, we can already conclude from the comparison of the 4D and 2D model that ``the rotational-torsional coupling'' cannot be the only 
reason for a potential disagreement of both descriptions, as certain types of couplings are considered in both models.


\section{A numerical approach to four-di\-men\-sio\-nal torsional control}
\label{sec:4d-mod}

\begin{figure}[tb!]
\vspace*{1\baselineskip plus 0.125\baselineskip minus 0.075\baselineskip}

\centering{\includegraphics[width=8cm]{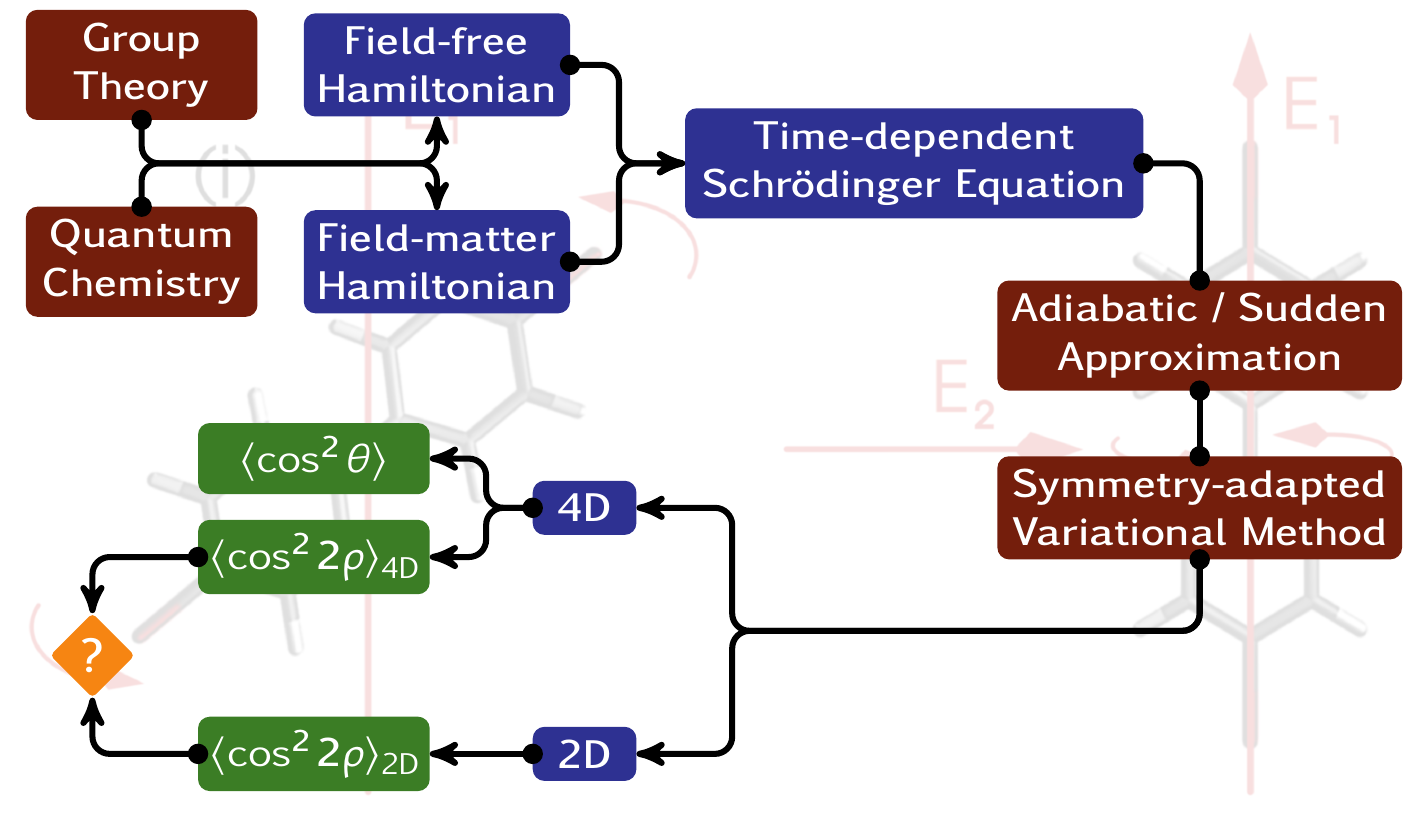}}
\caption{An illustration of our approach to solving the 4D Schr\"odinger equation(s) for the scenario we illustrate in Fig. \ref{mod-2step}.}
\label{4D-scheme}
\end{figure}

To simulate the two-step mechanism of molecular alignment from Fig. \ref{mod-2step} we numerically solve the time-dependent Schr\"odinger equation by transforming it into a 
matrix problem, using an expansion into energy-eigenfunctions.  To minimize the drawbacks of this ansatz, we make use of symmetry arguments, thus characterizing our 
method as a symmetry-adapted variational approach to torsional control. We illustrate the strategy we employ here in Fig. \ref{4D-scheme}. 

Using group theory, we derive the symmetry adapted  form of the matrix representations for the field free and field matter Hamiltonians. We obtain the required molecular 
parameters with quantum chemistry and use a symmetry adapted fitting procedure to implement the data in our numerical code. To solve the time-dependent Schr\"odinger 
equation, we use two different approximations, the adiabatic and the sudden approximation. Physically,  they reflect the two limiting cases of molecular alignment we are 
considering here, the adiabatic\cite{Friedrich.1991,Stapelfeldt.2003} 
and impulsive\cite{Seideman.2005}
regime. In both cases, we transform the time-dependent Schr\"odinger equation into a symmetry adapted matrix problem and calculate their solutions by diagonalizing the 
matrix representation of the respective Hamiltonian. Finally, we determine and compare the relevant expectation values, namely the alignment factor  $\braket{\cos^2\theta}$ 
and the torsional alignment factor $\braket{\cos^2 2 \rho}$ in the 4D model and the torsional alignment factor in  2D model, see Fig. \ref{mod-2step} for an illustration. 

\thisfloatsetup{capposition=beside,capbesideposition={inside,bottom},floatwidth=8.25cm}

In the present Section, we limit our discussion on the conceptional aspects that are necessary to understand our results from Sections \ref{sec:two-step}. For technical 
details of our numerical code see in particular Appendix \ref{app:matrix-ele} and \ref{app:code}. A critique of our approach we develop in Section \ref{sec:limits}.

\subsection{Adiabatic and non-adiabatic alignment as two limiting cases of the time-dependent Schr\"odinger equation}
\label{subsec:ad-nonad}

When the molecules are interacting with the first, off-resonant, moderately intense, nanosecond laser pulse, pendular states are created by the field.\cite{Friedrich.1991} 
If the laser pulse varies sufficiently slowly, we can assume the molecular eigenstates to be adiabatically evolving into the pendular states, using the adiabatic 
theorem of quantum mechanics.\cite{Messiah.1965}
Then, we can approximate $\epsilon_1 (t)\approx\epsilon^{\rm max}_1$ during the pulse and, as a consequence, the Hamiltonian for the molecule interacting with the first 
laser pulse 
\be
\label{h-fd}  
\hat{H}^{\rm fd} = \hat{H}^{\rm rt}+\hat{H}^{\rm int}_{1}
\ee
is time-independent, see Eqs. \eqref{h-rt}, \eqref{h-rot}, \eqref{h-tor}, \eqref{h-int-sce} and \eqref{alpha-zz} for the definition of $\hat{H}^{\rm rt}$ and 
$\hat{H}^{\rm int}_1$, respectively. The Hamiltonian $\hat{H}^{\rm fd}$ being time-independent allows us to find the {pendular states} 
$\Phi^{\rm fd}_{\mathtt{n}_{\rm fd}}$ and the corresponding field-dressed energies $E^{\rm fd}_{\mathtt{n}_{\rm fd}}$ from solving the eigenvalue equation
\be
\label{sg-adiab}
\hat{H}^{\rm fd}\Phi^{\rm fd}_{\mathtt{n}_{\rm fd}}
=
E^{\rm fd}_{\mathtt{n}_{\rm fd}}\Phi^{\rm fd}_{\mathtt{n}_{\rm fd}}\;.
\ee

Numerically, we calculate $\Phi^{\rm fd}_{\mathtt{n}_{\rm fd}}$ and $E^{\rm fd}_{\mathtt{n}_{\rm fd}}$ by expanding $\Phi^{\rm fd}_{\mathtt{n}_{\rm fd}}$ into 
a symmetry-adapted basis, see Subsection \ref{subsec:savm} of this Section and Appendices \ref{app:matrix-ele} and \ref{app:code} for a more elaborate outline. The 
expansion coefficients we then find from diagonalizing the matrix ${\bm H}^{\rm rt} + {\bm H}^{\rm int}_1$, where we define
\be
\label{h-int-ad}
{\bm H}^{\rm int}_1=-|\epsilon^{\rm max}_1|^2 \,{\bm W}_1
\ee
and the operator $\hat{W}_1$, corresponding to ${\bm W}_1$, is readily identified from Eqs. \eqref{h-int-sce} and \eqref{alpha-zz}. 

Once the molecules have adiabatically evolved into a pendular state, a second, off-resonant, femtosecond laser pulse is applied to the system. Thus, wave packets 
composed of pendular states are created by the field
\be
\label{wp-aus-nad}
\Psi(t)
=
\sum_{\mathtt{n}_{\rm fd}}c_{\mathtt{n}_{\rm fd}}(t_{0^+})
\exp\left(-\frac{\rm i}{\hbar}E^{\rm fd}_{\mathtt{n}_{\rm fd}} t\right)\Phi^{\rm fd}_{\mathtt{n}_{\rm fd}}\,,
\ee
with $t_{0^+}$ denoting the time at the end of the pulse, and $t\geq t_{0^+}$. 

To obtain the coefficients $c_{\mathtt{n}_{\rm fd}}(t_{0^+})$, we make use of the sudden approximation.\cite{Gershnabel.2008,Fleischer.2009} 
Within this approximation, the exact wave function at the end of the second laser pulse is approximated by 
\be
\label{psi-t0+}
\Psi(t_{0^+}) = \exp\left(-\frac{\rm i}{\hbar}\int_{t_{0^-}}^{t_{0^+}}\hat{H}^{\rm int}_{2} {\rm d}t \right)\Psi(t_{0^-}) \;,
\ee
where $\Psi(t_{0^-})$ denotes the wave function of the system before interacting with the laser pulse, and $\hat{H}^{\rm int}_2$ is defined in Eqs. \eqref{h-int-sce} and
\eqref{alpha-xx}. This allows us to introduce the operator 
\bse
\be
\label{h-int-nad}
\overline{\hat{H}^{\rm int}_{2}} =  \overline{\epsilon^2_2}\,\hat{W}_2\,,
\ee
where $\hat{W}_2$ has the same structure as $\hat{W}_1$ in Eq. \eqref{h-int-ad}, and we defined the integrated electric field strength
\be
\overline{\epsilon^2_2}=\int_{t_{0^-}}^{t_{0^+}}|\epsilon_2(t)|^2{\rm d}t\;.
\ee
\ese
As $\overline{\epsilon^2}_2$ is integrated over time, the pulse shape plays no role when the sudden-approximation is used to describe impulsive alignment. In our 
calculations we assumed throughout a Gaussian-like laser pulse
\be
\label{gauss-env}
\epsilon(t)=\epsilon_0\exp\left(-\frac{2\ln 2}{t^2_{\nicefrac{1}{2}}}t^2\right),
\ee
having the effective intensity
\be
\label{int-max}
{I}_{0}=\frac{1}{2\mu_0 c}|\epsilon_0|^2\;,
\ee
where, in Eq. \eqref{gauss-env} $t_{\nicefrac{1}{2}}$ is the time of the FWHM, not the total pulse length.

In practice, we obtain the coefficients in Eq. \eqref{wp-aus-nad} by expanding $\Psi(t_{0^+})$ in terms of symmetry-adapted basis functions, see Subsection 
\ref{subsec:savm} of this Section and Appendices \ref{app:matrix-ele} and \ref{app:code} for details. Then, solving Eq. \eqref{psi-t0+} is equivalent to finding the solution of 
the matrix equation\nolinebreak[4]
\be
\label{psi-t0+-num}
{\bm c}(t_{0^+}) = \exp\left(-\frac{\rm i}{\hbar} \overline{\epsilon^2}\,{\bm W} \right){\bm c}(t_{0^-})\,,
\ee
where the column-matrices ${\bm c}(t_{0^{\pm}}) $ contain the expansion coefficients of the wave packets. 

As wave functions are not directly accessible in an experiment, we need to calculate the expectation values of observable quantities. In case of alignment studies these 
observables are the alignment factors. The alignment along the main principal axis of the molecules is characterized by the expectation value
\bse
\be
\label{alg-theta}
{A}_{\theta} (t) 
=
\langle \cos^2\theta \rangle 
=
\bra{\Psi(t)}\cos^2\theta\ket{\Psi(t)} \;,
\ee
where
\be
{A}_{\theta} (t_{0^-}) = \bra{\Phi^{\rm fd}_0}\cos^2\theta\ket{\Phi^{\rm fd}_0} \;.
\ee
\ese
If $A_{\theta}$ is one, all molecules are aligned along the field axis; if $A_{\theta}$ is zero, all molecules in the probe are aligned perpendicular to the field; under thermal 
conditions the alignment factor $A_{\theta}$ is \nicefrac{1}{3}; see also Fig. \ref{mod-2step} for a graphical illustration of $A_{\theta}$.

Whether or not the control of the torsional degree of freedom by the laser fields was successful, we can learn from the torsional alignment factor
\bse
\be
\label{alg-tor}
{A}_{2\rho} (t) = \langle \cos^2 2\rho \rangle
=
\bra{\Psi(t)}\cos^2 2\rho\ket{\Psi(t)} \;,
\ee
where, again,
\be
{A}_{2\rho} (t_{0^-}) = \bra{\Phi^{\rm fd}_0}\cos^2 2\rho \ket{\Phi^{\rm fd}_0} \;.
\ee
\ese
In case the dihedral angle $\gamma = 2\rho$ is $90^{\circ}$ for all molecules in the probe, ${A}_{2\rho}=0$; if $\gamma$ is $0^{\circ}$ for all molecules, 
${A}_{2\rho}=1$; the equilibrium value of ${A}_{2\rho} $ is determined by the shape of the torsional potential $E^{\rm tor}_0(\rho)$; for a classical illustration of 
${A}_{2\rho}$ see Fig. \ref{mod-2step}.

The time-evolution of the expectation values $A_{\theta}$ and $A_{\rho}$ is the basis of our analysis of what the conditions are for the 2D model to be a reliable 
approximation. We present our results of the simulations for $A_{\theta}$ and $A_{\rho}$ for different molecules, see Fig. \ref{mol-gal}, and laser pulse intensities, 
\textit{c.f.} Eq. \eqref{int-max}, in Section \ref{sec:two-step}.


\subsection{The symmetry-adapted variational method}
\label{subsec:savm}

To actually solve Eqs. \eqref{sg-adiab} and \eqref{psi-t0+-num} numerically, we need to choose a basis. Throughout this work we employ the ansatz
\be
\label{ansatz-basis}
\Phi^{\rm rt} =\sum_{k_{\rho}}\sum_{J,k,m}c_{k_{\rho},J,k,m} \Phi_{k_{\rho}}(\rho)\cdotp\Phi_{J,k,m}(\theta,\phi,\chi)\;,
\ee
where the free rotor basis functions for describing the torsional degree of freedom $\rho$ are defined in Eq. \eqref{eig-plan-rot}, and
\addtocounter{equation}{-1}
\bse
\be
\label{sk-eig-fun}
{\Phi}_{J,k,m}
=
\sqrt{\frac{2J+1}{8\pi^2}}\left({\mak D}^{J}_{m,k}(\theta,\phi,\chi)\right)^*
\ee
\ese
are the rotational eigenfunctions of a symmetric top as a function of the Euler angles $\theta$, $\phi$, $\chi$, with ${\mak D}^J_{m,k}$ denoting the elements of the rotation 
matrix for the symmetric top quantum numbers $J,k,m$.\cite{Zare.1988} 

Contrary to earlier studies,\cite{Coudert.2011,Ortigoso.2013,Coudert.2015} we do not use a grid-based method. For two reasons:\cite{Reuter.2009}
First, due to singularities, a large number of basis functions are necessary to adequately represent the field-free Hamiltonian numerically. Using a grid-based method would 
therefore limit our studies to very low laser intensities to give reliable results. Second, when employing the grid method, the matrix representations of the Hamiltonians we use to 
describe the process of alignment contain non-vanishing elements between states of different $m$ and $k$, leading to numerical inaccuracies, known as $m$-mixing problem.

If we use the ansatz Eq. \eqref{ansatz-basis}, we can avoid the issue of $m$-mixing, but we still face the problem of unfeasible basis set sizes. One common approach to 
reduce the demand of numerical calculations is using molecular symmetry. We thus use the EMS group $\rm G_{16}(EM)$ we have introduced in Subsection 
\eqref{subsec:sym-coupling} of Section \ref{sec:coupling} to construct a symmetry-adapted basis out of the ``primitive'' basis functions Eq. \eqref{ansatz-basis}. 

The symmetry-adapted basis functions for the rotations are Wang-functions\cite{Wang.1929}
\bse
\label{sym-bas-rot}
\be
\Phi^{\pm}_{J,K,m} \equiv \frac{1}{\sqrt{2}}\left(\Phi_{J,K,m} \pm (-1)^J  \Phi_{J,-K,m}\right),\; K \equiv |k| \neq 0\;,
\ee
where $\Phi_{J,\pm K,m}$ denote the symmetric top eigenfunctions from Eq. \eqref{sk-eig-fun}. For $k=0$ holds 
\begin{align}
\Phi^{+}_{J,0,0}&\equiv \Phi_{J,0,0}\; \quad \text{if $J$ is even}\\
\Phi^{-}_{J,0,0}&\equiv \Phi_{J,0,0}\; \quad\text{if $J$ is odd.}
\end{align}
\ese
The symmetry-adapted basis for the torsion, on the other hand, is given by
\bse
\label{sym-bas-tor}
\begin{align}
\Phi^+_{K_{\rho}} & \equiv \frac{1}{\sqrt{\pi}} \cos\left(K_{\rho}\rho \right) \\
\Phi^-_{K_{\rho}} &\equiv \frac{1}{\sqrt{\pi}} \sin\left(K_{\rho}\rho \right)
\end{align}
for $K_{\rho} = |k_{\rho}|\neq 0$; for $k_{\rho}=0$ 
\be
\Phi^+_{0} \equiv \frac{1}{\sqrt{2\pi}}\;.
\ee
\ese

The functions Eqs. \eqref{sym-bas-rot} and \eqref{sym-bas-tor} transform irreducible in the group $\rm G_{16}(EM)$; their irreducible representations 
$\Gamma^{\rm rot}$ and $\Gamma^{\rm tor}$ are shown in Table IV of the work of \citeauthor{Merer.1973},\cite{Merer.1973}
which we also provide in the supplemental material. The exact eigenfunctions of any Hamiltonian we consider here are linear combinations of products of 
$\Phi^{\pm}_{J,K,m}$ and $\Phi^{\pm}_{K_{\rho}}$ having the same product symmetry $\Gamma^{\rm rt}=\Gamma^{\rm tor}\otimes \Gamma^{\rm rot}$. Thus, the 
matrix representation ${\bm H}$, written in the symmetry-adapted basis, decomposes into blocks according to theses symmetries.\cite{McWeeny.2002}
Numerically, a convenient way to transform $\bm H$ to the symmetry-adapted basis is to use special projection operators,\cite{McWeeny.2002}
see Appendix \ref{app:code} for our implementation. Once we have obtained the matrix $\bm H$ in the symmetry-adapted basis, we can diagonalize each of the symmetry 
blocks separately, reducing the numerical effort substantially.

Yet, not every product of rotational and torsional basis functions in Eq. \eqref{ansatz-basis} is symmetry-allowed; they need to be combined in a specific way. As we pointed out 
in Subsection \ref{subsec:sym-coupling} of Section \ref{sec:coupling}, the product representation $\Gamma^{\rm rt}$ is only allowed to contain single-valued irreducible 
representations,\cite{Bunker.1998}
otherwise the wave function $\Phi^{\rm rt}$ would be double-valued.\cite{Hougen.1964,Bunker.1998} 
The combinations that fulfill this condition are summarized in Table IV of the work of \citeauthor{Merer.1973}\cite{Merer.1973},
see also the supplemental material. Hence, only rotational and torsional basis states are compatible of which quantum numbers $K$ and $K_{\rho}$ both are either even or 
odd, reflecting the symmetry-induced coupling of rotations and torsions. This aspect is ignored by most studies on torsional 
alignment\cite{Madsen.2009,Madsen.2009b,Christensen.2014} 
and is also relevant for molecules with non-identical moieties.\cite{Soldan.1996}  
In the following, we only consider states of $\rm A_{1g}^+$-symmetry, \textit{i.e.} states having the symmetry of the rotational-torsional ground state.

\begin{table}[t!]
\caption{The rotational constants $\mak A$, $\mak B$ and the reduced rotational constant $\mak B_{\rm red}$ for the molecules we are investigating here; see also Fig. 
\ref{mol-gal}. For the torsional constant holds $\mak F = \mak A$. The point-group of the optimized structure is denoted as $\rm G^{\rm ref}$. The constants $\mak A$, 
$\mak B$ are given in units of $10^{-3}{\rm meV}$.}
\label{rotkonst-bsp}

\vspace*{0.5\baselineskip plus 0.125\baselineskip minus 0.075\baselineskip}
\renewcommand\arraystretch{1.25}
\begin{tabular}{@{\extracolsep{\fill}}lR{1.15cm}R{1.15cm}R{1.15cm}R{1.15cm}}
\textit{Molecule}
& ${\mak A}$  	& ${\mak B}$ &  $\mak B_{\rm red}$  
& $\rm G^{\rm ref}$\\
\hline
\ce{B2F4}  
& $21.720$ & $8.457$ &$0.195$ 
& $\rm D_{2h}$\\
Biphenyl 
& $12.000$ & $2.287$	& $0.087$  
& $\rm D_{2h}$\\
DBBP
& $11.960$ & $0.369$ & $0.015$ 
& $\rm D_{2h}$ \\
Biazu
& $5.965$ & $0.560$ & $0.047$ 
& $\rm D_{2d}$\\
AAC 
& $0.937$ & $0.357$ & $0.190$ 
& $\rm D_{2d}$\\
\end{tabular}
\vspace*{0.5\baselineskip plus 0.125\baselineskip minus 0.075\baselineskip}
\end{table}


\subsection{Getting the molecular parameters: quantum chemistry}
\label{sub-sec:qc}

The Hamiltonians for the field-free rotational-torsional motions, Eqs. \eqref{h-rt}, \eqref{h-rot} and \eqref{h-tor}, and for the field-matter interaction, Eqs. \eqref{h-int-sce},
\eqref{alpha-zz} and \eqref{alpha-xx}, both depend on parameters being characteristic for each molecule: the rotational constants $\mak A$ and $\mak B$; the torsional 
constant $\mak F$ and the torsional potential $E^{\rm el}$ as a function of the torsion angle $\rho$; and the components of polarizability tensor $\alpha_{qq}$, also being a 
function of $\rho$. To obtain all relevant quantities, we performed quantum chemical calculations using {Density Functional Theory} with B3LYP as 
correlation-exchange functional and an {aug-cc-pVTZ} basis set. The program package of our choice was \textsc{QChem}.\cite{Shao.2014}

The rotational constants $\mak A$ and $\mak B$ we obtained from a geometry optimization. We either optimized the planar structure, having $\rm D_{2h}$ point-group 
symmetry, or the orthogonal structure with $\rm D_{2d}$ point-group symmetry, depending on which of the configurations is lower in energy for the respective molecule. 
Table \ref{rotkonst-bsp} shows the results for the molecules from Fig. \ref{mol-gal}.  There, we also display the reduced rotational constant $\mak B_{\rm red}$, 
\textit{c.f.} \eqref{def-Bred}: Clearly, for most of the molecules, $\mak B_{\rm red}$ is small; with the exception of AAC and \ce{B2F4}, the conditions Eqs. 
\eqref{approx-rot-konst} are reasonable approximations. The results from Table \ref{rotkonst-bsp} suggest that the influence of the field-free coupling is small, and 
therefore has only little influence on the success of torsional alignment, as we shall also see in Section \ref{sec:two-step}.

To calculate the torsional potential, we performed a potential scan by varying the torsional angle $\rho$ in increments of $5^{\circ}$ without allowing the other internal 
coordinates to relax. As reference structure we used either the $\rm D_{2d}$ or $\rm D_{2h}$ configuration, see Table \ref{rotkonst-bsp}. Not allowing the molecule to 
relax, guaranteed a $\rm D_2$-symmetry while the molecules undergo torsion, which is important to ensure that the polarizability tensor and the moment of inertia tensor 
remain diagonal simultaneously. Analogously, we calculated the components of the polarizability tensor by varying the torsional angle $\rho$ in increments of $5^{\circ}$. 
We would like to stress that a scan with allowing the structure of the molecules to relax has lead to only small corrections of both, the polarizabilities and the torsional potential,
which can be neglected.

To actually implement the quantum chemical result for $E^{\rm el}(\rho)$ and $\alpha_{qq}(\rho)$, we interpolate the data and expand them in terms of analytical functions. 
The details of this procedure we describe in the next two Subsections. A critique of our approach to obtaining the molecular parameters quantum chemically we offer in 
Subsection \ref{subsec:crit-dft} of Section \ref{sec:limits}.


\subsection{Symmetry-adapted torsional potentials}
\label{sub-sec:-tor}

Once we have calculated the molecule-specific quantities we need to represent them numerically. As it allows us to calculate the matrix elements of the Hamiltonian 
$\hat{H}^{\rm rt}$ analytically, we expand the torsional potential in terms of a Fourier series
\be
\label{def-pot-tor-num}
E^{\rm el}_0(\rho)
=
\sum_{{s}=0}^{\infty} {\mak V}_{{s}}\cos ({s}\rho)
+
\sum_{{a}=1}^{\infty} {\mak V}_{{a}} \sin ({a}\rho)\;.
\ee
The symmetry of the field-free rotational-torsional Hamiltonian limits the number of non-zero coefficients in the expansion Eq. \eqref{def-pot-tor-num}. As $\hat{H}^{\rm rt}$ is 
invariant in $\rm G_{16}(EM)$, the torsional potential must transform according to the total symmetric representation $\rm A^+_{1g}$. The only functions fulfilling this 
conditions are
\be
\cos (4n\rho) \sim {\rm A^+_{1g}}\;;
\ee
all remaining terms in Eq. \eqref{def-pot-tor-num} are necessarily vanishing. 

Hence, we can expand the torsional potential according to
\be
\label{def-pot-tor}
E^{\rm el}_0(\rho)
=
{\mak V}^{\rm tor}_{0} \sum_{n=0}^{N} {\mak V}_{n}\cos (4n\rho)\;,
\ee
where ${\mak V}^{\rm tor}_{0}$ denotes the torsional potential for $\rho = 0$. We obtained the expansion coefficients ${\mak V}_{n}$ in Eq. \eqref{def-pot-tor} by 
numerical integration, after interpolating our quantum chemical data using the spline function of  \textsc{MatLab}. For Eq. \eqref{def-pot-tor} to be exact, $N$ must be infinite. 
Yet, our calculations showed that we can adequately reproduce our quantum chemical data, if we set $N=7$. A complete list of the expansion coefficients for the molecules from 
Fig. \ref{mol-gal} we display in Table \ref{tor-pot}. Since the planar configuration is not necessarily the configuration highest in energy, we show the quantity
\be
\label{def-dE-tor}
\Delta E^{\rm el}_0 = \max[E^{\rm el}(\rho)] - \min[E^{\rm el}(\rho)]
\ee
in Table \ref{tor-pot} as well, illustrating the electronic energy difference the molecules undergo during the torsion.

The expansion of the torsional potential in terms of analytical functions allows us to calculate the elements of the matrix representing the Hamiltonian for the pure torsion, Eq. 
\eqref{h-tor}, analytically. For further details, in particular the explicit form of these elements, see Appendix \ref{app:matrix-ele}.

\begin{table}[tb!]
\caption{List of the parameters ${\mak V_n}$, ${\mak V}^{\rm tor}_0$ and $\Delta E^{\rm el}_0$ for the molecules we study here; see also Fig. \ref{mol-gal}. The parameter 
$\Delta E^{\rm el}_0$ is defined in Eq. \eqref{def-dE-tor}; for the meaning of the parameters $\mak V_n$ and ${\mak V}^{\rm tor}_0$ see Eq. \eqref{def-pot-tor}. The values
for ${\mak V}^{\rm tor}_0$ and $\Delta E^{\rm el}_0$ are given in units of $\rm meV$.}
\label{tor-pot}

\vspace*{0.5\baselineskip plus 0.125\baselineskip minus 0.075\baselineskip}
\renewcommand{\arraystretch}{1.5}
\begin{tabularx}
{\textwidth}{@{\extracolsep{\fill}}lC{0.95cm} R{0.65cm}R{0.65cm}R{0.65cm}R{0.65cm} R{0.65cm}R{0.65cm}R{0.65cm}R{0.65cm}R{0.95cm}}
\textit{Molecule} 
& ${\mak V}^{\rm tor}_0$  & 
${\mak V}_{0}$ &  ${\mak V}_{1}$ &  ${\mak V}_{2}$ &  ${\mak V}_{3}$ & ${\mak V}_{4}$  & ${\mak V}_{5}$  & ${\mak V}_{6}$  & ${\mak V}_{7}$  
& $\Delta E^{\rm el}_0$ \\
\hline
\ce{B2F4}
& $2.499$ 
& $0.589$ & $0.515$ & $0.061$ & $-0.068$ & $0.019$ & $-0.042$ & $0.092$ &  $-0.008$ 
& $3.080$\\
Biphenyl
& $62.463$ 
& $0.711$ & $-0.416$ & $0.581$ & $0.141$ & $0.004$ & $0.002$ & $0.006$ & $0.003$ 
& $100.463$\\ 
DBBP
& {$53.066$} 
& {$0.853$} & {$-0.645$} & {$0.650$} & {$0.163$} & {$-0.023$} & {$0.015$} & {$-0.013$} & {$0.003$} 
& {$103.015$} \\
Biazu
& $666.739$ 
& $0.255$ & $0.384$ & $0.250$ & $0.078$ & $0.023$ & $0.005$ & $0.003$ & $0.000$ 
& $666.751$\\
AAC 
& $54.054$ 
& $0.404$ & $0.098$ & $0.322$ & $0.163$ & $-0.078$ & $0.065$ & $0.008$ & $-0.005$ 
& $54.054$ \\
\end{tabularx}
\vspace*{0.5\baselineskip plus 0.125\baselineskip minus 0.075\baselineskip}
\end{table}

\subsection{A symmetry-adapted interaction Hamiltonian}
\label{subsec:sym-int-g16}

Analogously to the torsional potential, we can use molecular symmetry to find an adequate numerical representation of our quantum chemical results for the molecular 
polarizability. It is a general result from the theory of (E)MS groups that the  space-fixed components of the polarizability must remain invariant under each operation of the (E)MS 
group. As the polarization vector of the external field transforms invariantly in the MS group too, the Hamiltonian Eq. \eqref{h-int} has to transform according to the total 
symmetric representation in the (E)MS group of the molecule under consideration.\cite{Bunker.1998,Bunker.2005,Grohmann.2011}

Using this argument, we can draw some general conclusions about the symmetry conditions for each term in Eqs. \eqref{alpha-zz} and \eqref{alpha-xx}. Particularly, it holds
\bse
\label{sym-rules-polar}
\begin{align}
\Gamma\big[\alpha^{(0,0)} \big] &= \Gamma_{\rm ts} \\
\Gamma\big[\alpha^{(2,K)} \big] &= \breve\Gamma\big[{\mak D}^{2}_{m,k} + {\mak D}^{2}_{m,-k} \big]\;,
\end{align}
where $\Gamma_{\rm ts}$ denotes the total symmetric representation of the (E)MS group of the molecule, and $\breve\Gamma$ is the contragredient representation of
$\Gamma$. Here, we used the general theorem that the direct product of two irreducible representations $\Gamma_i$, $\Gamma_j$ of any given group contains the 
total symmetric representation only if $\Gamma_i=\breve\Gamma_j=\big(\Gamma^{-1}_j\big)^{\rm T}$.\cite{Hamermesh.1989}  
It holds furthermore
\be
\Gamma\big[\alpha^{(2,0)} \big] = \Gamma\big[{\mak D}^2_{m,0} \big]= \Gamma_{\rm ts} \qquad m=0,\pm 2\;,
\ee
\ese
which follows directly from the transformation properties of the Wigner matrices in the MS group.\cite{Bunker.1998}.

Returning to the specific case of $\rm G_{16}$-type molecules, we can show with the help of the character table displayed in Table IV of the work of 
\citeauthor{Merer.1973}\cite{Merer.1973}  
\bse
\label{wig-mat-sym}
\begin{align}
{\mak D}^2_{m,0}										
&
\sim {\rm A_{\rm 1g}^+} \qquad m=0,\pm 2\\
{\mak D}^2_{m,2}+{\mak D}^2_{m,-2}	
&
\sim {\rm B_{\rm 1g}^+}\qquad m=0,\pm 2 \;.
\end{align}
\ese
Consequently, for the molecule-fixed, irreducible components of $\bm\alpha$ must hold
\bse
\label{polar-irred-sym-b2f4}
\begin{align}
\alpha^{(J,0)}(\rho) 
&
\sim {\rm A_{\rm 1g}^+} \qquad J = 0,2	\\ 
\alpha^{(2,2)}(\rho) 
&
\sim {\rm B_{\rm 1g}^+} \;
\end{align}
\ese
for $\hat{H}^{\rm int}_i,i=1,2,3$ to be invariant in $\rm G_{16}$. Here, we used that for all irreducible representations of $\rm G_{16}$ holds 
$\breve\Gamma_j\simeq\Gamma_j$.

Equations \eqref{sym-rules-polar} clearly show us: quantum chemical results for the polarizability can not be modeled in an arbitrary fashion. For $\rm G_{16}$-type
molecules they have to be written as 
\be
\label{polar-irred-sym}
\alpha^{(J,K)}(\rho) \approx \alpha^{(J,K)}_{0}	 \sum_{n=0}^{N}{\mak P}^{(J,K)}_n \cos\left((4n+K)\rho\right)\;,
\ee
since
\addtocounter{equation}{-1}
\bse
\begin{align}
\cos(4n\rho)			& \sim {\rm A_1^+}	\\
\intertext{and}
\cos((4n+2)\rho) 	& \sim {\rm B_1^+}	\;
\end{align}
\ese
if $n$ is integer. In Eq. \eqref{polar-irred-sym}, $\alpha^{(J,k)}(\rho=0)\equiv\alpha^{(J,K)}_{0}$ with $J=0,2$ and $K=0,2$, respectively. Furthermore, our quantum 
chemical results show that in good approximation for all molecules we are studying here, we can truncate the expansion Eq. \eqref{polar-irred-sym} at $N=2$. All relevant 
parameters we need to calculate the polarizabilities for the molecules from Fig. \ref{mol-gal} are summarized in Table \ref{polar-const}. 
Using the model for the interaction we developed here, we are able to consistently describe the alignment of $\rm G_{16}$-type molecules.

\begin{table}[t!]
\caption{List of the parameters ${\mak P}^{(J,K)}_n$ and $\alpha^{(J,K)}_{0}$ for the molecules we study here; see Fig. \ref{mol-gal}. For the meaning of the 
parameters ${\mak P}^{(J,K)}_n$ and $\alpha^{(J,K)}_{0}$ see Eq. \eqref{polar-irred-sym}. The values for the $\alpha^{(J,K)}_{0}$ are given in units of 
$10^{-40}\nicefrac{\rm Cm^2}{\rm V}$.}
\label{polar-const}

\vspace*{0.5\baselineskip plus 0.125\baselineskip minus 0.075\baselineskip}
\renewcommand{\arraystretch}{1.5}
\begin{tabularx}{\textwidth}
	{@{\extracolsep{\fill}}lC{1.25cm}R{0.75cm}R{0.75cm}R{0.75cm}C{1.25cm}R{0.75cm}R{0.75cm}R{0.75cm}C{1.25cm}R{0.75cm}R{0.75cm}R{0.75cm}}
\textit{Molecule}
& $\alpha^{(0,0)}_{0}$
& ${\mak P}^{(0,0)}_0$ &  ${\mak P}^{(0,0)}_1$ &  ${\mak P}^{(0,0)}_2$
& $\alpha^{(2,0)}_{0}$
& ${\mak P}^{(2,0)}_0$ &  ${\mak P}^{(2,0)}_1$ &  ${\mak P}^{(2,0)}_2$
& $\alpha^{(2,2)}_0$
& ${\mak P}^{(2,2)}_0$ &  ${\mak P}^{(2,2)}_1$ &  ${\mak P}^{(2,2)}_2$	\\
\hline
\ce{B2F4}
& $7.226$  
& $1.001$ &$-0.001$ & $0$ 
& $0.514$
& $1.020$ & $-0.021$ & $0.001$ 
& $0.952$
& $0.997$ & $-0.004$ & $0$ \\
Biphenyl 
& $39.194$
& $0.978$ &$-0.024$ & $-0.002$ 
& $15.295$
& $0.868$ & $0.133$ & $-0.002$ 
& $8.847$
& $-1.005$ & $0.003$ & $0.001$ \\
DBBP
& $54.555$
& $0.968$ &$0.034$ & $-0.002$ 
& $29.706$
& $0.892$ & $0.110$ & $-0.002$ 
& $8.615$
& $-1.005$ & $0.003$ & $0.001$ \\
Biazu
& $86.980$
& $0.949$ & $0.051$ & $-0.001$ 
& $57.774$
& $0.871$ & $0.124$ & $0.042$ 
& $15.081$ 
& $-1.005$ & $0.002$ & $0.001$ \\
AAC 
& $119.260$
& $0.980$ &$0.023$ & $-0.002$ 
& $33.590$
& $0.831$ & $0.179$ & $-0.008$ 
& $45.851$
& $-1.004$ & $-0.003$ & $0$ \\
\end{tabularx}
\vspace*{0.5\baselineskip plus 0.125\baselineskip minus 0.075\baselineskip}
\end{table}

\subsection{Additive and non-additive models for the molecular polarizability}
\label{subsec:pol-add}

In many studies on torsional control, however, a simplified model for the interaction was used, which is based on the additivity scheme of molecular 
properties.\cite{Ramakrishna.2007,Parker.2011,Parker.2012,Floss.2012,Ashwell.2013,Ashwell.2013b,Coudert.2011,Ortigoso.2013,Coudert.2015}
If we employ the additive model, we assume the polarizability of the molecule to be a sum of the polarizabilities of molecular subunits.\cite{Bonin.1997} 
Using this scheme, the Hamiltonian $\hat{H}^{\rm int}$, \textit{c.f.} Eq. \eqref{h-int}, for a circular-polarized laser pulse in the four-dimensional case is explicitly given 
by\cite{Ortigoso.2013}
\be
\label{h-int-add}
\hat{H}^{\rm int} 
=
- \frac{|{\epsilon}(t)|^2}{8}\left( \alpha^{(0,0)}_0 - \alpha^{(2,0)}_0 \cos^2\theta +\alpha^{(2,2)}_0 \sin^2\theta\cos2\chi \cos 2 \rho\right)\;,
\ee
see Eqs. \eqref{trans-cart-irred} in Appendix \ref{app:der-H-int} for the definitions of the irreducible components of the molecular polari\-zability. 

By comparing Eqs. \eqref{h-int}, \eqref{alpha-qq} and \eqref{polar-irred-sym} with Eq. \eqref{h-int-add} we conclude that we obtain the Hamiltonian within the additive 
model by only taking the leading terms of the expansions from Eqs. \eqref{polar-irred-sym} into account. Consequently, the additive model is a good approximation if the 
Fourier series are converging reasonably fast. From this it also follows that the field-induced rotational-torsional coupling is minimized if the additive model is applied. Then, only 
one term of the polarizabilities depends on $\rho$, limiting the possible excitations due to the external fields.

As Table \ref{polar-const} shows, some molecules we are considering here meet the conditions prescribed by the additive model. The component $\alpha^{(2,2)}$ of 
\ce{B2F4}, for example, can be written in very good approximation as
\be
\alpha^{(2,2)}
\approx
\alpha^{(2,2)}_0\cos2\rho\;,
\ee
while the change of the other two components, compared to the change in $\alpha^{(2,2)}$, is negligible and they can be therefore considered to be 
constant. For the other molecules, however, the torsional dependence of $\alpha^{(0,0)}$ and $\alpha^{(2,0)}$ need to be taken into account. As a general trend, one
might say, the additive model is the worse the more polarizable the molecules under consideration are.


\section{On the two-step model}
\label{sec:two-step}

Recently, we have presented some of our main results on the systematic comparison of the 4D model with the conventional 2D approach to torsional 
control.\cite{Grohmann.2017}
In the following, we cast a more detailed glance on our findings. Not only we give more examples that underline our recently published interpretations; we also argue why the 
broad conclusions of earlier studies\cite{Coudert.2011,Ortigoso.2013,Coudert.2015}  
are limited to the scenario they consider, and why their calculations could be generally flawed due to a lack of convergence. To address the critique raised in these works, we 
systematically study the influence of the field-free and field-induced rotational-torsional coupling on the rotational-torsional alignment for the molecules from Fig. \ref{mol-gal}. 
Moreover, we discuss in detail what the conditions are for the 2D model to be a reasonable approximation to the 4D model, and we illustrate why the theoretical description of 
the polarizability is closely related to answering this question. We therefore provide the theoretical basis why in certain cases we have to extend the conventional 2D model 
towards a generalized 2D model. We close this Section with a detailed theoretical analysis of our results in order to underscore why our conclusions are general.


\subsection{General results from specific examples? Our approach}
\label{subsec:procedure}

But how can we, at all, draw general conclusions? The greatest challenge in molecular physics is the complexity of molecules, making every molecule a specific example and
formulating general rules that apply to every molecule difficult, if not impossible. Limiting the theoretical framework to the closed-system semi-rigid-rotor approach, we are able 
to fully characterize different molecules by a small set of numbers: the rotational constants ${\mak A},{\mak B}$, the torsional potential $E^{\rm el}_0(\rho)$, and the 
components of the molecular polarizability $\alpha^{(J,K)}$, see Tables \ref{rotkonst-bsp}, \ref{tor-pot}, and \ref{polar-const}, respectively. However, using this rather simple 
approach, we ignore a number of phenomena that may have an impact on torsional control, depending on the experimental setup, which we discuss in Section \ref{sec:limits}.

Moreover, to directly compare the molecules, we adjust the torsional barrier and the field strength for each molecule to a reference system, which we choose to be \ce{B2F4}. For 
all remaining molecules from Fig. \ref{mol-gal}, we scale the effective torsional barrier
\be
{\mak V}^{\rm eff} \equiv \frac{{\mak V}_0^{\rm tor}}{\mak F}
\ee
such that it is identical with the effective barrier of \ce{B2F4}. Accordingly, we adapt the effective field strength
\begin{subequations}
\begin{equation}
\label{P-ad}
{\tt P}^{(2,2)}_1
\equiv
\alpha^{(2,2)}_0 \cdotp |\epsilon^{\rm max}_1|^2
\end{equation}
for the nanosecond pulse ${\bm E}_1$ and
\begin{equation}
\label{P-nad}
{\tt P}^{(2,2)}_2
\equiv
\alpha^{(2,2)}_0 \cdotp \overline{\epsilon^2_2}
\end{equation}
\end{subequations}
for the femtosecond pulse ${\bm E}_2$. We pursued a similar strategy in earlier works.\cite{Leibscher.2003,Grohmann.2011,Floss.2012}

We stress, however, that in contrast to symmetric tops and linear molecules, it is not possible to define a dimensionless form of the Schr\"odinger equation that is identical for all
molecules. The explicit shape of the torsional potential $E^{\rm el}_0(\rho)$, Eq. \eqref{def-pot-tor}, the coordinate dependence of the polarizabilities $\alpha^{(J,K)}$, Eq.
\eqref{polar-irred-sym}, and the ratio of the rotational constants $\mak A$ and $\mak B$ is different for all molecules we consider here. We are therefore not able to 
completely eliminate the molecule-specificity of our results. Yet, \rm{as we show hereafter,} we still can identify fundamental mechanisms that are decisive for answering the 
question if the 2D model is a good approximation to the 4D approach to torsional control. In this Section, we limit our discussion to some illustrative results; in the supplemental 
material, we provide more examples that strengthen the arguments we present in the following.


\thisfloatsetup{capposition=beside,capbesideposition={inside,bottom},floatwidth=10cm}
\begin{figure}[t!]
\vspace*{1\baselineskip plus 0.125\baselineskip minus 0.075\baselineskip}

\centering{\includegraphics[width=10cm]{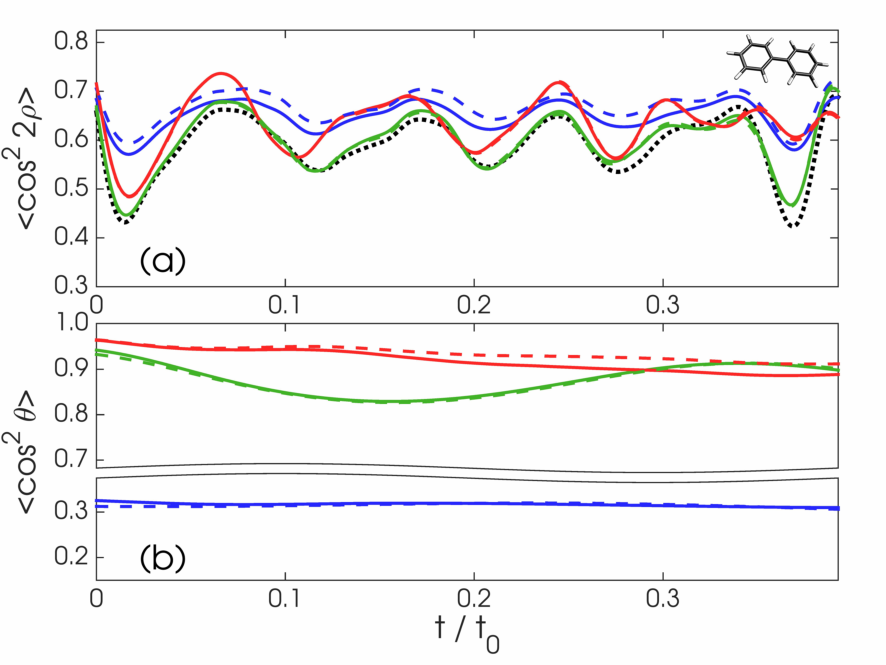}}

\caption{Alignment factors  $\langle \cos^2 2 \rho \rangle$ (a) and $\langle \cos^2 \theta \rangle$ (b) of Biphenyl after interacting with a short, $x$-polarized pulse with 
intensity $I_2 =5.4\, \nicefrac{\rm TW}{\rm cm^2}$ and duration $\tau= 150$ fs. The solid curves display the results of four dimensional simulations in the presence of a 
$z$-polarized pulse with $I_1= \textcolor{blue}{0},\textcolor{green}{59.5},\textcolor{red}{297.5}\, \nicefrac{\rm GW}{\rm cm^2}$. Time is given in units of 
$t_0=\nicefrac{\hbar}{\mak A}= 54.8$ ps. For all calculations $J^{\rm max}=20$.}

\label{BP_coup}
\end{figure}

\subsection{Torsional alignment in four dimensions: a second look}
\label{subsec:second-look}

As a first example, we discuss the rotational-torsional alignment of Biphenyl. In Fig. \ref{BP_coup}, we show the rotational (lower panel) and the torsional (upper panel) alignment 
factors, ${A}_{\theta}= \langle \cos^2 \theta \rangle$ and 
${A}_{2\rho}=\langle \cos^2 2\rho \rangle$, from 4D calculations for three different adiabatic pulse strengths, 
$I_1= \textcolor{blue}{0},\,\textcolor{green}{59.5},\,\textcolor{red}{297.5}\, \nicefrac{\rm GW}{\rm cm^2}$, 
and for a femtosecond laser pulse with intensity 
$I_2 =5.4\, \nicefrac{\rm TW}{\rm cm^2}$.
We compare them with 2D calculations for $I_2 =5.4\, \nicefrac{\rm TW}{\rm cm^2}$, which are depicted by black dotted lines.

The best agreement between the 2D and 4D simulations we obtain for moderate adiabatic pulse strengths (green line in the upper panel of Fig. \ref{BP_coup}). Here, the 2D 
model almost completely reproduces the torsional alignment $A_{2\rho}$ obtained from 4D calculations.  We can also see that {the variation of} $A_{2\rho}$ is reduced if no 
adiabatic field is applied (blue line in the upper panel of Fig. \ref{BP_coup}), and thus, no attempt is made to align the molecules along their main principal axis. If, on the other 
hand, the intensity of the adiabatic pulse is very high (red line in upper panel of Fig. \ref{BP_coup}), the agreement between 2D model and 4D is again less pronounced than for 
an adiabatic pulse with moderate intensity, {a result that we have also observed for B$_2$F$_4$.} \cite{Grohmann.2017}
We find it important to note, however, that the change in $A_{2\rho}$ due to the interaction with the laser pulses is highest for the strongest adiabatic pulse. Here, the 2D 
model underestimates the degree of torsional alignment. Contrary to earlier studies\cite{Coudert.2015}
we therefore conclude that less congruence between 2D and 4D simulations does not necessarily correspond to a worse alignment within the 4D model.
 
As we consider Biphenyl, one of the most intensively studied molecules when it comes to strong field control of torsions, we should comment more extensively on some of the 
differences of our study compared to earlier works.\cite{Coudert.2011,Ortigoso.2013,Floss.2012,Coudert.2015,Ashwell.2013,Ashwell.2013b} 
These disparities derive in parts from different quantum chemical results on the torsional potential, a different definition of the torsional constant, and a different description of 
the molecular polarizabilities. To describe the molecular polarizabilities, for example, all of the cited studies use the additive model, leading to very different degrees of torsional 
excitation, as we discuss in Subsection \ref{subsec:add-mod} of this Section. Further, the shape of the torsional potential can substantially differ if different quantum chemical 
methods with different basis sets are employed.\cite{Ashwell.2013,Ashwell.2013b} 
Hence, to compare our studies to earlier works, these differences in methodology have to be taken into account.

For the rotational motion, our results show that the adiabatic pulse effectively aligns the molecules along their principal axis, see green and red lines in lower panel of Fig. 
\ref{BP_coup}. Even for moderate field intensities (green line), the molecules show almost perfect alignment. Moreover, we see that the rotational alignment factors change 
only little in time, irrespective of the pulse strength. Thus, the rotational motion perpendicular to the main principal axis occurs on a timescale that is significantly longer than 
the timescale on which the torsional dynamics takes place.

Consequently, in case of Biphenyl, we observe the same behavior we have seen earlier for \ce{B2F4}, see Fig. 2 in Ref. \citenum{Grohmann.2017}.
For moderate adiabatic pulses, the presumption of the two-step mechanism illustrated in Fig. \ref{mod-2step} is acceptable: the first laser pulse effectively controls the rotation,
before the femtosecond laser pulse selectively excites the torsional motion of the molecule. We have found similar results for almost all molecules from Fig. \ref{mol-gal}; an 
exception is Biazu, as we shall see in Subsection \ref{subsec:add-mod} of this Section.

%
%

\thisfloatsetup{capposition=beside,capbesideposition={inside,bottom},floatwidth=10cm}
\begin{figure}[tb!]
\vspace*{1\baselineskip plus 0.125\baselineskip minus 0.075\baselineskip}

\centering{\includegraphics[width=10cm]{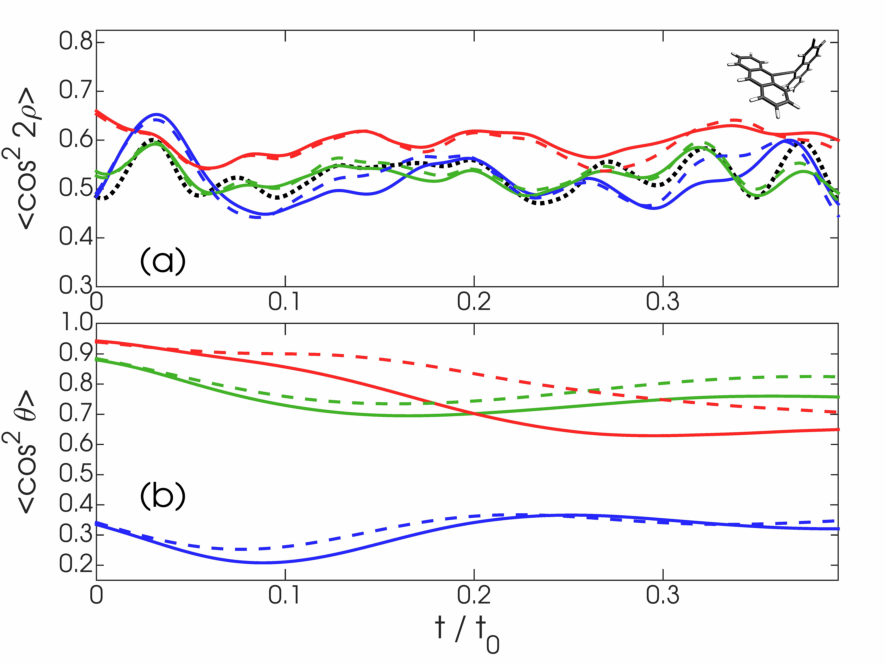}}

\caption{Alignment factors  $\langle \cos^2 2 \rho \rangle$ (a) and $\langle \cos^2 \theta \rangle$ (b) of AAC after interacting with a short, $x$-polarized pulse with 
intensity $I_2 =1.04\, \nicefrac{\rm TW}{\rm cm^2}$ and duration $\tau= 150$ fs. The solid curves display the results of four dimensional simulations in the 
presence of a $z$-polarized pulse with $I_1= \textcolor{blue}{0},\textcolor{green}{0.9},\textcolor{red}{4.5} \, \nicefrac{\rm GW}{\rm cm^2}$. Time is 
given in units of $t_0= \nicefrac{\hbar}{\mak A}= 702.2$ ps. For all calculations $J^{\rm max} =20$.}

\label{AAC_coup}
\end{figure}

\subsection{How important are field-free rotational-torsional couplings?}
\label{subsec:couplings-res}

In the following, we examine in more detail the argument that rotational-torsional couplings are the reason why in previous 
studies\cite{Coudert.2011,Ortigoso.2013,Coudert.2015} 
the torsional alignment has been seen to be reduced in 4D simulations when compared to 2D simulations. Yet, as we have pointed out in Section \ref{sec:coupling} and in the 
previous Subsection, the mechanisms of rotational-torsional couplings are complex; here, we unravel the influence of the field-free coupling on the rotational and torsional 
alignment of the molecules from Fig. \ref{mol-gal}. Researching this type of coupling is crucial as it is inherent to the molecule. Being completely determined by the reduced 
rotational constant, it cannot be controlled, or even modified by the external laser field. To quantify the impact of the field-free rotational-torsional coupling, we have run 
simulations applying the conditions Eqs. \eqref{approx-rot-konst} in order to eliminate the coupling and compared them to simulations including the full coupling.

Consider AAC as a first example. Among the molecules we have studied, it has one of the largest reduced rotational constants (${\mak B}_{\rm red}=0.19$, see Table 
\ref{rotkonst-bsp}), which is why we expect the influence of the field-free coupling on the rotational-torsional alignment to be most distinct. Figure \ref{AAC_coup} shows the 
rotational (lower panel) and the torsional (upper panel) alignment factors, ${A}_{\theta}$ and ${A}_{2\rho}$, for three different adiabatic pulse strengths, 
$I_1= \textcolor{blue}{0},\textcolor{green}{0.9},\textcolor{red}{4.5}\, \nicefrac{\rm GW}{\rm cm^2}$, 
and for a femtosecond laser pulse with intensity 
$I_2 =1.04\, \nicefrac{\rm TW}{\rm cm^2}$. 
Calculations including the field-free coupling are depicted by solid lines; calculations without field-free coupling correspond to dashed lines. Clearly, the field-free coupling has 
only little influence on the alignment, in particular on the torsional alignment factor. For the rotations, simulations with and without the field-free rotational-torsional coupling 
differ more. Here, the influence of the field-free coupling is most distinct for adiabatic pulses with high intensity (red lines), while for the torsion the field-free 
coupling is visible the most in case no adiabatic pulse is applied (blue lines). For rotations, the influence of the coupling is negative, \textit{i.e.} it reduces the alignment factor 
$A_\theta$ compared to simulations neglecting the coupling. Moreover, we see that the effect of the field-free coupling on the rotational-torsional alignment becomes more 
influential as time evolves. This effect was also observed for other types of couplings.\cite{Ramakrishna.2005b,Ramakrishna.2006,Ashwell.2013b} 
For B$_2$F$_4$, which has a comparable rotational constant, we have obtained similar results as for AAC, see Ref. \citenum{Grohmann.2017}.

For Biphenyl, we observe an even less pronounced effect; compare dashed and solid lines in Fig. \ref{BP_coup}. Having a very small reduced rotational constant 
(${\mak B}_{\rm red}=0.087$, see Table \ref{rotkonst-bsp}), the field-free coupling has almost no influence on the alignment factors for rotations and torsion alike. The same 
effect we have seen for DBBP, see Fig. 3 in Ref. \citenum{Grohmann.2017}, and we can also observe it for other molecules, see supplemental material. Thus, while the field-free 
coupling has indeed a negative effect on the rotational alignment, the effect is rather small, even for molecules with large ${\mak B}_{\rm red}$.

\thisfloatsetup{capposition=beside,capbesideposition={inside,bottom},floatwidth=10cm}
\begin{figure}[t!]
\vspace*{1\baselineskip plus 0.125\baselineskip minus 0.075\baselineskip}

\centering{\includegraphics[width=10cm]{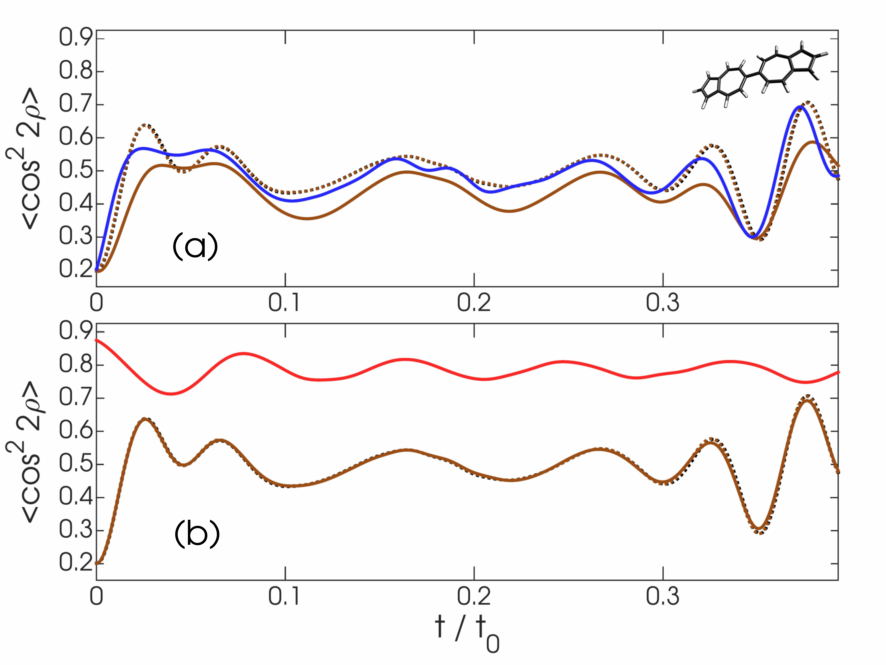}}

\caption{\textit{(a):} Torsional alignment factor  $\langle \cos^2 2 \rho \rangle$ for Biazu after interacting with a femtosecond laser pulse of intensity 
$I_{2}=3.42\, \nicefrac{\rm TW}{\rm cm^2}$ and duration $\tau= 150$ fs in case no nanosecond pulse is applied first. Calculations with the full form of the polarizabilities are 
depicted in blue, calculations using the additive model are represented by brown lines; 2D simulations with quantum chemical polarizabilities are pictured by black dotted lines, 
2D simulations using the additive model are illustrated by dotted brown lines. Time is given in units of $t_0= \nicefrac{\hbar}{\mak A}= 110.4$ ps. \textit{(b):} Analogue 
calculations for a nanosecond pulse with intensity $I_1 = \textcolor{red}{85}\, \nicefrac{\rm GW}{\rm cm^2}$.
}

\label{biazu_add}
\end{figure}

\subsection{The crucial role of field-induced couplings: the additive model and torsional alignment}
\label{subsec:add-mod}

The \textit{prima facie} presumption is therefore that the disagreement between 2D and 4D simulations, if it occurs, is a result of the field-induced coupling. As we explain
in Subsection \ref{subsec:field-coupling} of Section \ref{sec:coupling} and \ref{subsec:pol-add} of Section \ref{sec:4d-mod}, this type of rotational-torsional coupling is directly 
related to the polarizability of the molecule, and it is minimized if we use the additive model instead of the full quantum chemical model to describe the molecular polarizabilities. 
In what follows, we compare the impact both models have on the rotational-torsional alignment. We discuss Biazu as an example. Table \ref{polar-const}, line 4 shows us that for 
this molecule in particular, the additive model is a bad approximation. We therefore expect the effect of the field-induced coupling to be most visible. 

First, we consider the scenario where no adiabatic pulse is applied and the intensity of the femtosecond pulse is $I_{2}=3.42\, \nicefrac{\rm TW}{\rm cm^2}$. The upper 
panel of Fig. \ref{biazu_add} displays the torsional alignment obtained from 4D (solid brown line) and 2D (dotted brown line) simulations employing the additive model and 
compares them to 4D calculations with the full form of the polarizabilities (blue line).  We observe that the torsional alignment factor obtained from 4D simulations with additive 
model and the full form of the polarizabilities differ: here, the additive model slightly underestimates the degree of torsional alignment. Moreover, the 2D simulations with the full 
form of the polarizability almost  coincides with the corresponding 4D simulations, an exception to the results from other molecules, as mentioned in Section 
\ref{subsec:second-look}. This is at first glance surprising, since  the 2D model assumes perfectly aligned molecules while in the 4D simulations, the angular distribution of the 
molecules is isotropic. We attribute this effect to the strong $\rho$-dependence of the term $\alpha^{(0,0)}$, see line 4 in Table \ref{polar-const}, since this term leads to 
excitation of torsion independent from the rotational state of the molecules, see Eqs. \eqref{alpha-zz} and \eqref{alpha-xx}.

Comparing the 2D with employing (brown dotted line) and without employing (black dotted line) the additive model with 4D simulations (solid brown line) based on the additive 
model, we also learn that the 2D simulations clearly overestimate the degree of torsional alignment. Yet, we also observe that the 4D simulations using quantum chemical 
polarizabilities and the 2D simulations are in better agreement than the 2D simulations and the 4D simulations based on the additive model. Thus, if the additive model is 
employed and the field-induced coupling is minimized, the torsional alignment is underestimated indeed. These results support the argument of earlier 
studies,\cite{Coudert.2011,Ortigoso.2013,Coudert.2015}   
which considered exactly this case.

\begin{figure}[t]
\vspace*{1\baselineskip plus 0.125\baselineskip minus 0.075\baselineskip}

\centering{\includegraphics[width=10cm]{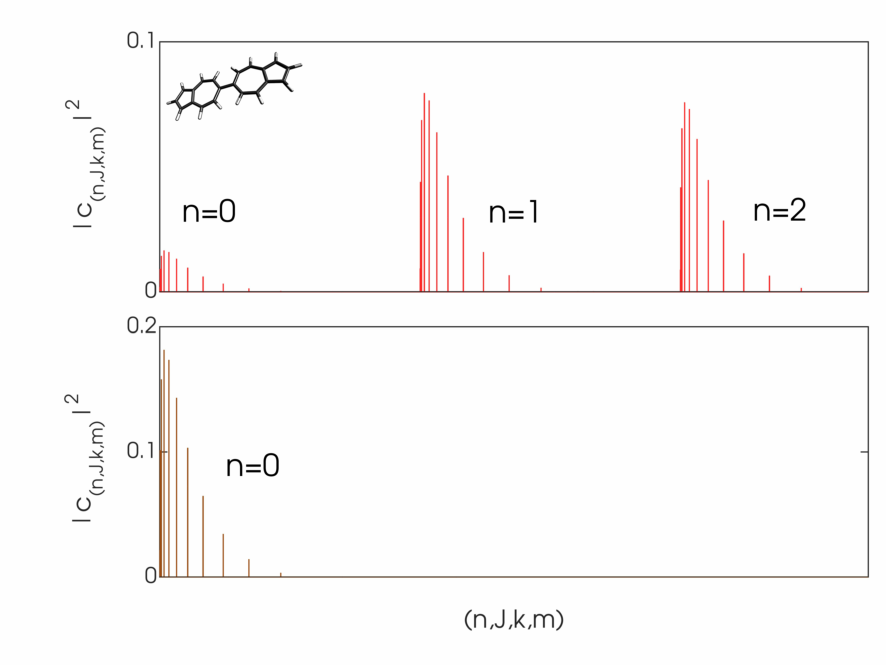}}

\caption{Expansion coefficients from Eq. \eqref{ansatz-basis} for the lowest pendular state of Biazu after interacting with a nanosecond laser pulse having the intensity 
$I_1 = 85\,\nicefrac{\rm TW}{\rm cm^2}$ \textcolor{brown}{with} and \textcolor{red}{without} employing the additive model for the molecular polarizabilities.}

\label{Biazu_coeff}
\end{figure}

The picture changes, however, if we turn to the case in which the molecule interacts with a nanosecond pulse while a short pulse excites a rotational-torsional wavepacket. In Fig.
\ref{biazu_add}, lower panel, we show our results for $I_1 = \textcolor{red}{85}\, \nicefrac{\rm GW}{\rm cm^2}$ and $I_{2}=3.42\, \nicefrac{\rm TW}{\rm cm^2}$.
Considering 4D simulations using the full form of the polarizabilities (red line) and comparing them with respective simulations without applying an adiabatic pulse
(blue line in the upper panel of Fig. \ref{biazu_add}), the torsional alignment factor is already increased at $t=0$. In this case, the pendular state created by the nanosecond 
pulse contains not only excited rotational states; the first and second excited torsional states are also populated, as it can be seen in the upper panel of Fig. \ref{Biazu_coeff}. 
The adiabatic excitation of torsional  states is a result of the strong $\rho$-dependence of the polarizabilities, in particular of the terms $\alpha^{(0,0)}$ and $\alpha^{(2,0)}$, 
see line 4 in Table \ref{polar-const}. Thus, continuing our discussion from Section \ref{subsec:field-coupling}, aligning Biazu adiabatically with a strong nanosecond laser pulse 
represents a case where the $\rho$-dependence of $\alpha^{(2,0)}$ is so strong that the molecule cannot be aligned without exciting torsional states. An analogue effect of 
adiabatic torsional alignment, we also observed for AAC, see Fig. \ref{AAC_coup}, and DBBP see Fig. 4 in Ref. \citenum{Grohmann.2017}. 

Furthermore, these results support our conclusion that within the additive model for the molecular polarizabilities, which neglects the $\rho$-dependence of  $\alpha^{(2,0)}$, 
the field-induced coupling is minimized. When employing the additive model, no excited torsional states contribute to the pendular state, see lower panel of Fig. 
\ref{Biazu_coeff}, and thus no adiabatic torsional alignment occurs, as it can be seen from the brown line in Fig. \ref{biazu_add}, right panel. Notably, if we apply the additive 
model and therefore minimize the field-induced coupling, the 2D model (brown dotted lines in Fig. \ref{biazu_add}) does not overestimate but it underestimates the degree of 
torsional alignment. Hence, contrary to earlier findings,\cite{Coudert.2015}
2D calculations using the additive model underestimate the degree of torsional alignment in certain cases. 

Based on our results, we can moreover relate the validity of the 2D model and the additive model. Using the 2D model, we assume that the molecules a perfectly aligned 
without exciting any torsional states. This assumption is only valid if the $\rho$-dependence of $\alpha^{(2,0)}$ can be neglected, as it is done within the additive model. If we 
apply the additive model, 2D and 4D simulations agree almost perfectly; see left and middle panel of Fig. \ref{biazu_add}. Consequently, if the additive model is a good 
approximation to the molecular polarizabilities, the 2D model reproduces the torsional alignment obtained from a 4D calculation. This coincidence of 2D and 4D simulations 
based on the additive model we observe for all molecules we have studied; see supplemental material and Fig. 4 in Ref. \citenum{Grohmann.2017}.
By tendency, the correlation of both approaches is the more pronounced the more intense the adiabatic laser pulse is.

Summarizing Subsections \ref{subsec:second-look}, \ref{subsec:couplings-res}, and \ref{subsec:add-mod}, we identify four main results giving some indication about the 
nature of the rotational-torsional couplings: (1) the dominant coupling effect is the field-induced rotational-torsional coupling; (2) the effects originating from the field-induced 
coupling are not necessarily negative, but they rather assist the torsional alignment; (3) if the field-induced coupling is minimized, \textit{e.g.} by employing the additive model, 
good, if not excellent agreement between 2D and 4D model is expected; and (4) if the additive model fails because of a strong $\rho$-dependence of the polarizabilities, 
additional effects occur, namely adiabatic torsional alignment during the nanosecond pulse.

\thisfloatsetup{capposition=beside,capbesideposition={inside,bottom},floatwidth=10cm}


\subsection{Couplings as numerical artifacts? On convergence}
\label{subsec:convergence}

\begin{figure}[t]
\vspace*{1\baselineskip plus 0.125\baselineskip minus 0.075\baselineskip}

\centering{\includegraphics[width=10cm]{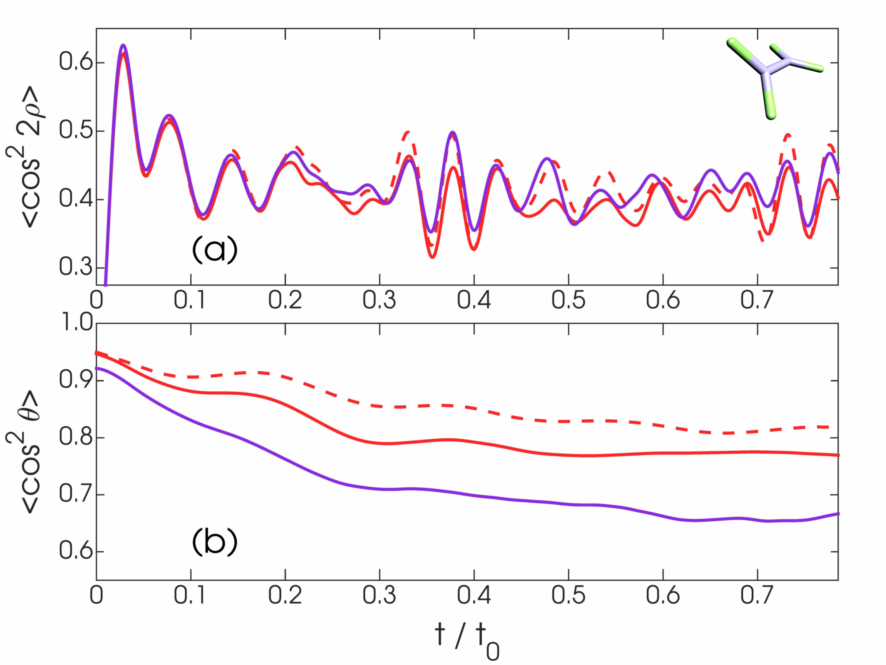}}

\caption{Alignment factors  $\langle \cos^2 2 \rho \rangle$ (a) and $\langle \cos^2 \theta \rangle$ (b) of B$_2$F$_4$ after interacting with a short, $x$-polarized pulse with 
intensity $I_2 =50\, \nicefrac{\rm TW}{\rm cm^2}$ and duration $\tau= 150$ fs. The solid curves display the results of four dimensional simulations in the presence of a 
$z$-polarized pulse with $I_1=5 \, \nicefrac{\rm TW}{\rm cm^2}$ for $J^{\rm max}= \textcolor{red}{20}$ and for $J^{\rm max}=\textcolor{violet}{10}$; dashed lines 
represent calculations without field-free rotational-torsional couplings. Time is given in units of $t_0= \nicefrac{\hbar}{\mak A}= 30.3$ ps.}

\label{b2f4_conv}
\end{figure}

Before we provide a more elaborated theoretical explanation of our results, we discuss a further important aspect: convergence. Our studies have unearthed that convergence is 
crucial when calculating the alignment factors $A_{\theta}$ and $A_{2\rho}$. Using the example of \ce{B2F4}, Fig. \ref{b2f4_conv} shows calculations for different basis set 
sizes, including (solid lines) and excluding (dashed lines) the field-free rotational-torsional coupling. In all calculations, $I_1=5 \, \nicefrac{\rm TW}{\rm cm^2}$  and  $I_2 =50\, 
\nicefrac{\rm TW}{\rm cm^2}$; red lines represent converged calculations, \textit{i.e.} $J^{\rm max}=20$ and ${K}^{\rm max}_{\rho}=250$, purple lines correspond to 
calculations with $J^{\rm max}=10$ and ${K}^{\rm max}_{\rho}=50$.

\thisfloatsetup{capposition=beside,capbesideposition={inside,bottom},floatwidth=10cm}

What we can observe here is a distinct correlation between the basis set size and the degree of the rotational-torsional alignment: the larger the basis set size, the less the 
degree of alignment is reduced as time evolves. Consequently, the degree of alignment is underestimated if the basis set is too small. The effect is more dominant for rotations 
than for the torsion. 

We can understand this result if we recall that pendular states corresponding to a high degree of alignment in $\theta$ are very narrow in the angular space, thus requiring a 
large number of field-free energy eigenstates for an adequate numerical representation. Additionally, the number of energy eigenstates in the rotational manifold is, in general, 
much larger than for the torsion, because there are three rotational degrees of freedom we have to represent numerically. Hence, if we wish to describe a molecule that is highly 
aligned along its molecular axis, we need a large rotational-torsional basis.

To conclude that calculations with a small basis set overestimate the rotational-torsional coupling is, however, wrong: For small basis sets, the results from simulations 
including the field-free coupling are indistinguishable from those neglecting the coupling; see purple lines in Fig. \ref{b2f4_conv}. This result, too, is what you would expect: 
Small basis sets are known to be inappropriate for describing energy spectra that consist of groups of levels close in energy but with large differences between different 
groups. However, this is exactly the case for field-dressed states, in which the field-free rotational-torsional coupling leads to small energy splittings. Considering that in earlier 
studies on four-dimensional rotational-torsional alignment only basis sets with $J^{\rm max}=10$ were used,\cite{Coudert.2011,Ortigoso.2013,Coudert.2015}
we conclude that in these works the field-free coupling is not adequately described. Very recent works on the rotational-torsional alignment of biphenyl-like systems in 
electronically excited states also point to the importance of convergence when describing rotational-torsional couplings.\cite{Omiste.2017}
Taking furthermore into account our results from Subsection \ref{subsec:second-look}, \ref{subsec:couplings-res} and \ref{subsec:add-mod} of this Section, it is more likely that 
the negative results in Refs. \citenum{Coudert.2011,Ortigoso.2013,Coudert.2015} are due to the lack of convergence, rather than the field-free coupling of the rotations and the 
torsion.

Additionally, we see how unreliable, in general, our simulations become if we choose a small, yet numerically feasible basis set size. In all of our calculations, we had to use 
large basis sets with at least $J^{\rm max}=20$ and ${K}^{\rm max}_{\rho}=250$ to obtain converged results. Our calculations therefore took rather long and, more 
importantly, had a high demand in memory. Truly converged results we only obtained for \ce{B2F4}; for all remaining molecules, we were still able to observe slight changes in 
the alignment when enlarging the basis set. Consequently, calculations for molecules like DBBP, AAC, Biphenyl and Biazu with unscaled torsional barrier that are reliable are 
numerically unfeasible. To study the torsional alignment of these systems, our conclusion is, we need to develop alternative strategies in order to reduce dimensionality and 
hence the numerical effort. Taking into account that most of the systems being studied till date are of lower symmetry, advancing such strategies becomes even more important. 
In case the symmetry group of the molecule is smaller, the number of basis states that can be coupled by the field is larger, and thus, larger basis sets are required to describe 
pendular states of a given symmetry accurately. Here, we do not discuss how to develop these methods, but we will readdress this problem in a future publication.


\subsection{Why the conventional two-dimensional model fails}
\label{subsec:2D-fail}

After having discussed our results in detail, we now analyze the underlying mechanisms making the 2D model a good or bad approximation to the 4D model. Therefore, we also 
explore what processes in general might be responsible if the 2D model fails to describe torsional control appropriately, which might allow us to go beyond our particular 
empirical findings and to draw some general conclusions under which models of reduced dimensionality are appropriate.  

To do so, we first need to reflect the implicit presuppositions the conventional 2D model makes. One premise of the 2D model is that the molecule under consideration is 
perfectly aligned along its axis of torsion, i.e. the molecule-fixed ${\bm e}_Z$-axis is parallel or anti-parallel to the space-fixed ${\bm e}_z$-axis. What remains are the two 
coordinates $\chi$ and $\rho$, the rotation about the main principal axis and the torsion of the two molecular moieties. 

The second assumption being made within the 2D conventional approach is that neither the torsion nor the rotation about the ${\bm e}_Z$-axis shall be excited during the 
alignment. This argument is reflected by the conditions conventional studies on torsional alignment are starting from: They assume the initial state of the molecules to be 
adequately represented by the ground state of the 2D field-free Hamiltonian, $\hat{H}^{\rm 2D}$, \textit{c.f.} Eqs. 
\eqref{h-2d-ff}.\cite{Hoki.2001,Kroner.2003,Ramakrishna.2007,Parker.2011,Grohmann.2007,Floss.2012,Parker.2012,Ashwell.2013,Ashwell.2013b} 
For this assumption to be right, the rotational projection quantum numbers $k$ and $m$ as well as the torsional quantum number ${\mathtt n}_{\rho}$ need to
be conserved during the process of alignment. To judge this assumption, we therefore have to identify the terms in the Hamiltonians Eqs. \eqref{h-rt} and \eqref{h-int-sce} of 
which excitations of these kind could originate from.

The third assumption is what we call the frozen mode approximation.  When employing the conventional 2D model, it is anticipated that during the process of torsional control 
the rotations perpendicular to the torsional axis can be considered to be fixed rather than adiabatically separated. Within this picture, the torsion $\rho$ and the rotation $\chi$ 
perpendicular to the main principal axis are too fast for the modes described by $\theta$ and $\phi$ to adapt to new configurations in $\rho$ and $\chi$. Only if this assumption 
is reasonable, it is legitimate to ignore motions along $\theta$ and $\phi$. 

In the following, we discuss the three assumptions---perfect alignment along the main principal axis, conservation of the quantum numbers $k$, $m$, ${\mathtt n}_{\rho}$, 
and the frozen mode approximation---separately. The first assumption, our results confirm, is reasonable: In all of our simulations, we observe that it is possible to almost 
perfectly align the molecules adiabatically with moderate intense laser pulses, see Figs. \ref{BP_coup} and \ref{AAC_coup} in this Section, the supplemental material, and Figs. 2 
and 3 in Ref. \citenum{Grohmann.2017}. 
Although it might be wrong to model the interaction with the first laser pulse as an adiabatic process, see Subsection \ref{subsec:crit-adiab} of Section \ref{sec:limits}, we see 
no argument to assume that it is fundamentally impossible to align the molecules along their principal axis.

The second assumption, namely the condition that no torsion or rotation about the axis of torsion is excited by the first laser pulse, is, however, dubious. The field-free and the 
field-induced coupling both prevent the quantum numbers $k$ and ${\mathtt n}_{\rho}$ to be conserved during the process of adiabatic alignment. To illustrate why, we begin 
with recasting Eq. \eqref{h-rt} according to
\be
\label{ham-rot-tor-2D-4D}
\hat{H}^{\rm rt}
= 
\hat{H}^{\rm red}_{\rho}
+
\hat{H}^{\rm 2D}\,,
\ee
where
\addtocounter{equation}{-1}
\bse
\be
\label{H-red}
\hat{H}^{\rm red}_{\rho}
=
\frac{{\mak B}_{X^2+Y^2}(\rho)}{\hbar^2}\left(\hat{J}^2 - \hat{J}_Z^2\right) 
+
\frac{{\mak B}_{X^2-Y^2}(\rho)}{\hbar^2}\left(\hat{J}_+^2+\hat{J}_-^2\right)
\ee
and $\hat{H}^{\rm 2D}$ is defined in Eq. \eqref{h-2d-ff}. In Eq. \eqref{H-red}, we used the identity
\label{J_x-y-z-2}
\be
\hat{J}_X^2 + \hat{J}_Y^2 = \hat{J}^2-\hat{J}_Z^2
\ee
and we introduced the molecule-fixed raising and lowering operators
\be
\label{J_+-}
\hat{J}_{\pm} = \hat{J}_X \pm {\rm i} \hat{J}_Y\,.
\ee
\ese
Consequently, whereas the operator $\hat{H}^{\rm 2D}$ is independent of the Eulerian angles $\phi$ and $\theta$, the operator $\hat{H}^{\rm red}_{\rho}$ depends on 
{all} four coordinates $\theta,\phi,\chi,$ {and} $\rho$ and thus, leads to excitations in all degrees of freedom. 

To quantify this effect, we recall the results of the discussion from Subsection \ref{subsec:field-free-coupling} of Section \ref{sec:coupling}: the smaller the reduced rotational 
constant ${\mak B}_{\rm red}$, \textit{c.f.} Eq. \eqref{def-Bred}, the less the rotations and the torsion are coupled in the field-free case. As the 2D model relies on this 
decoupling, the assumption that the torsional state ${\mathtt n}_{\rho}$ is retained during the alignment of the main principal axis is therefore, too, the better, the smaller 
${\mak B}_{\rm red}$. 

For the rotational quantum numbers $m$ and $k$ to be conserved, the operator $\hat{H}^{\rm red}_{\rho}$ must commute with the angular momentum operators 
$\hat{J}_z$ and $\hat{J}_Z$. While this holds true for the space-fixed $z$-component of the angular momentum $\bm\hat{J}$, the molecule-fixed component $\hat{J}_Z$ does 
not commute with $\hat{H}^{\rm red}_{\rho}$ due to the second term on the right-hand side of Eq. \eqref{H-red}. Thus, $m$ is a conserved quantum number, but $k$ is not; 
the raising- and lowering operators $\hat{J}_{\pm}$ increase or decrease the value of $k$. For rigid molecules, this effect is known as asymmetry-splitting. What we can again 
learn from the discussion in Subsection \ref{subsec:field-free-coupling} of Section \ref{sec:coupling} is that for $\rm G_{16}$-type molecules, the asymmetry-splitting and 
reduced rotational constant $\mak B_{\rm red}$ are also intertwined: the smaller $\mak B_{\rm red}$, the smaller the asymmetry-splitting. Thus, how the field-free coupling 
and the validity of the 2D model correlate is completely determined by the reduced rotational constant: the smaller ${\mak B}_{\rm red}$, the less the effects that prohibit the 
2D model to be a good approximation to the 4D simulations.

Correspondingly, to discuss how the field-induced coupling and the validity of the 2D model relate to each other, we analyze the structure of the field-matter Hamiltonian 
$\hat{H}^{\rm int}$ for the adiabatic alignment, Eqs. \eqref{h-int-sce} and \eqref{alpha-zz}. Here, two sources might jeopardize the presumptions of the 2D model: (1) the 
change of the polarizability as the torsional angle changes, and (2) the contribution of the term containing $\alpha^{(2,2)}$ to the excitation by the adiabatic laser pulse.

The first mechanism we have already illuminated in Subsections \ref{subsec:second-look} and \ref{subsec:add-mod} of this Section. The $\rho$-dependence of the three 
irreducible components of the polarizability, $\alpha^{(0,0)}$, $\alpha^{(2,0)}$ and $\alpha^{(2,2)}$, leads to torsional excitation; the more the polarizability changes if the 
molecules undergo torsion, the more likely torsional states are excited by the adiabatic pulse. Thus, even if the field-free rotational-torsional coupling was insignificant, the 
torsional quantum number ${\mathtt n}_{\rho}$ was not a true quantum number for the pendular states. 

For the part of the field-matter interaction that manipulates the rotations, we again find that $m$ is a conserved quantum number while $k$ is not. From Eq. \eqref{alpha-zz} 
and Eqs. \eqref{small-wigner-2}, we conclude that $\hat{H}^{\rm int}_1$ commutes with $\hat{J}_z$, because $\alpha_{zz}$ only contains rotation matrices to $m=0$ and is 
therefore independent of $\phi$. The last term on the right-hand side of Eq. \eqref{alpha-zz}, however, shows us that $k$ is changed by the interaction with the
adiabatic laser pulse; terms containing ${\mal D}_{0,\pm 2}^2$ either raise or lower the quantum number $k$. Consequently, the larger $\alpha^{(2,2)}$, the less the 
presumption of the 2D model is fulfilled that the initial state can be described by $k=0$. Interestingly, this finding is again related to the asymmetry of a rigid rotor---for 
symmetric tops $\alpha^{(2,2)}$ vanishes. Therefore, we conclude, the more the molecule can be considered as an asymmetric top, the less the presumptions of 
the conventional 2D model are fulfilled.

The {third assumption} we have identified  is the frozen mode approximation: the motions perpendicular to the torsional axis have to be much slower than the motions about 
the torsional axis. In a simplified approach, we can test this condition by calculating the timescale of the torsion  
\bse
\be
\label{def-t0}
t_{\mak A} \equiv \frac{\hbar}{\mak A} = t_0
\ee
and of the rotation perpendicular to the torsion axis
\be
t_{\mak B} \equiv \frac{\hbar}{\mak B} \;.
\ee
\ese
Comparing both timescales, we obtain 
\be
\label{ratio-time}
\frac{t_{\mak A}}{t_{\mak B}} = 2{\mak B}_{\rm red}\;,
\ee
where we used the definition of the reduced rotational constant, Eq. \eqref{def-Bred}. Consequently, the larger $\mak A$ with respect to $\mak B$, \textit{i.e.} the 
smaller the reduced rotational constant, the more the timescales of both motions are separated \textit{and} the less the expectation value $A_{\theta}$ changes on the 
timescale of the torsion. Clearly, our results support this simple argument; see Figs. \ref{BP_coup} and \ref{AAC_coup} in this section, the supplemental information and Figs. 2 
and 3 in Ref. \citenum{Grohmann.2017}.
The rotational alignment factor $A_{\theta}$ changes only little on the timescale of the torsion, but it varies the more the larger the reduced rotational constant is. We would like
to point out, however, that this argument is not rigorous. For asymmetric tops with observable torsion, timescales in the sense of the period of corresponding wave-packets 
cannot be strictly defined. Yet, as our results show, Eq. \eqref{ratio-time} is a sufficient guide to decide how reasonable the frozen mode approximation is.

In summary, our theoretical analysis shows that three aspects are relevant to decide if the 2D model is a good approximation: (1) the magnitude of the reduced rotational 
constant $\mak B_{\rm red}$; (2) the quality of the additive model for describing the polarizability of the molecule; and (3) how much the molecule can be considered to be an 
asymmetric top. Thus, these conclusions support what our results in Subsection \ref{subsec:second-look}, \ref{subsec:couplings-res} and \ref{subsec:add-mod} of this Section 
show.


\subsection{Why we need an extended 2D model in some cases}
\label{subsec:2D-ext}

We realize that the conditions we have identified for the 2D model to be a good approximation place us in a dilemma if we would like to choose the conventional 2D model to 
describe the torsional control of molecules. On the one hand, it seems the less the additive model is valid, the more elaborate the theoretical model has to be to adequately 
describe torsional control. On the other hand, a strong $\rho$-dependence of the polarizability also corresponds to a high controllability of the torsion with moderate field 
strengths. This is also reflected by the systems that have been studied so far. Experimentalists usually study substituted biphenyls, for which the additive model is in particular a 
bad approximation. However, it is the distinct $\rho$-dependence of the molecular polarizability that makes them suitable candidates for experimental studies on torsional 
control.

To resolve this discrepancy, we think it is necessary to modify or to  extend the conventional 2D model. We believe the most promising way to describe these systems 
appropriately is to steer a middle way: the excitation of the torsion by the first laser pulse is calculated by a modified 2D model based on adiabatic separation, while for 
describing the excitations by the second laser pulse and the subsequent propagation in time the conventional 2D model with a modified initial state is used. Still, calculations 
based on this extended 2D model would be less time-consuming than those within the 4D approach, but the new model would, at least in parts, reflect the effect of the adiabatic 
alignment on the torsion. Moreover, decomposing the process of torsional alignment into two lower-dimensional problems also avoids the complications resulting from a lack of 
convergence; see the discussion in Subsection \ref{subsec:convergence} of this Section.


\section{A critique of our approach}
\label{sec:limits}

As every scientific method, the approximations and techniques we used to calculate the alignment of $\rm G_{16}$-type molecules are limited. In the following, we scrutinize 
which phenomena our model does not include, and what are the limitations of the theoretical methods we have used. Hereby, we intend to facilitate comparing our results with 
experiments and other theoretical studies. We explain why the models and methods we are using are legitimate approximations for the scenario we consider here, and we sketch 
out how, if necessary, they can be extended.

\subsection{Failure of the adiabatic approximation}
\label{subsec:crit-adiab}

To describe the alignment by the nanosecond laser pulse, we assume the non-rigid molecules to remain in a defined quantum state, correlating unambiguously with the 
field-free ground state. This assumption may be wrong: As studies on rigid molecules have shown, adiabatic alignment is impossible in some 
cases.\cite{Omiste.2011,Omiste.2012,Nielsen.2012}
Due to crossings of {field-dressed states} even at low laser intensities, the conditions for an adiabatic passage are not fulfilled. In these cases, the interaction of the molecules
with the first laser pulse has to be modeled by a time-dependent or a diabatic model.\cite{Omiste.2011,Omiste.2012,Nielsen.2012}

One way to find signs for state crossings is to analyze the expansion coefficients $c_{k_{\rho},J,k,m}$ as a function of the laser intensity. If they change drastically under a 
small change of the field strength, and thus the pendular state changes its character, it is likely a state crossing occurred. This method is not exact; only a time-dependent
model and experiments can tell.  But this test gives us at least a broad idea if and at which laser intensities a non-adiabatic treatment might be necessary. Consequently, we 
have analyzed the expansion coefficients of each pendular state for all molecules from Fig. \ref{mol-gal}. We did not find evidence for a failure of the adiabatic approximation. 
Yet, as we do not have experimental data at hand, we cannot judge if our analysis is correct. In any case, being aware of this effect is important when analyzing experimental and 
theoretical results on the rotational-torsional alignment.

\subsection{Effects of molecular symmetry}
\label{subsec:crit-sym}

We limit our considerations to states of one irreducible representation, \textit{i.e.} the symmetry of the field-free rotational ground state 
$\Gamma^{\rm rt}={\rm A}^+_{1g}$. Yet, as recent studies have shown,\cite{Fleischer.2007,Fleischer.2008,Fleischer.2009,Grohmann.2011,Grohmann.2012,Floss.2012}
the {alignment dynamics strongly depends on the symmetry of the initial molecular state}. Therefore, we cannot judge if the agreement of the 2D and 4D model is depending  
on symmetry. Possibly, for other symmetries results for the torsional alignment obtained from the 2D and 4D model, respectively, coincide less.

When speaking about symmetry, we have to consider a further argument: the {molecular symmetry is changed if the adiabatic alignment was successful}. Intuitively, this 
argument is clear: As the number of degrees of freedom is lower than in a full 4D treatment, the number of symmetry operations, leaving the Hamiltonian of the 2D model 
invariant, are lower as well. As we argue in Subsection \ref{subsec:sym-coupling} of Section \ref{sec:coupling}, a detailed analysis is complicated. In parts, we have already 
developed a theory consistently describing the symmetry of scenario shown in Fig. \ref{mod-2step}. Our preliminary conclusions is: the symmetry of the 2D model is indeed 
lower; the number of feasible operations are reduced in case the 2D model is a valid approach.\cite{Grohmann.2012}

Finally, we point to the fact that most molecules in experimental and theoretical studies have smaller MS groups than $\rm G_{16}$-type 
molecules\cite{Parker.2011,Floss.2012,Parker.2012,Ashwell.2013,Ashwell.2013b,Madsen.2009,Madsen.2009b,Hansen.2012,Christensen.2014}; 
see Ref. \citenum{Soldan.1996} for a systematic classification of non-rigid molecules with observable torsions. One consequence of the lower symmetry is a higher computational
demand in 4D calculations. As more basis states have the same symmetry for a given value of $J^{\rm max}$ and $N_{\rho}$, more states are needed to accurately represent 
the pendular states that are created by the first laser pulse.\cite{McWeeny.2002}
In conjunction with our insights from Subsection \ref{subsec:convergence} concerning the importance of convergence, we thus conclude that 4D simulations for these type of 
molecules are numerically not feasible, supporting our argument for the need of extended 2D models. 

Furthermore, for these molecules our conclusion cannot be directly transfered. Unlike for $\rm G_{16}$-type molecules, we cannot define one reduced rotational constant 
$\mak B_{\rm red}$, as these molecules lack of  a symmetric-top configuration. In addition, setting up the field-free Hamiltonian is more complicated for these molecules; kinetic 
coupling terms are unavoidable in the 4D case when using the IAM,\cite{Soldan.1996}
making the analysis of the field-free coupling more difficult and its influence might be more pronounced as it is for $\rm G_{16}$-type molecules. 

However, we simply cannot judge on the importance of all symmetry related aspects from our calculations. Whether they are relevant at all, or what their impact on the validity of 
the 2D model is,  future investigations have to show.

\subsection{What about temperature?}
\label{subsec:crit-temp}

The simulations we presented in the preceding Sections are only valid if the temperature of the molecular probe is $0 {\rm K}$. Experiments on molecular alignment, however, 
always take place at finite temperatures, sometimes even at room temperature. And despite of recent advances on cold molecules, it is still very difficult to 
prepare polyatomic molecules in a well-defined quantum state.\cite{Schnell.2009}
Thus, to describe our scenario more realistically, we would have to replace the pure initial states by a thermal ensemble and solve the Schr\"odinger equations for every 
state that is populated (significantly) at the respective temperature.\cite{Grohmann.2011,Ashwell.2013b}
The correct alignment factors, we then obtain by thermally averaging the alignment factors for each populated initial state, having regard of the correct statistical weight of each 
initial rotational-torsional state\cite{Grohmann.2011,Grohmann.2012}
(which is ignored in many studies on torsional alignment\cite{Madsen.2009,Madsen.2009b,Parker.2011,Parker.2012,Ashwell.2013,Ashwell.2013b,Christensen.2014}).

Earlier studies on torsional control have identified temperature as an important factor for the failure of the 2D approach to torsional alignment.\cite{Coudert.2011}
If we thus ignore temperature, we run into danger to miss the relevant point of finding the conditions for the 2D model to be a reliable approximation. However, our data suggest 
that concluding temperature is related to the validity of the 2D model is ambivalent. Admittedly, it is correct that the degree of alignment is reduced as temperature 
increases.\cite{Seideman.2005} 
But this is an (almost) universal phenomenon, in alignment studies in particular and in molecular quantum dynamics in general. Since our simulations show that the relevant 
coupling mechanism is field-induced, we can always use (a combination of) laser pulses to control it. Moreover, for the 2D and 4D simulations to agree less at 
higher temperatures, the premises of the 2D model have to be fulfilled less at higher temperatures, \textit{i.e.} for initial states with higher $k$ and ${\tt n}_{\rho}$. Besides 
on the field-induced coupling, which is controllable, only the field-free coupling could be the origin for this increasing disagreement. Yet, as we shown in particular in Subsection 
\ref{subsec:couplings-res} of Section \ref{sec:two-step}, in many cases this influence of the field-free coupling is negligible. Consequently, it is not clear, why temperature, in 
general, should have an effect so destructive that the torsional alignment vanishes.

\subsection{Couplings with other modes}
\label{subsec:crit-coup}

Yet, temperature is not the only phenomenon having a negative impact on the degree of molecular alignment. Roconvibronic couplings are known for having a similar 
effect: as the {rotational-torsional motions are coupled with other molecular degrees of freedom}, they lead generally to a decrease of alignment as time goes by. For diatomic 
and symmetric-top molecules, for example, it was demonstrated that rovibronic couplings reduce the degree of alignment on a nanosecond to microsecond 
timescale.\cite{Hamilton.2005,Owschimikow.2010} 
Thus, if  the intention of the experiment is to control the torsion for this long, the model we developed here needs to be extended.\cite{Bunker.1998}
Our studies, however, are limited to time-scales being too short for rotational-torsional couplings to be relevant. 

On a related note, we consider the molecules to be non-interacting with each other and/or the environment. Approximately, this scenario is realizable under certain conditions, 
yet not achievable for many interesting applications of torsional control. In case it is necessary, our treatment has to be extended to an open-system approach, as it was recently 
formulated for torsional control.\cite{Ramakrishna.2006,Pelzer.2008,Ashwell.2013b}
It was shown, however, that the timescales upon which interactions with the environment typically occur are much longer than the timescale $t_0$, \textit{c.f.} Eq. 
\eqref{def-t0}. Thus, we conclude that for our simulations the impact of environmental effects are negligible.

\subsection{Is strong-field ionization not a problem?}
\label{subsec:crit-ion}

When a molecule interacts with an off-resonant laser pulse, alignment is not the only phenomenon that may occur. At laser intensities on the order of 
$10^{14}\,\nicefrac{\rm W}{\rm cm^2}$, tunnel ionization might take place as well. Although being known theoretically for a long time, not much is known about if and when 
tunnel ionization is important in the context of molecular alignment. Only recently a systematic theoretical study on linear molecules was published, which has discovered a 
universal relation between the alignment intensity dependence and the dependence of the threshold intensity.\cite{Szekely.2016}
Although these findings are limited to the adiabatic regime and cannot be directly applied to the control of internal motions, they show that the maximal degree of alignment is
often achieved at intensities well below the ionization threshold. And yet, tunnel ionization is a phenomenon that always can occur in strong field processes. Thus, the question if 
it is relevant for the studied molecule has to be answered case-by-case.

\subsection{Failure of the electric dipole approximation}
\label{subsec:crit-dip}

The Hamiltonian we employed to describe the field-matter interaction, Eq. \eqref{h-int}, is based on the semi-classical electric dipole approximation,\cite{Seideman.2005}
which assumes the laser field to be constant over the size of the molecule. Recently, also X-ray pulses were used to control the alignment of molecules\cite{Ho.2009},
and the control of molecular motions with X-ray laser pulses is a rapidly growing field in molecular physics. Here, however, the dipole-approximation fails and the theory of 
alignment has to be modified.\cite{Buth.2008}

As a consequence, not the molecular polarizability but the dipole moment is the relevant quantity for describing the field-matter interaction. As dipole moments obey different 
symmetry rules than polarizabilities,\cite{Bunker.1998} 
our whole discussion on the field-induced rotational-torsional coupling needs to be adjusted, beginning with the symmetry-adapted Hamiltonian, see Subsection 
\ref{subsec:sym-int-g16} of Section \ref{sec:4d-mod}. 

Moreover, as the symmetry of the overall system is lower,\cite{Bunker.1998} 
the computational demands are higher, making a theoretical treatment possibly unfeasible, see also Subsection \ref{subsec:crit-sym} of this Section. Yet, what follows from 
these changes for the validity of the 2D model if X-ray pulses are used to control the torsion, only further studies can explore.

\subsection{A very simple propagator}
\label{subsec:crit-prop}

Crucial to an accurate solution of the time-dependent Schr\"odinger equation is an appropriate choice of the propagator. The impulsive approximation we employed for 
describing the interaction with the femtosecond laser pulse is one of the simplest approaches to this problem. It is only valid if the length of the laser pulse is much shorter than 
the typical timescale of the motion the laser is supposed to manipulate; the smaller $t_0$, \textit{c.f.} Eq. \eqref{def-t0}, the worse the approximation. Although this 
approximation was very successful in past studies\cite{Leibscher.2003,Fleischer.2008,Fleischer.2007,Fleischer.2007b} 
the shorter timescale of the torsion might make this approximation less reliable. 

We are aware that in earlier studies, more accurate propagators have been used, such us the split operator 
technique.\cite{Parker.2011,Parker.2012,Ashwell.2013,Ashwell.2013b}
Yet, these methods involve calculating products of matrix exponentials for every time-step of the interaction. Considering the larger number of basis states we had to use, see
Subsection \ref{subsec:convergence} of Section \ref{sec:two-step}, employing these type of propagators were too time-consuming. In general, calculating matrix exponentials 
was one of the critical points of implementing our approach. We readdress this problem briefly in Appendix \ref{app:code}.

Moreover, we add for consideration that the potential mistakes we commit by choosing the sudden approximation are systematic; we use the sudden approximation for the 
4D and 2D model alike. To conclude that our arguments---which we are able to develop based on theoretical considerations, see Subsection \ref{subsec:2D-fail} of Section 
\ref{sec:two-step}---might be generally flawed, is therefore not appropriate. However, for accurate predictions of the torsional alignment, using more elaborate propagators might 
be necessary.

\subsection{Why DFT?}
\label{subsec:crit-dft}

Experts of quantum chemistry may wonder, and legitimately so, why we employed a method of comparably low level of theory to calculate the molecular properties. We chose 
{density functional theory} mainly for practical reasons. For molecules like Biazu or AAC, see Fig. \ref{mol-gal}, calculating the potential energy surfaces is computationally still 
demanding and time-consuming. Additionally, we had to calculate the polarizability of the molecules from Fig. \ref{mol-gal} as well, which is on the \textit{state-of-the-art} level 
of theory, in general and for larger molecules in particular, computationally still inaccessible, see below. 

More sophisticated methods may lead to completely different potentials, as especially low barrier heights are causing practical problems when using standard quantum chemical 
approaches.\cite{Kochikov.2003,Grohmann.2012} 
And as our simulations show, these modifications in the potential indeed change the time-evolution of the alignment factors. Yet, how the alignment dynamics changes is 
potential-specific, and thus particular to a given molecule. We are therefore not able to give a general conclusion on the influence of different potential forms, and we leave a 
detailed discussion of the quantum chemical nuances to our colleagues from electronic structure theory.   

Furthermore, we stress that inaccurate potentials (and polarizabilities) are, too, systematic errors. As they apply equally to both, 2D and 4D simulations, we are not 
expecting them not to change the main findings of our study. To reproduce experiments on the torsional alignment of a given molecule as good as possible, however, accurate 
calculations might be necessary.

\subsection{Accurate polarizabilities are difficult to calculate}
\label{subsec:crit-pol}

Beyond that calculating polarizabilities is in particular a problem. While the electronic energies of a molecule, and thus its torsional potential, can directly be optimized by 
quantum chemical procedures, obtaining accurate polarizabilities is still difficult.\cite{Mitroy.2010}
Within the \textsc{QChem} package, a direct method is used, based on a time-dependent Hartree-Fock procedure.\cite{Sekino.1986,Kussmann.2007}
These methods are limited; sometimes they substantially fail to reproduce the polarizabilities of a molecule.\cite{Bishop.1990,Mitroy.2010}

Moreover, we only use the electronic part of the polarizability. Although electronic polarizabilities are indeed dominating the molecular polarizability, cases are known of 
which contributions due to vibrational and rotational motions are significant.\cite{Bonin.1997}
All the more we find it worth to mention that these type of corrections are often ignored in quantum chemical calculations.\cite{Shao.2014}

For comparing our results with experimental studies, another aspect is important to consider: Here, we only used static polarizabilities, as it is commonly done in theoretical 
studies.\cite{Ramakrishna.2007,Grohmann.2011,Floss.2012,Parker.2011,Parker.2012,Ashwell.2013,Ashwell.2013b} 
Yet, in Eq. \eqref{h-int} the dynamic polarizability, which depends on the frequency of the laser, is the relevant molecular property. The frequency dependence is usually small
and contingent on the particular laser that is used to create alignment. Thus, we ignore it here. When simulating a specific experiment with a specific light source, 
however, it should be taken into account.

As for the torsional potential and the rotational constants, this discussion does not allow for concluding that our insights about which mechanisms decide the question if the 2D 
model is appropriate are wrong. We only want to sensitize the reader for necessary modifications of our theory if specific experimental setups are used.


\section{Conclusion: 2D models are valid approximations. And we need them anyway}
\label{sec:conclusions}

In this work, we have analyzed the requirements the 2D model, commonly used for describing the two-step mechanism of torsional control from Fig. \ref{mod-2step}, has to 
meet for being an adequate approximation to the 4D semi-rigid-rotor model. Recently, it has been argued that the rotational-torsional couplings which are not included in the 2D 
model, destroy the torsional alignment. To address this critique, we have systematically studied the nature of the couplings and examined how they influence the 
rotational-torsional dynamics of $\rm G_{16}$-type molecules. 

Here, we have investigated the impact of the field-free and the field-induced coupling on the rotational-torsional dynamics in general, and how these couplings relate to the 
validity of the 2D model in particular. We have found that the field-free coupling is completely determined by the reduced rotational constant $\mak B_{\rm red}$, see Eq. 
\eqref{def-Bred}. It is therefore inherent to the molecule and cannot be controlled by external fields. The field-induced coupling, however, is directly linked to the dependence of 
the polarizability on the torsion angle $\rho$: the more the polarizability anisotropies $\alpha^{(0,0)}$, $\alpha^{(2,0)}$ and $\alpha^{(2,2)}$
change as the molecule undergoes torsion, the larger the field-induced coupling. Consequently, if the 
prominent\cite{Ramakrishna.2007,Parker.2011,Floss.2012,Parker.2012,Ashwell.2013,Ashwell.2013b,Coudert.2011,Ortigoso.2013,Coudert.2015}
additive model is employed for modeling the molecular polarizabilities, the field-induced rotational-torsional coupling is minimized. Our simulations have shown that the effect of 
the field-free rotational coupling is generally rather small. The field-induced coupling, however, is essential for inducing torsional alignment.

Moreover, we have found that, by tendency, the 2D model can reliably reproduce the results from 4D simulations if the adiabatic pulse is of moderate intensity. Typically, the 2D 
model slightly overestimates the torsional alignment in agreement with earlier studies.\cite{Coudert.2011,Ortigoso.2013,Coudert.2015} 
If the intensity of the adiabatic pulse is high, our 4D simulations reveal an additional effect which is neglected in the conventional 2D model: adiabatic torsional alignment 
caused by the excitation of torsional states due to high field-induced coupling during the first pulse.

On a related note, we have found that the validity of the 2D model correlates with the validity of the additive model: As the field-induced coupling is minimized, the 2D model 
reproduces the results from 4D simulations the better (if not perfectly), the more the additive model is a good approximation to the molecular polarizabilities. 

The results of our theoretical analysis suggest that it is possible to realize an extended 2D model, relying on adiabatic separation of the motions perpendicular and parallel to the 
torsional axis. Such a model is the more  appropriate, the smaller the reduced rotational constant ${\mak B}_{\rm red}$. This condition is in particular fulfilled for substituted 
biphenyls, a subclass of molecules that is often used in experiments,\cite{Madsen.2009,Madsen.2009b,Hansen.2012,Christensen.2014}
illustrating the practical relevance of modifying the conventional 2D approach. 

Our insights, however, are limited: rotational-torsional motions on longer timescales, non-adiabatic effects during the alignment by the first laser pulse, couplings with other 
modes, \textit{e.g.} vibrations or the environment as well the temperature effects have not been considered so far. Moreover, we study a class of molecules having a specific 
molecular symmetry group. Thus, the conclusions we made for those $\rm G_{16}$-type molecules might be incorrect for molecules with other symmetries.  Finally, we only 
take into account states of one symmetry, namely the symmetry of the rotational-torsion ground state. We shall investigate the rotational-torsional alignment of states with 
different symmetry, which are excited at higher temperatures, in a future publication.

And yet, simulations of applications for which torsional control is relevant have to rely on simplified models. As we have demonstrated here, convergence is very important for 
obtaining reliable results, otherwise the rotational and torsional alignment is underestimated. The main reason why we were able to perform our 4D simulations with sufficiently 
large basis sets is the high symmetry of the molecules we considered---a condition that is no longer fulfilled for most experimentally studied molecules. To simulate the torsional 
alignment of these species, it seems, using lower dimensional models is unavoidable. Our studies suggest that 2D models---either in the conventional or an extended form---are 
able to reliably reproduce simulations based on a 4D semi-rigid-rotor model. Future investigations have to show if our faith in this conclusion is justified.

\renewcommand{\thesection}{X}

\section{Appendices}

\subsection{Derivation of the interaction Hamiltonian}
\label{app:der-H-int}

To derive the Hamiltonian for the interaction with an off-resonant laser pulse, Eqs. \eqref{h-int}, we first need to express the space-fixed components 
of the molecular polarizability, $\alpha_{qq'}, q,q'=x,y,z$, in terms of the molecular-fixed components $\alpha_{QQ'}, Q,Q'=X,Y,Z$. If $\bm\alpha(\rho)$ is diagonal in the 
molecular-fixed frame, which is true for the molecules we are considering, the space-fixed-components of the molecular polarizability can be written as
\be
\label{mol-raum-ten}
\alpha_{qq'} = \sum_{Q}{\mak S}_{Qq}{\mak S}_{Qq'}\alpha_{QQ} \quad q=x,y,z;\, Q=X,Y,Z\;,
\ee
where $\mak S_{Qq}$ denote the direction cosines as a function of the Euler angles $\phi$, $\theta$, $\chi$.\cite{Zare.1988}

To evaluate the matrix elements of $\hat{H}^{\rm int}_i$ in the basis Eq. \eqref{ansatz-basis}, it is convenient to use the irreducible tensor method. Here, instead of the nine 
Cartesian components of $\bm\alpha$, nine irreducible components are used. For a diagonal $\bm\alpha$ in the molecule-fixed frame, only three irreducible components are 
relevant; they can be written as\cite{Bunker.1998}
\bse
\label{trans-cart-irred}
\begin{align}
\alpha^{(0,0)}
&= 
\frac{1}{\sqrt{3}}\left(\alpha_{XX}+\alpha_{YY}+\alpha_{ZZ}\right)
\\
\alpha^{(2,0)}
&=
\frac{1}{\sqrt{6}}\left(2\alpha_{ZZ}-\alpha_{XX}-\alpha_{YY}\right)
\\
\alpha^{(2,2)}
&=
\frac{1}{\sqrt{2}}\left(\alpha_{XX}-\alpha_{YY}\right) \;.
\end{align}
\ese
Using Eq. \eqref{mol-raum-ten}, we find the diagonal elements of $\bm\alpha$ in the space-fixed coordinate system to be
\be
\label{alpha-diag-space}
\alpha_{qq} 
= 
\frac{\alpha^{(0,0)}}{\sqrt{3}} \left({\mak S}_{Xq}^2 + {\mak S}_{Yq}^2 + {\mak S}_{Zq}^2 \right)
+
\frac{\alpha^{(2,0)}}{\sqrt{6}} \left(2{\mak S}_{Zq}^2 - {\mak S}_{Xq}^2 - {\mak S}_{Yq}^2  \right)
+
\frac{\alpha^{(2,2)}}{\sqrt{2}} \left({\mak S}_{Xq}^2 - {\mak S}_{Yq}^2\right)\;,
\ee
which we can simplify to
\be
\label{alpha-diag-space-fin}
\alpha_{qq} 
=
\frac{\alpha^{(0,0)}}{\sqrt{3}} 
+
\frac{\alpha^{(2,0)}}{\sqrt{6}} \left(3{\mak S}_{Zq}^2 - 1\right) 
+ 
\frac{\alpha^{(2,2)}}{\sqrt{2}} \left({\mak S}_{Xq}^2 - {\mak S}_{Yq}^2\right)\;,
\ee
if we take into account the orthogonality-relations of the direction cosines\cite{Zare.1988}
\be
\label{ort-cond-direct}
\sum_{q}{\mak S}_{qQ}{\mak S}_{qQ'} = \delta_{QQ'}
\quad\text{and}\quad
\sum_{Q}{\mak S}_{qQ}{\mak S}_{q'Q} = \delta_{qq'}
\ee
into account. When treating molecules without observable torsion, the first term on the right-hand side of Eq. \eqref{alpha-diag-space-fin} is neglected; it leads to an 
angle-independent shift, having no consequences for the alignment. For molecules with torsion, however, this term generally depends on the contorsional variables and has to 
be included.
 
Using the explicit definition of the direction cosines,\cite{Wilson.1980} we obtain after some manipulations
\bse
\label{alpha-space-fixed-diag}
\begin{align}
\alpha_{xx} 
=&
\frac{\alpha^{(0,0)}}{\sqrt{3}} 
+
\frac{\alpha^{(2,0)}}{\sqrt{6}} 
\left(-{{\mak D}}_{0,0}^2+\frac{3}{\sqrt{6}}\left({\mak D}_{2,0}^2+{\mak D}_{-2,0}^2 \right)\right)\nonumber\\
&
+
\frac{\alpha^{(2,2)}}{\sqrt{2}} 
\left\{\frac{1}{\sqrt{6}}\left({\mak D}_{0,2}^2+{\mak D}_{0,-2}^2 \right)+
\frac{1}{2}\left({\mak D}_{2,2}^2+{\mak D}_{-2,-2}^2+{\mak D}_{2,-2}^2+{\mak D}_{-2,2}^2 \right)\right\}\\
\alpha_{zz} 
=&
\frac{\alpha^{(0,0)}}{\sqrt{3}}
+
\frac{2\alpha^{(2,0)}}{\sqrt{6}}{\mak D}_{0,0}^2
+
\frac{\alpha^{(2,2)}}{\sqrt{3}}\left({\mak D}_{0,2}^2+{\mak D}_{0,-2}^2 \right)\,.
\end{align}
\ese
The Wigner matrices ${\mak D}^J_{m,k}$, which we have introduced in Eqs. \ref{alpha-space-fixed-diag}, are defined, in general, as
\be
\label{def-wigner}
{\mak D}^J_{m,k} = \exp\left(-{\rm i}m\phi\right)\cdotp {\mak d}^{J}_{m,k}(\theta)\cdotp \exp\left(-{\rm i}k\chi\right)\;.
\ee
The small Wigner matrices ${\mak d}^{J}_{m,k}(\theta)$ in Eq. \eqref{def-wigner} are tabulated in common textbooks about angular momenta; see for example the 
book of \citeauthor{Zare.1988}.\cite{Zare.1988} Here, we employed the matrices for $J=2$\nolinebreak[4]
\bse
\label{small-wigner-2}
\begin{align}
{\mak d}^{2}_{0,0}(\theta) 
&= 
\frac{1}{2}\left(3\cos^2\theta -1\right)\\
{\mak d}^{2}_{2,0}(\theta) 
&=
\sqrt{\frac{3}{8}}\sin^2\theta 
\\
{\mak d}^{2}_{2,\pm 2}(\theta) 
&=
\frac{1}{4}\left(1\pm\cos\theta \right)^2
\end{align}
\ese
and their symmetry properties
\be
{\mak d}^{J}_{m,k}(\theta) = (-1)^{k-m}\,{\mak d}^{J}_{k,m}(\theta) = {\mak d}^{J}_{-m,-k}(\theta)\;.
\ee
Taking into account the explicit definition of the Wigner-matrices, Eq. \eqref{def-wigner}, we obtain the expression Eqs. \eqref{alpha-zz} and \ref{alpha-xx}.

If the molecules were perfectly aligned, $\theta=\{ 0,\pi\}$ and consequently for $\alpha_{xx}$ holds
\be
\alpha_{xx}
= 
\frac{\alpha^{(0,0)}}{\sqrt{3}}
-
\frac{\alpha^{(2,0)}}{\sqrt{6}} 
+ 
\frac{\alpha^{(2,\bar{2})}}{2\sqrt{2}}\left(\exp(2{\rm i}\phi)\exp(2{\rm i}\chi) +  \textbf{c.c.}\right)\,,
\ee
where we have used the explicit definitions of the Wigner matrices Eqs. \eqref{small-wigner-2}. Since ${\bm e}_Z$ and ${\bm e}_z$ are parallel, $\phi$ is redundant and we 
may set $\phi = 0$ to obtain after some algebra 
\be
\alpha_{xx} = \frac{1}{2}\left(\alpha_{XX}+\alpha_{YY}\right) + \frac{1}{2}\left(\alpha_{XX}-\alpha_{YY}\right)\cos(2\chi)\,,
\ee
where the definitions of the irreducible polarizabilities, Eq. \eqref{trans-cart-irred}, were used. If we introduce
\bse
\label{irred-polar-2d}
\begin{align}
\tilde{\alpha}^{(0,0)} &= \frac{1}{2}\left(\alpha_{XX}+\alpha_{YY}\right) \\
\tilde{\alpha}^{(2,{2})} &= \frac{1}{2}\left(\alpha_{XX}-\alpha_{YY}\right)\;,
\end{align}
\ese
we obtain as a final result
\be
\label{H-int-2d}
\hat{H}^{\rm int}_{2} (t_2)
=
 -\frac{\left|\epsilon_2 (t_2) \right|^2}{4} \left(\tilde\alpha^{(0,0)} + \tilde\alpha^{(2,\bar{2})}\cos(2\chi)  \right)\;,
\ee
which is identical to the Hamilton for a linear-polarized laser pulse within the two-dimensional treatment; see Ref. \citenum{Floss.2012}.


\subsection{On matrix elements}
\label{app:matrix-ele}


To obtain the coefficients $c_{k_{\rho},J,k,m}$ in Eq. \eqref{ansatz-basis} in the field-free case, we have to diagonalize the matrix representation ${\bm H}^{\rm rt}$ of the 
operator $\hat{H}^{\rm rt}$ in the basis Eq. \eqref{ansatz-basis}. We can express the matrix $\bm H^{\rm rt}$ symbolically as 
\be
\label{h-rt-voll}
{\bm H^{\rm rt}} = {\bm H}^{\rm rot}_{\rho} + {\bm H}^{\rm tor}\otimes{\bm 1}^{\rm rot}\,,
\ee
where ${\bm H}^{\rm rot}_{\rho}$ and ${\bm H}^{\rm tor}$ are the matrix representation of the operators
\addtocounter{equation}{-1}
\bse
\begin{align}
\label{H-rot-rho}
\hat{H}^{\rm rot}_{\rho} 
=& 
\frac{{\mak B}_{X^2+Y^2}}{\hbar^2}\left(\hat{J}^2-\hat{J}_Z^2\right) 
+
\frac{{\mak B}_{X^2-Y^2}}{\hbar^2}\left(\hat{J}_+^2+\hat{J}_-^2\right)
+
\frac{\mak A}{\hbar^2}\hat{J}_Z^2\\
\intertext{and}
\label{H-tor}
\hat{H}^{\rm tor}
=&
\frac{\mak A}{\hbar^2}\hat{J}^2_{\rho}
+
{\mak V}^{\rm tor}_0 \sum_{n=0}^{N} {\mak V}_{n}\cos (4n\rho)\;,
\end{align}
\ese
respectively, and ${\bm{1}}^{\rm rot} $ is the identity matrix written in the symmetric top basis. For all molecules considered here, it is sufficient to truncate the sum in Eq. 
\eqref{H-rot-rho} at $N=6$. In Eq. \eqref{H-rot-rho}, we used again the identity Eq. \eqref{J_x-y-z-2} and the definition of the molecule-fixed raising and lowering operators, Eq. 
\eqref{J_+-}.

We thus have to calculate the matrix elements of the operators $\hat{J}^2$, $\hat{J}_Z$ and $\hat{J}_{\pm}$ in the basis Eq. \eqref{ansatz-basis}
to evaluate the elements of $\bm H^{\rm rot}_{\rho}$. They are given by\cite{Zare.1988}
\bse
\label{mat-ele-rot}
\begin{align}
\label{mat-ele-j2}
\Big(\hat{J}^2\Big)_{\{k'_{\rho};J',k',m'\},\{k_{\rho};J,k,m\}}
&= 
\hbar^2\, J(J +1)\, \delta_{k'_{\rho},k_{\rho}}\;\delta_{J',J}\delta_{k',k}\delta_{m',m}\\
\label{mat-ele-jz}
\Big(\hat{J}_Z\Big)_{\{k'_{\rho};J',k',m'\},\{k_{\rho};J,k,m\}} 
&= 
\hbar^2\, J(J +1)\, \delta_{k'_{\rho},k_{\rho}}\delta_{J',J}\delta_{k',k}\delta_{m',m}\\
\label{mat-ele-jpm}
\Big(\hat{J}_{\pm}\Big)_{\{k'_{\rho};J',k',m'\},\{k_{\rho};J,k,m\}}
&= 
\hbar^2\, {\mal C}_{k\mp 2,k} \,\delta_{k'_{\rho},k_{\rho}}\,\delta_{J',J}\delta_{k\mp 2,k}\delta_{m',m}\,,
\end{align}
with
\be
\small
{\mal C}_{k\mp 2,k} =\sqrt{J(J+1)-(k\mp 1)(k\mp 2)}\sqrt{J(J+1)-k(k\mp 1)}\,.
\ee
\ese
In a full treatment, however, the matrix ${\bm H}^{\rm rot}_{\rho}$ contains non-vanishing matrix elements not only between different rotational, but also between 
different torsional basis states, as the functions ${\mak B}_{X^2\pm Y^2}$ in Eq. \eqref{B-const-rho} both depend on $\rho$. The matrix elements of these functions, 
written in the basis of free rotor eigenfunctions Eq. \eqref{eig-plan-rot}, are given by
\be
\label{rot-konst-int}
\left({\mak B}_{X^2+Y^2}(\rho)\right)_{\{k'_{\rho};J',k',m',\},\{k_{\rho};J,k,m,\}}
=
\int_{0}^{2\pi} {\mak B}_{X^2\pm Y^2}\exp(-{\rm i}k_{\rho}' \rho)\exp({\rm i}k_{\rho} \rho) {\rm d}\rho\, \delta_{J',J}\delta_{k',k}\delta_{m',m} \;;
\ee
they must be calculated numerically. To do so, we make use of the expansion Eq. \eqref{exp-rot-konst}, since then we only have to calculate matrix elements of the type
\be
\label{mat-ele-rot-konst}
\left(\cos(2p\rho) \right)_{k'_{\rho},k_{\rho}}
=
\frac{1}{2} \delta_{k'_{\rho}+2p,k_{\rho}} + \frac{1}{2} \delta_{k'_{\rho}-2p,k_{\rho}}\,,
\ee
if we take
\be
\cos^{p} x 
= 
\frac{1}{2^p} \sum_{o = 0}^p 
\binom{\,p\,}{\,o\,}
\cos(({p} - 2 o) x)
\ee
into account. Thus, we can reduce Eq. \eqref{rot-konst-int} to an algebraic problem, which is numerically more efficient to solve than numerical integration. Furthermore, taking 
advantage of the expansion Eq. \eqref{exp-rot-konst} allows us to systematically improve our approach, if necessary.

The matrix ${\bm H}^{\rm tor}$ is the free planar rotor representation of the Hamiltonian for the pure torsion $\hat{H}^{\rm tor}$, see Eq. \eqref{H-tor} and Eq. 
\eqref{eig-plan-rot}, respectively. The matrix elements of ${\bm H}^{\rm tor}$ for the potential Eq. \eqref{def-pot-tor} in the basis Eq. \eqref{eig-plan-rot} are given by
\be
\label{mat-ele-htor-frei}
H^{\rm tor}_{k_{\rho}',k_{\rho}} 
= 
{\mak A} k_{\rho}^2\delta_{k_{\rho}',k_{\rho}}
+
\sum_{n=0}^6
\frac{{\mak V}_n}{2}\big(\delta_{k_{\rho}',k_{\rho}+4n} + \delta_{k_{\rho}',k_{\rho}-4n}\big)\;,
\ee
completing the list of matrix elements we have to evaluate for calculating the matrix representation of the field-free Hamiltonian, Eqs. \eqref{h-rt}, \eqref{h-rot} and 
\eqref{h-tor}.


As we pointed out in Sec. \ref{subsec:ad-nonad}, we need to calculate the matrix representation of the operator $\hat{W}$, \textit{c.f.} Eqs. \eqref{h-int-ad} and 
\eqref{h-int-nad}, to quantify the field-matter interaction. If we write $\hat{W}$ in the basis Eq. \eqref{ansatz-basis}, it contains matrix elements of the type
\bse
\label{mat-ele-rad}
\be
\left(\alpha^{(J'',K'')}\right)_{k_{\rho}',k_{\rho}}\cdotp \left({\mak D}^{J''}_{m'',k''} \right)_{\big\{J',k',m'\big\},\big\{J,k,m\big\}}\;,  
\ee
with $K''=|k''|$. In Eq. \eqref{mat-ele-rad},
\be
\label{mat-ele-alph}
\left(\alpha^{(J'',k'')}\right)_{k_{\rho}',k_{\rho}}
= 
\frac{1}{2}\sum_{n=0}^{\infty}{\mak P}^{(J'',k'')}_{n}\left(\delta_{k_{\rho}',k_{\rho}+(4n+K'')} + \delta_{k_{\rho}',k_{\rho}-(4n+K'')}\right)\;.
\ee
For the integrals over the Wigner matrices holds\cite{Zare.1988}
\be
\label{mat-win-d}
\setlength{\arraycolsep}{3pt}
\left({{\mak D}}^{J''}_{m'',k''} \right)_{\big\{J',k',m'\big\},\big\{J,k,m\big\}}
=
(-1)^{k+m}\sqrt{2J+1}\sqrt{2J'+1}
\begin{pmatrix}
J'	&	J''	&	J\\
m'	& m''	& -m
\end{pmatrix}
\begin{pmatrix}
J'	&	J''	&	J\\
k'	& 	k'' & -k
\end{pmatrix}
\ee
with $\left( :::\right)$ denoting a so-called $3j$-symbol. They are non-zero only if\cite{Zare.1988}
\begin{align}
\label{sel-rul}
 |J-J''|\leq J' & \leq J+J'' \\  
 k'' + k' - k &=  0 \\
 m''+ m' - m &=  0\,.
\end{align}
\ese

Finally, to calculate the relevant alignment factors, we have to evaluate the matrix representations of $A_{\eta}={\braket{\cos^2\eta}}$, with $\eta= \theta$, $2\rho$. 

For $A_{\theta}$ we employ\cite{Zare.1988} 
\be
\label{mat-A-theta}
\cos^2 \theta = \frac{1}{3}+\frac{2}{3} {\mak D}_{0,0}^2
\ee
and we use the results from Eq. \eqref{mat-win-d} to determine the matrix elements of the Wigner matrices.

For the alignment factor $A_{2\rho}$, we first recall that
\be
\label{mat-A-rho-1}
\cos^{2}2\rho = \frac{1}{2} + \frac{1}{2}\cos 4\rho \;.
\ee
If we then use the basis Eq. \eqref{ansatz-basis}, the relevant matrix elements read 
\be
\label{mat-A-rho-2}
\left(\cos 4 \rho\right)_{\{k'_{\rho},J',k',m'\},\{k_{\rho},J,k,m\}} 
=
\frac{1}{2}\left(\delta_{k_{\rho}',k_{\rho}+4} + \delta_{k_{\rho}',k_{\rho}-4}\right) \delta_{J',J} \delta_{k',k}\delta_{m',m} \;.
\ee


\subsection{On our code}
\label{app:code}

As the mechanism we study is composed of two steps, we are able to decompose our numerical code into two (almost) independent parts as well. Consequently, we have created 
two separate programs, one for calculating the adiabatic alignment and one for simulating the non-adiabatic alignment of molecules with feasible torsion in the electronic 
round-state. Both codes can be run (almost) independently. In what follows, we explain here how the code is structured to allow the reader to judge our strategy.
To implement our code, we have used the software \textsc{MatLab}.
\thisfloatsetup{capposition=beside,capbesideposition={inside,bottom},floatwidth=10cm}
\begin{figure}[tb!]
\vspace*{1\baselineskip plus 0.125\baselineskip minus 0.075\baselineskip}

\centering{\includegraphics[width=10cm]{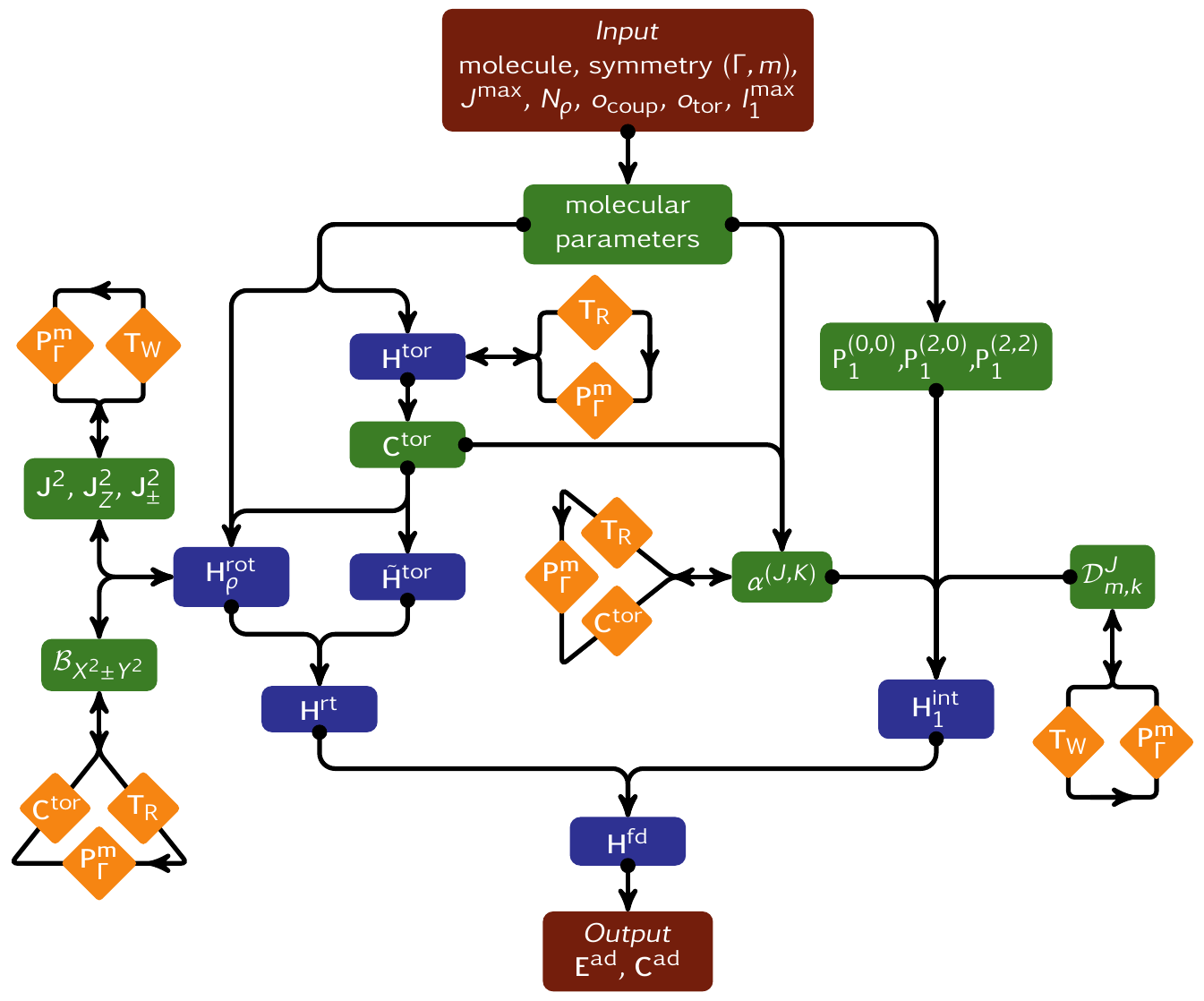}}

\caption{An illustration of our code for calculating the adiabatic alignment; see text for a detailed description.}
\label{Code-ad}
\end{figure}
%

How we calculate the pendular states numerically, we show diagrammatically in Fig. \ref{Code-ad}. First, the molecule, the irreducible representation of the initial state 
$\Gamma$, the symmetry of $m$ (\textit{i.e.} even or odd), and the size of the basis set, determined by $J^{\rm max}$ and $N_{\rho}$, has to be 
specified. The parameter $o_{\rm tor}$ specifies the energy above which no more torsional states are taken into account (see below); it is a multiple of the barrier height 
$V_{\rm B}=\max(E_0(\rho))-\min(E_0(\rho))$. The parameter $o_{\rm coup}$ defines the order at which expansions Eqs. \eqref{exp-rot-konst} are truncated.

In a second step, the molecular data is called. The data for the rotational constants ${\mak A}$, ${\mak B}$, the parameter describing the torsional potential $V_{\rm B}$, 
$V_0^{\rm tor}$, ${\mak V}_i$ ($i=0,...,7$), and the parameter for the polarizabilities $\alpha_0^{(J,K)}$ and ${\mak P}_i^{(J,K)}$ ($i=0,...,2$) are stored an external 
subroutine called ``molecular parameters''. 

Afterwards, the matrix representation of the field-free Hamiltonian $\hat{H}^{\rm rt}$, Eq. \eqref{h-rt}, is calculated. We begin with setting up the matrix ${\bm H}^{\rm tor}$
in the complex free rotor basis, see Eq. \eqref{eig-plan-rot}; we use Eq. \eqref{mat-ele-htor-frei} to calculate its elements. We then transform ${\bm H}^{\rm tor}$ to the real 
free rotor basis according to Eq. \eqref{sym-bas-tor}, before we project out the states of the irreducible representation $\Gamma$ and $m$-symmetry, see Table IV of the work of 
\citeauthor{Merer.1973} for the conditions for $J,K$ and $K_{\rho}$.\cite{Merer.1973}
Next, we calculate the eigenstates for the pure torsion and obtain the eigenvector matrix $\bm C^{\rm tor}$, which we use to calculate matrix representation of the Hamiltonian 
for the pure torsion written in its eigenbasis, $\tilde{\bm H}^{\rm tor}$. The size of the torsional basis is steered by the parameter $o_{\rm tor}$; all basis states having a 
higher eigenenergy than $o_{\rm tor}\cdotp V_{\rm B}$ are discarded.

To calculate the matrix representation of $\hat{H}^{\rm rot}_{\rho}$, we first set up the matrix representation of the functions $\mak B_{X^2\pm Y^2}$, Eq. 
\eqref{B-const-rho}, in the complex free rotor basis, see Eq. \eqref{eig-plan-rot}. To explicitly calculate $\mak B_{X^2\pm Y^2}$, we use Eqs. \eqref{exp-rot-konst}; the 
expansion is truncated at order $o_{\rm coup}$. We then (i) change to the real free rotor basis according to Eq. \eqref{sym-bas-tor}; (ii) project out the states of the irreducible 
representation $\Gamma$ and $m$-symmetry; and (iii) transform to the torsional eigenbasis. Simultaneously, we calculate the matrix representation of  the operators 
$\hat{J}^2$, $\hat{J}^2_Z$ and $\hat{J}^2_{\pm}$ in the basis Eq. \eqref{sk-eig-fun} using Eqs. \eqref{mat-ele-rot}. We transform the resulting matrices to the Wang basis, 
Eq. \eqref{sym-bas-rot}, and project out all states of the irreducible representation $\Gamma$ and $m$-symmetry. The final form of the matrix $\bm H^{\rm rt}_{\rho}$ we 
obtain by calculating the direct products of the matrix representations of $\mak B_{X^2\pm Y^2}$, $\hat{J}^2$, $\hat{J}^2_Z$ and $\hat{J}^2_{\pm}$ according to Eq. 
\eqref{h-rot}. Calculating the matrix representation of $\hat{H}^{\rm rt}$ in the symmetry-adapted basis according to Eq. \eqref{h-rt-voll} completes the calculation of the 
field-free rotational-torsional Hamiltonian.

To obtain the matrix representation for the interaction Hamiltonian $\hat{H}^{\rm int}_1$, we first calculate the effective pulse strengths ${\tt P}^{(J,K)}_1$ according to Eq. 
\eqref{P-ad}. Subsequently, (i) we set up the matrix representations of the polarizabilities $\alpha^{(J,K)}$  in the complex free rotor basis, Eq. \eqref{eig-plan-rot}, using 
Eqs. \eqref{mat-ele-alph}; (ii) we transform the resulting matrices to the real basis Eq. \eqref{sym-bas-tor}; and (iii) we project out every state having the right symmetry 
($\Gamma$, $m$). Accordingly, we first calculate the matrix representation of the Wigner-matrices in the symmetric-top basis Eq. \eqref{sk-eig-fun} using Eq. 
\eqref{mat-win-d}, and transform it into the symmetry-adapted basis. Then, we calculate the matrix representation of $\hat{H}^{\rm int}_1$ according to Eqs. \eqref{h-int-sce} 
and \eqref{alpha-zz}.

In the last step, we calculate the matrix ${\bm H}^{\rm fd}={\bm H}^{\rm rt}+{\bm H}^{\rm int}_1$ and diagonalize it. As a result, we obtain the adiabatic eigenenergies 
$\bm E^{\rm ad}$ and eigenvector matrix $\bm C^{\rm ad}$.

\thisfloatsetup{capposition=beside,capbesideposition={inside,bottom},floatwidth=8cm}
\begin{figure}[tb!]
\vspace*{1\baselineskip plus 0.125\baselineskip minus 0.075\baselineskip}

\centering{\includegraphics[width=8cm]{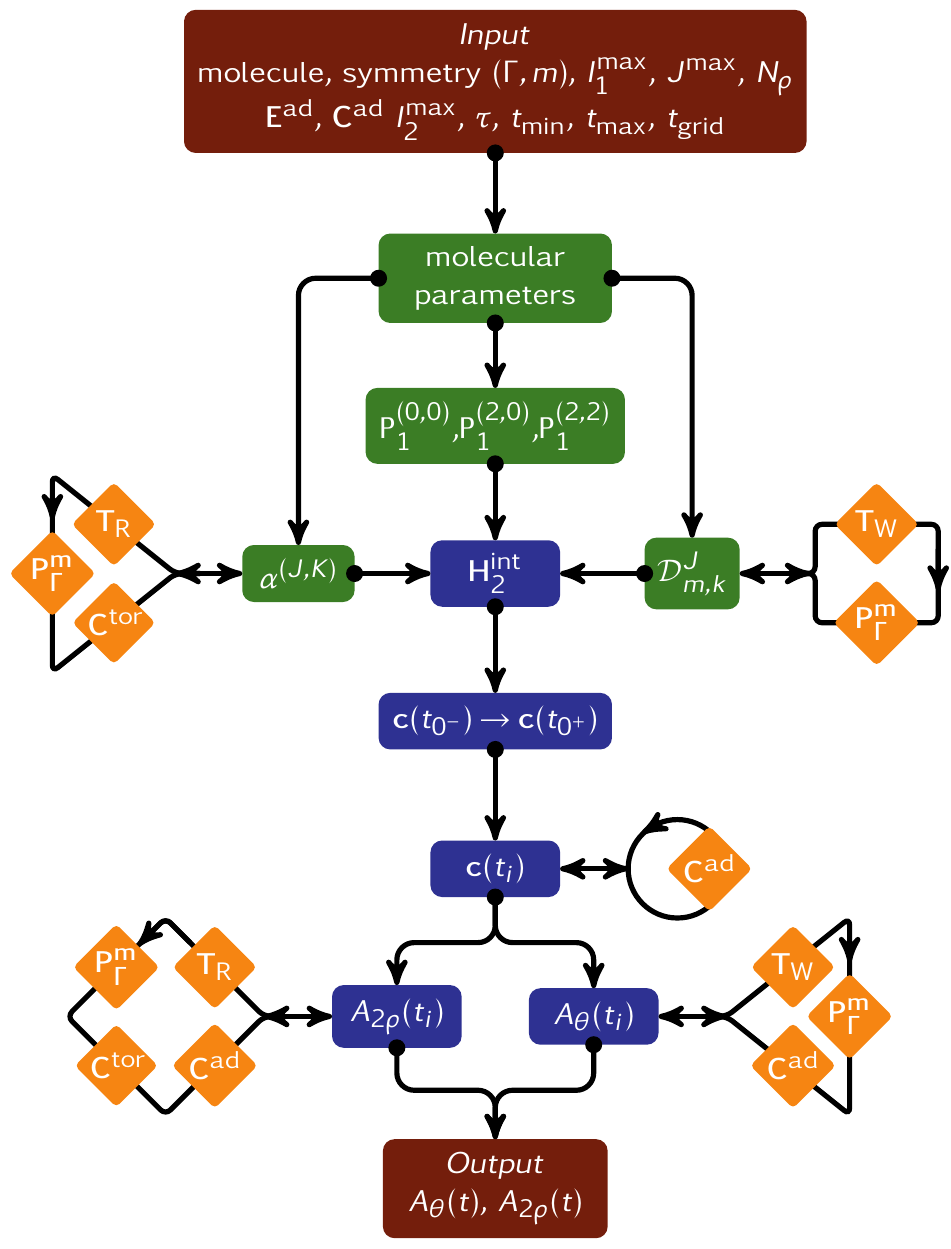}}
\caption{An illustration of our code for calculating the alignment factors after excitation with the second laser pulse; see text for a detailed description.}
\label{4D-code-nad}
\end{figure}


To calculate the impulsive alignment due to the second laser pulse, we begin with specifying the molecule, the symmetry ($\Gamma$, $m$), and the intensity of the first laser 
pulse $I^{\rm max}_1$. \textsc{MatLab} then loads the file generated by the code used for calculating the adiabatic alignment, see above.. The 
file contains the parameters $J^{\rm max}$ and $N_{\rho}$, which specify the basis set size, and the adiabatic energies $\bm E^{\rm ad}$ and the pendular states 
$\bm C^{\rm ad}$ in the symmetry-adapted basis. In case no adiabatic field is applied, $\bm E^{\rm ad}$ and $\bm C^{\rm ad}$ correspond to the field-free eigenenergies 
and eigenvectors, respectively. As input is furthermore required: the strength of the second laser pulse $I^{\rm max}_2$, the pulse length $\tau$, the start and end point of the 
propagation $t_{\rm min}$ and $t_{\rm max}$, respectively, and the size of the time-grid $t_{\rm grid}$.

After calling the molecular parameters $\alpha_0^{(J,K)}$ and ${\mak P}_i^{(J,K)}$ ($i=0,...,2$), the effective interaction strengths ${\tt P}^{(J,K)}_2$ are calculated 
according to Eq. \eqref{P-nad}. Subsequently, we calculate the matrix representation of the interaction $\hat{H}^{\rm int}_2$, \textit{c.f.} Eqs. \eqref{h-int-sce} and 
\eqref{alpha-xx}; it works completely analogues to calculating $\hat{H}^{\rm int}_1$, see above.

Next, we calculate the expansion coefficients of the wave packet at the end of the pulse according to Eq. \eqref{psi-t0+-num}. It is the most demanding step in terms of 
memory, as \textsc{MatLab} is not able to calculate the matrix exponential in sparse form. It is therefore unavoidable to use symmetry within all calculations.

To calculate the coefficients at time-step $t_i$, we solve Eq. \eqref{wp-aus-nad} numerically. Therefore, we need to transform every quantity of interest to the pendular state
basis, using the matrix $\bm C^{\rm ad}$. Once we obtain the coefficients ${\bm c}(t_i)$, we calculate the expectation values $A_{\theta}(t)$ and $A_{2\rho}(t)$. The matrix 
representation of the torsional alignment factor, we set up first in the complex free rotor basis, see Eq. \eqref{eig-plan-rot}, using Eqs. \eqref{mat-A-rho-1} and 
\eqref{mat-A-rho-2}. We then transform the matrix to the symmetry-adapted basis, see Eq. \eqref{sym-bas-tor}, and delete all states with wrong symmetry. As a last step, we 
transform the matrix representation of $\cos^2(4\rho)$ to the pendular states basis. Analogously, we calculate $A_{\theta}(t)$ first in the in the symmetric-top basis Eq. 
\eqref{sk-eig-fun} using Eq. \eqref{mat-win-d}, transform it to the symmetry-adapted basis, and use ${\bm C}^{\rm ad}$ to obtain $\cos^2\theta$ written in the pendular 
state basis.

Finally, we obtain the alignment factors $A_{\theta}$ and $A_{2\rho}$ as a function of time. They are the output of the code.


\section*{Acknowledgements}

We acknowledge support by the US Department of Energy (Award No. DE-FG02-04ER15612) and the Deutsche Forschungsgemeinschaft (project LE 2138/2-1 and GR
4508/1-1). 

This research was supported in part through the computational resources and staff contributions provided for the \textsc{Quest} high performance computing facility at 
Northwestern University, which is jointly supported by the Office of the Provost, the Office for Research, and Northwestern University Information Technology.

For help and discussions on the quantum chemical calculations, we thank Dr. Partha Pal. For proposing suitable candidates for our study, we are grateful to Dr. Benjamin Ashwell.
We acknowledge the support of Dr. Joshua Szekely,  Dr. Partha Pal and Mr. Tom Purcell when familiarizing with the \textsc{Quest} facility. We thank Dr. Jean Christophe Tremblay 
for a critical discussion on the relation of the rotational-torsional coupling and the convergence of our calculations, which has encouraged us to rethink our conclusions. Finally, we 
are grateful to Mr. Tom Purcell and Dr. Erik Hoy for proofreading our manuscript.

\renewcommand\refname{\sf\bfseries\Large References\vspace*{-0.15\baselineskip}}
\singlespacing
\renewcommand{\bibfont}{\normalfont\footnotesize}
\addcontentsline{toc}{section}{{}{\sf\bfseries References}}
\printbibliography

\end{document}